\documentclass[final,1p,times]{elsarticle}

\usepackage{amssymb}
\usepackage{subfigure}
\usepackage{subfigmat}
\usepackage{color}
\usepackage{makecell}
\usepackage{float}					
 \usepackage{amsthm}
 \usepackage{amsmath}
\usepackage{multirow}
 \usepackage{tikz}
 \usepackage{pgfplots}
 \usepackage{lipsum}
 \usepackage{caption}
\usepackage{xfrac}	
\usepackage[toc,page]{appendix}

\usepackage{ulem}

\journal{}

\begin{document}

\begin{frontmatter}



\title{Calibration of projection-based reduced-order models \\ for unsteady compressible flows}

\author[label1]{Victor Zucatti}
\ead{victor.zucatti@gmail.com}
\author[label1]{William Wolf}
\ead{wolf@fem.unicamp.br}
\author[label2]{Michel Bergmann}
\ead{michel.bergmann@inria.fr}

\address[label1]{University of Campinas, Campinas, Brazil, 13086-860}
\address[label2]{Institut de Mathématiques de Bordeaux, Talence, France, 33405}

\begin{abstract}
An analysis of calibration for reduced-order models (ROMs) is presented in this work. The Galerkin and least-squares Petrov-Galerkin (LSPG) methods are tested on compressible flows involving a disparity of temporal scales. A novel calibration strategy is proposed for the LSPG method and two test cases are analyzed. The first consists of a subsonic airfoil flow where boundary layer instabilities are responsible for trailing-edge noise generation and the second comprises a supersonic airfoil flow with a transient period where a detached shock wave propagates upstream at the same time that shock-vortex interaction occurs at the trailing edge. Results show that calibration produces stable and long-time accurate Galerkin and LSPG ROMs for both cases investigated. In order to reduce the computational costs of the LSPG models, an accelerated greedy missing point estimation (MPE) algorithm is employed for hyper-reduction. For the first case investigated, LSPG solutions obtained with hyper-reduction show good comparison with those obtained by the full order model. However, for the supersonic case the transient features of the flow need to be properly captured by the sampled points of the accelerated greedy MPE method. Otherwise, the dynamics of the moving shock wave are not fully recovered.
The impact of different time-marching schemes is also assessed and, differently than reported in literature, Galerkin models are shown to be more accurate than those computed by LSPG when the non-conservative form of the Navier-Stokes equations are solved. For the supersonic case, the Galerkin and LSPG models (without hyper-reduction) capture the overall dynamics of the detached and oblique shock waves along the airfoil. However, when shock-vortex interaction occurs at the trailing-edge, the Galerkin ROM is able to capture the high-frequency fluctuations from vortex shedding while the LSPG presents a more dissipative solution, not being able to recover the flow dynamics.
\end{abstract}


\end{frontmatter}


\section{Introduction}

The higher computational power achieved in the last fifty years allowed the application of time-accurate numerical simulations of complex engineering problems. However, despite the improvement in computer performance, accurate numerical solutions of unsteady flows are still costly. In such problems, high resolution numerical schemes are typically employed to resolve the broad range of spatial and temporal scales. On one hand, small time steps are required to capture the higher frequencies of the flow. On the other hand, simulations need to be carried out for long periods to obtain meaningful statistics related to the low frequencies.

Direct numerical simulation (DNS), large eddy simulation (LES) and detached eddy simulation (DES) are the typical high-fidelity methodologies applied in the study of unsteady flows. While DNS resolves all spatial and temporal scales associated with the flow, LES is able to resolve the larger, more energetic scales, modeling the smaller, more isotropic and universal scales of turbulence. The computational cost is reduced in DES since it combines LES with a Reynolds-averaged Navier-Stokes (RANS) approach to solve flows over more complex configurations. All these methodologies are associated with high computational costs, especially in applications of low Mach number flows involving propagation of acoustic waves \cite{wolf_azevedo_lele_2012, ricciardiAIAA2018} or supersonic flows with shock-turbulence interaction \cite{olsonPoF2013}.

It is in this context that reduced-order models (ROMs) stand out when compared to conventional methods used for computational fluid dynamics. The application of ROMs allows the construction of simpler models by dimensionality reduction of the problem, what leads to lower cost simulations \citep{Bergmann_01}. It should be clear that ROMs will not replace traditional CFD methods, but improve the arsenal of tools available for solving complex engineering problems. Reduced-order models must be stable and accurate for long-time integration and they should be able to recover the main physical features of the unsteady flows investigated. Models with quick turnaround solution find application in preliminary design, optimization and flow control, to name a few. In recent years, ROMs have attracted the attention from mathematicians, physicists and engineers interested in the solution of complex non-linear dynamical systems such as those found in fluid flows \cite{Amsallem_02,Kevin02_comp,Bergmann_05,grimberg2020stability}. 

Projection of the flow governing equations into a low-dimensional subspace is probably the most widely found class of ROMs in the literature. Among these methods, we mention Galerkin \citep{Rowley2004} and least-squares Petrov-Galerkin (LSPG) projections \citep{Kevin01}. Succinctly, order reduction is made possible by first extracting a low-dimensional basis from data previously collected using a full order model (FOM), for instance, DNS, LES or experiments. This step is important to feed the ROM and obtain physical insight of the problem. The low-dimensional subspace is usually obtained via proper orthogonal decomposition (POD) \cite{Sirovich_snap1, Bergmann_02}. Finally, a system of non-linear ordinary differential equations with fewer degrees of freedom is obtained in the projection step while tying the problem to physical grounds.

Recently, data-driven ROMs based on regression have also been gaining attention \cite{Karthik2018, Wan2018, Vlachas2018} and unsteady flow problems have been successfully modeled with techniques based on sparse regression \cite{Brunton3932} and deep neural networks \cite{lui_wolf_2019}. Most of these methods do not impose any constraints from the governing equations, differently from Galerkin-type methods. It is shown in \cite{zucatti_01} that projection-based ROMs overcome regression methods for unsteady flows involving non-periodic patterns. In such cases, the regression schemes suffer from overfitting while the projection-based ROMs are able to represent the physical features from chaotic flows.

Stability and accuracy issues are common problems when designing reduced-order models, being well documented in literature \cite{Rowley2004,Cazemier1998,Noack_01_pressure}. Typically, POD-basis truncation is pointed out as the culprit since it acts as a filter that eliminates high-frequency modes responsible for energy dissipation. However, the wavenumber interactions appear due to the non-linear convection operators which are responsible for the creation of high frequencies and the growth of instabilities. This phenomenon is also a problem in full order models such as LES, which require subgrid scale modeling to represent filtered small-scale flow structures. A large number of methods have been proposed in the last 20 years to overcome such issues and render ROMs more robust to complex engineering problems.

Artificial viscosity models are probably the most popular methods to overcome unstable ROMs since they can be directly adapted from the CFD community. This class of closure models is based in adding the lacking dissipation effects of truncated POD modes using corrective terms in the dynamical system \cite{bergmann_enablers}. The simplest model considers a change in the viscosity coefficient that impacts equally on all POD modes. This idea can be easily adapted so the viscosity can impact differently on each individual mode; for example, Smagorinsky models introduce an artificial viscosity that is variable in both space and time \cite{iliescu_01_burgers}. Overall, little overhead is added to the ROM although the dissipation intensity is a free parameter to be not always easily determined.



Adopting different inner products may also be a solution. According to \cite{iollo_01}, the importance of the smaller scales can be strengthened adopting an $H^1$ Sobolev norm. Nonetheless, this approach also has a free parameter to be defined. Studies comparing different physics-based closure models are not common, but can be found for the simulation of the one-dimensional Burgers equation \cite{iliescu_01_burgers} and the 3D flow past a cylinder at $Re = 1000$ \cite{iliescu_02_turbulent}.

Physics-based closure models may show inadequate performance, especially when modeling complex problems that require several POD modes. In some cases, the higher POD modes may not be well-resolved and there is no benefit in adding further modes to the ROM. In the previous case, the ROM can show an inferior performance if too many modes are used. Errors from the numerical schemes should also be taken into account: errors in derivatives and integration, discrete grid sampling, and non-satisfied boundary conditions are examples of common sources of error when developing ROMs. Unsurprisingly, these can be specially tragic when the flows involve a broad range of spatial and temporal scales. The importance of high-order schemes for numerical derivation was pointed out in a previous study of reduced-order modeling for convective heat transfer \cite{zucatti_01}. 

A different way to tackle the problem would be to model ROM inconsistencies using a global black box approach instead of trying to account for error sources individually. The ROM should be capable of recovering the temporal modes in the training window, and hopefully beyond, but this can be a challenging task even for simple flow configurations. This can be imposed by minimizing the error between the POD and ROM temporal modes which leads to a non-linear minimization problem \cite{galletti_01}. A cheaper linear approximation can be considered \cite{couplet2005} and it is the method used in the present study. This methodology was previously used in reduced-order modeling of unsteady flows around airfoils at low Reynolds numbers \cite{Bourguet_02,Bourguet_01}. This class of techniques is normally referred to as calibration (as opposed to closure) and further technical details will be discussed in section \ref{section_calibration_lls}.

The performance of different calibration methods is studied in \cite{favier_these,favier_01} for a cylinder wake flow at $Re = 200$ using six POD modes. For this case, linear least squares was shown to be $3000$ times cheaper than non-linear optimization which was, however, slightly more accurate. In the same reference, a second problem was investigated consisting of a separated flow around an ONERA-D airfoil at $Re = 10^5$ where a $60$-mode POD basis was obtained from PIV snapshots. The initially unstable ROM was stabilized using the linear least-squares methodology, but the model results were not accurate. Moreover, the iterative non-linear optimization could not provide a stable solution for this case.

An adaptation of the energy conserving sampling and weighting (ECSW) method \cite{ecws} was recently presented for Galerkin and LSPG ROMs. The method imposes conservation by element weighting in order to compensate for missing energy contributions and it was applied to solve convection-dominated problems \cite{grimberg2020stability, grimberg_scitech}. In \cite{grimberg_scitech}, instead of using the POD temporal modes, data calibration was accounted for by directly using a subset of the FOM snapshots also used in the POD-basis construction.
Another alternative, grounded in differential geometry, is able to model the effects of truncated modes by a minimal rotation of the projection subspace \cite{Balajewicz2016}. This method avoids adding supplementary terms to the systems of differential equations. Although this approach was applied to a non-conservative compressible Navier-Stokes framework, showing both stable and accurate results, the optimal POD basis representation is lost. Furthermore, minimization of the error between the POD and ROM temporal modes is also enforced.

%

Imposing the error minimization of temporal modes as a constraint may be undesirable for a large number of applications, such as flow control and many-query problems, because in these applications one may desire to have the existing features of the POD spatial modes with a different temporal evolution. In general, ROMs are built from snapshots of a single parameter space realization. Therefore, physical structures of different parameters may (unsurprisingly) be ill-represented or absent. Fragility to parameter variation is well-documented \cite{bergmann_enablers, Amsallem_01} and possible solutions usually involve POD-basis interpolation techniques \cite{Amsallem_01}. This being said, calibration tailored for a specific spatial and temporal evolution would most probably  be inadequate when applied in a different context and, for this reason, we focus on calibration for accurate long-time prediction.

In this work, calibration of ROMs is assessed for projection-based methods and we present a new calibration approach for the LSPG scheme. In order to reduce the computational costs of this method, hyper-reduction is applied. Calibration of Galerkin and LSPG ROMs is tested for the compressible flow past an airfoil where boundary layer hydrodynamic instabilities lead to secondary tones in the acoustic radiation. Then, the methods are tested for a supersonic transient flow past an airfoil where the performance of calibrated ROMs is assessed. The comparison of calibrated Galerkin and LSPG ROMs for unsteady compressible flow problems involving a disparity of temporal scales is one of the contributions of this work together with new insights on the use of hyper-reduction for transient problems.


\section{Reduced-order modeling}

\subsection{Proper orthogonal decomposition}
\label{subsection:POD}

In the ROMs studied in this work, proper orthogonal decomposition (POD) is applied to compute low-dimensional subspaces of specific volume, velocity and pressure fields given by $\zeta$, $u$, $v$ and $p$, respectively. These variables are chosen because the flows of interest are compressible and more details are provided in Section \ref{NS:equations}. The unsteady flow fields can be decomposed as follows
\begin{equation} 
	\mathbf{q} (\mathbf{x},t) = \mathbf{\bar{q} (x)}  + \sum_{i = 1}^{N} \mathbf{\Phi_i(x) a_i(t)}
	\label{velocity} \mbox{ ,}
\end{equation}

\noindent
where $\mathbf{q}=\left\{\zeta,u,v,p\right\}^{\top}$, $\mathbf{\bar{q} (x)}$ is the mean flow, $ \mathbf {\Phi_i} $ are the orthonormal spatial eigenfunctions, $ \mathbf {a_i} $ represent the temporal modes and $\left\{\cdot\right\}^{\top}$ is the transpose of $\left\{\cdot\right\}$. The parameter $N$ is the number of data sets extracted from the numerical simulation and $i$ represents the mode index. 

The POD consists of looking for the deterministic functions $\mathbf{\Phi_i}$ that are most similar in an averaged sense to the realizations $\mathbf{q} (\mathbf{x},t)$. An alternative technique introduced in \cite{Sirovich_snap1} is adopted here and the resulting constrained optimization problem reduces to the following Fredholm integral eigenvalue problem
\begin{equation}
    \int_{0}^{T} C_{ij} \, a_i (t') \, dt' = \lambda_i \, a_i (t) \mbox{ ,}
    \label{eq:fredholm}
\end{equation}
\noindent
where the temporal covariance matrix $\mathbf{C}$ is defined by
\begin{equation}
    C_{ij} = \frac{1}{T} \int_{\Omega} \mathbf{q} (\mathbf{x},t_i) \, \mathbf{q} (\mathbf{x},t_j) \, d \mathbf{x} 
    \approx \frac{1}{T} \langle \mathbf{q} (\mathbf{x},t_i) , \mathbf{q} (\mathbf{x},t_j) \rangle_{\mathbf{\Pi}}
    \mbox{ .}
\end{equation}


\noindent
In the previous equation, $\mathbf{\Pi}$ is a symmetric positive definite matrix defining the inner-product $\langle \mathbf{q} (\mathbf{x},t_i) , \mathbf{q} (\mathbf{x},t_j) \rangle_{\mathbf{\Pi}}=\mathbf{\langle q_i , q_j \rangle_{\Pi} = q_i^T \Pi q_j}$. Here, matrix $\mathbf{\Pi}$ is diagonal with non-zero elements defined as $\mathbf{\Pi_{ii} = A_i}$, where $\mathbf{A_i}$ is the area associated to the $i$-th vector element. This choice of inner-product \cite{Rowley2004, Bergmann_02} is equivalent to the quadrature rule $\int_{\Omega} \mathbf{u \, v \, dx} \approx \sum_{i=1}^{N_g} \mathbf{u_i v_i A_{i}}$, where $N_g$ is the total number of grid points. The covariance matrix $\mathbf{C}$ is symmetric positive semidefinite and, therefore, allows the use of singular value decomposition to compute the eigenvalues and eigenvectors (modes) of the POD reconstruction. Such modes are calculated so that the reconstruction is optimal in the sense of truncated mean quadratic error. The idea of writing a temporal covariance matrix comes from the fact that solution cost grows rapidly for large computational grids. This is an issue especially in multidimensional problems.

In Eq. \ref{eq:fredholm}, $a_i$ are the $i-th$ time-dependent POD eigenfunctions, also called POD temporal modes \cite{Bergmann_02}, that form an orthogonal set satisfying the condition 
\begin{equation}
    \frac{1}{T} \int_{0}^{T} a_i (t) \, a_j (t) \, dt = \lambda_i \, \delta_{ij} \mbox{ .}
\end{equation}
The associated eigenvectors $\mathbf{\Phi_i}$, also called empirical eigenfuctions or POD spatial modes, form a complete orthogonal set and are normalized so that they can verify $\mathbf{\langle \Phi_i , \Phi_j \rangle_{\Pi}} = \delta_{ij}$. These eigenvectors will be used in the Galerkin and LSPG projections to reconstruct the system of ordinary differential equations that, in turn, will determine the evolution of temporal modes. The spatial basis functions $\mathbf{\Phi_i}$ can then be calculated from the realizations $\mathbf{q_i}$ and the temporal modes $\mathbf{a_i}$ with
\begin{equation}
    \mathbf{\Phi_i} (\mathbf{x}) = \frac{1}{T \lambda_i} \int_{0}^{T} q (\mathbf{x},t) \, a_i (t) \, dt \mbox{ .}
\end{equation}

Finally, reconstruction of the fluctuation fields of specific volume, velocity and pressure can be approximated by
\begin{equation}
	\mathbf{{q}'} (\mathbf{x},t) \approx \sum_{i=1}^{M} \mathbf{\Phi_i} (\mathbf{x}) \, \mathbf{a_i (t)} 
	\mbox{ ,}
	\label{fluctuations}
\end{equation}
\noindent
where $M$ is the number of modes used in the reconstruction of fluctuation fields. In practical ROM applications, one seeks $M \ll N$. The POD method is widely used in literature and we refer to the following references for further information and applications \cite{Bergmann_01, Rowley2004, Sirovich_snap1, Bergmann_02, Noack_01_pressure}.

\subsection{Projection methods}

Let us consider the system of non-linear partial differential equations $\mathbf{F(q)}$ defined in a connected open region $\Omega \subset \mathbb{R}^{N}$ whose boundary $\Gamma$ is well defined
\begin{equation}
	\begin{cases}
		\mathbf{F(q}) = \frac{d \mathbf{q}}{dt} - \mathbf{G(q)} = \mathbf{0} \quad in \quad \Omega \\
		\mathbf{q}(t = t_0) = \mathbf{q_0} \\
		\mathbf{q = g} \quad on \quad \Gamma \mbox{ .}
	\end{cases}
	\label{dynamical_system}
\end{equation}

\noindent
In the system above, $\mathbf{q}$ is a function of space and time, and the non-linear operator $\mathbf{G(q)}$ is given by the convective and diffusive terms appearing in the mass, momentum and energy equations, herein referred to as Navier-Stokes equations. Let $\mathbf{\Phi}$ define an orthonormal basis obtained by POD. The state variable $\mathbf{q}$ is then approximated as the linear combination of this basis vector as
\begin{equation}
	\mathbf{q} \approx \mathbf{\hat{q}} = \mathbf{\bar{q}} + \sum_{i = 1}^{M} \mathbf{\Phi_i} \, \mathbf{a_i} \mbox{ ,}
\end{equation}
where the explicit dependencies on space and time are omitted for simplicity.

In general, after discretization and approximation, $ \mathbf{F} (\mathbf{q}) \approx \mathbf{R} (\mathbf{\hat{q}}) \neq \mathbf{0}$ for the physical problem being solved. A solution is sought by enforcing the residual $\mathbf{R(\hat{q})}$ orthogonality as
\begin{equation}
	\sum_{i=1}^{M} \mathbf{\langle \Psi_i , R (\hat{q}) \rangle_{\Pi} = 0} \mbox{ ,}
\end{equation}
\noindent
where $\mathbf{\Psi_i}$ is the test space. A projection method is generally called Galerkin (Petrov-Galerkin) when the test and solution bases are equal (different),  i.e., $\mathbf{\Psi = \Phi}$ ($\mathbf{\Psi \neq \Phi}$) for the following projection
\begin{equation}
	\sum_{i=1}^{M} \mathbf{\langle \Psi_i , R} (\mathbf{\bar{q}} + \sum_{j=1}^{M} \mathbf{\Phi_j a_j ) \rangle_{\Pi} = 0} \mbox{ .}
\end{equation}

Boundary conditions must be implicitly satisfied by the POD solution basis, otherwise the problem may lead to an ill-conditioned or ill-posed reduced-order model. Homogeneous Dirichlet or Neumann boundary conditions can be inherited by the spatial modes $\mathbf{\Phi}$ from the snapshot collection \cite{Bergmann_01}.

\subsubsection{Galerkin projection}

Galerkin projection is the most popular alternative for reduced-order modeling of time dependent problems. This can be attributed to its implementation simplicity and solid mathematical foundation. Applying the Galerkin projection method ($\mathbf{\Psi = \Phi}$) to Eq. \ref{dynamical_system} we obtain

\begin{equation}
	\sum_{i = 1}^{M} \mathbf{\langle \Phi_i} , \sum_{j=1}^{M} \mathbf{\Phi_j \dot{a}_j \rangle_{\Pi}} = \sum_{i = 1}^{M} \langle \mathbf{\Phi_i}, \mathbf{G} (\mathbf{\bar{q}}+\sum_{j=1}^{M} \mathbf{\Phi_j a_j) \rangle_{\Pi}} \mbox{ .}
\end{equation}
This equation can be further simplified since the functions $\mathbf{\Phi_i}$ are orthonormal. Hence, a system of ordinary differential equations arises for the temporal modes as
\begin{equation}
	\sum_{i = 1}^{M} \mathbf{\dot{a}_i} = \sum_{i = 1}^{M} \langle \mathbf{\Phi_i}, \mathbf{G} (\mathbf{\bar{q}}+\sum_{j=1}^{M} \mathbf{\Phi_j a_j )\rangle_{\Pi}} \mbox{ ,}
	\label{galerkin_ode}
\end{equation}

\noindent
with initial conditions obtained by projection of a single snapshot in the vector basis
\begin{equation}
	\sum_{i = 1}^{M} \mathbf{a_i} (t_0) = \sum_{i = 1}^{M} \langle \mathbf{\Phi_i (x)}, \mathbf{q_0 \rangle_{\Pi}}
	\label{galerkin_ode_a0} \mbox{ .}
\end{equation}

The previous system of ordinary differential equations represents the ROM associated to the FOM and can be solved using a time-marching method. The right-hand side of Eq. \ref{galerkin_ode} should not scale with the full-order model so as to achieve reduced computational cost. Following the POD-Galerkin approach, the ROM obtained for the non-conservative form of the compressible Navier-Stokes equations (see details in Section \ref{NS:equations}) can be written as
\begin{equation} 
	\sum_{i = 1}^{M} \mathbf{\dot{a}_i} = \sum_{i=1}^{M} \mathbf{e_i} + \sum_{i,j=1}^{M} \mathbf{A_{ij}} \mathbf{a_j}  + \sum_{i,j,k=1}^{M} \mathbf{N_{ijk}} \mathbf{a_j a_k} \mbox{ ,}
	\label{galerkin_ode_cav}
\end{equation}
where the ODE coefficients $\mathbf{e}$, $\mathbf{A}$ and $\mathbf{N}$ can be found in \ref{appendix_galerkin}. It is worth mentioning that these coefficients are time-independent and, thus, need to be calculated only once, in a pre-processing step.

\subsubsection{Least-squares Petrov-Galerkin}

Test and trial bases are different when a Petrov-Galerkin approach is used ($\mathbf{\Psi \neq \Phi}$). In this work, we employ the least-squares Petrov-Galerkin (LSPG) method that has shown promising results for reduced order modeling \cite{Kevin01,Quarteroni2016}. According to these references, the Petrov-Galerkin approach improves robustness compared to the Galerkin technique, despite the absence of a priori stability guarantees for general non-linear problems \cite{Kevin02_comp}. The least-squares variant is obtained by solving the fully discrete residual (i.e., the residual after temporal and spatial discretization) minimization problem at each $n$-th time-step as
\begin{equation}
	\operatorname*{minimize}_{\mathbf{\hat{q}}} f^n  \mathbf{(R(\hat{q}))}
	\mbox{ .}
	\label{eq:regression_obj}
\end{equation}

\noindent
The objective function $f^n \mathbf{(R(\hat{q}))}$ is defined in the following special form
\begin{equation}
    f^n  \mathbf{(\hat{q})} = 
    \frac{1}{2} \mathbf{ \| R^n (\hat{q}) \|_{\Pi}^{2}} =
    \frac{1}{2} \mathbf{\langle R^n (\hat{q}), R^n (\hat{q}) \rangle_{\Pi}}
    \mbox{ ,}
\end{equation}

\noindent
or equivalently
\begin{equation}
     \sum_{i=1}^{M} \mathbf{a_i^{n}} = \ arg \ min \ \| \mathbf{R} (\mathbf{\bar{q}} + \sum_{i=1}^{M} \mathbf{\Phi_i a_i^n)} \|_{\mathbf{\Pi}}^{2}
     \mbox{ ,}
     \label{eq:regression_arg_min}
\end{equation}
where the initial conditions are also given by Eq. \ref{galerkin_ode_a0}.

Optimality conditions are derived from Taylor's theorem and can be determined by examining the gradient $\nabla f^n \mathbf{(\hat{q})}$ and Hessian $\Delta f^n \mathbf{(\hat{q})}$ matrices \cite{book_nonlinearOpt}. The derivatives of $f^n  \mathbf{(\hat{q})}$ can be expressed in terms of the Jacobian $\mathbf{J(\hat{q}) = \frac{\partial R}{\partial \hat{q}}}$. Applying the first-order necessary condition $\mathbf{\nabla}  f^n(\mathbf{\hat{q}}) = 0$ yields
\begin{equation}
    \nabla f(\mathbf{\hat{q}}) = 
    \mathbf{\langle J(\hat{q}) , R(\hat{q}) \rangle_\Pi} =
    \mathbf{\Bigg \langle \frac{\partial R(\hat{q})}{\partial \hat{q}} , R(\hat{q}) \Bigg \rangle_\Pi = 0}
    \mbox{ ,}
\end{equation}

\noindent
and applying the chain-rule to the previous equation we have
\begin{equation}
    \mathbf{\Bigg \langle \frac{\partial R(\hat{q})}{\partial a} \Phi, R(\hat{q}) \Bigg \rangle_\Pi = 0}
    \mbox{ .}
    \label{eq:lspg_ṕrojection}
\end{equation}
Therefore, in the LSPG method, the discrete test basis $\mathbf{\Psi}$ is given by
\begin{equation}
	\mathbf{\Psi} = 
	\mathbf{J}= 
	\mathbf{\frac{\partial \mathbf{R(\hat{q})}}{\partial \mathbf{a}} \Phi}
	\mbox{ .}
\end{equation}

Proof of the equivalence of Eqs. \ref{eq:regression_obj} and \ref{eq:lspg_ṕrojection} is available in \cite{grimberg2020stability}. Algorithms that follow the line-search framework are commonly used for solving problems such as Eq. \ref{eq:regression_obj}. The main idea in the line-search approach is to chose a direction $\mathbf{{}^{(k)} p}$ leading to a decrease of the objective function $f^n$ when moving from the current iterate $\mathbf{{}^{(k)} a_ {i}^{n}}$ to a new one $\mathbf{{}^{(k+1)} a_ {i}^{n}}$. These algorithms halt when the accuracy criteria has been satisfied or when further progress is impossible.

The steepest descent algorithm is a first-order line-search method where the steps taken are proportional to the negative gradient $\mathbf{{}^{(k)}p = - \nabla {}^{(k)}f}$. The slow first-order convergence of this method is counterbalanced by requiring only gradients. A second-order alternative is Newton's method which requires the calculation of second derivatives. When applied in the solution of Eq. \ref{eq:regression_obj}, it results in the following iterations
\begin{subequations}
    \begin{align}
        \begin{split}
            \Delta \mathbf{{}^{(k)}f {}^{(k)}p} &= - \nabla \mathbf{{}^{(k)} f} \mbox{ ,}
        \end{split}\\
        \begin{split}
            \mathbf{{}^{(k+1)}a_{i}^{n}} &= \mathbf{{}^{(k)}a_{i}^{n} + {}^{(k)}\alpha {}^{(k)}p}
            \mbox{ ,}
        \end{split}
    \end{align}
    \label{eq:newton_method}
\end{subequations}
\noindent
where $k = 1, \dots , K$ with $K$ satisfying the convergence criterion. The step length ${}^{(k)}\alpha \in \mathbb{R_{+}^{*}}$ can simply be set to unity, ${}^{(k)}\alpha = 1$, or computed using a line search in the direction $\mathbf{{}^{(k)}p}$ satisfying, for example, Wolfe or Goldstein conditions \cite{book_nonlinearOpt}. The systematic choice of the step length is crucial, since it should be cheap and significantly reduce the objective function $f$. Also, it should be noted that problems can emerge when using second derivatives, especially when the Hessian is not a symmetric positive definite matrix. 

The Gauss-Newton algorithm, an alternative to Newton's method, is capable of avoiding some, but not all, of the issues potentially emerging when using the Hessian. In this method, second derivatives are neglected leading to an approximated Hessian using only the Jacobian $\Delta f \approx \mathbf{\langle J, J \rangle_{\Pi}}$. LSPG reduced-order models combined with a Gauss-Newton solver have been widely and successfully used by \cite{Kevin02_comp,Amsallem_02,Kevin01,Kevin03}.  Applying the modified Hessian to Eq. \ref{eq:newton_method} results in the following iterations for the search direction $\mathbf{{}^{(k)} p}$
\begin{equation}
    \mathbf{\langle {}^{(k)} J , {}^{(k)} J \rangle_{\Pi} {}^{(k)}p = - \langle {}^{(k)} J , {}^{(k)}R \rangle_{\Pi}}
    \mbox{ .}
\end{equation}
However, the Gauss-Newton method does not address one of the main problems of Newton's method and iterations may fail when the Hessian is near, or exactly, rank-deficient. The Levenberg-Marquardt method is employed in this work and it considers $\Delta f \approx \mathbf{\langle J, J \rangle_{\Pi}} + \lambda \mathbf{I}$ to ensure full rank, where $\mathbf{I}$ is the identity matrix and $\lambda \geq 0$ is a scalar. This algorithm can be seen as a combination of both Gauss-Newton and steepest descent: when $\mathbf{\mathbf{\langle J, J \rangle_{\Pi}}} \gg \lambda \mathbf{I}$ the search direction is similar to the direction given by the Gauss-Newton algorithm and, when $\mathbf{\mathbf{\langle J, J \rangle_{\Pi}}} \ll \lambda \mathbf{I}$, the method is similar to the steepest descent.

A detailed theoretical and computational comparison of Galerkin and LSPG projection methods is provided by \cite{Kevin02_comp}. A number of interesting outcomes from this previous reference are outlined in the following. The LSPG solution may converge to that obtained from Galerkin projection in some situations, for example, as the time step converges to zero. Also, the regression problem from Eq. \ref{eq:regression_obj} can be linear or non-linear depending on the time integration scheme employed (explicit or implicit) and the set of equations being solved. Non-linear least-squares increases the ROM cost considerably and may become an issue. Finally, finding that the error does not decrease as the time step approaches zero, but is optimal for an intermediate value, is at the same time a surprising and inconvenient finding.
Despite these issues, the LSPG method has demonstrated in many situations to be a better alternative to the Galerkin method for general non-linear dynamical systems in spite of the higher level of complexity. As also shown in \cite{Kevin02_comp}, the method is more robust (without hyper-reduction) than Galerkin projection and it has desirable stability properties if implemented together with implicit time-marching schemes.

\subsubsection{Hyper-reduction}

Projection-based reduced-order modeling may fail to produce significant computational time gains. Problems containing strong non-linearities (i.e. non-polynomial) or non-affine parameter dependence impede pre-computation of Galerkin coefficients. Consequently, full order scaling inner-products must be systematically and consistently calculated during time integration. Similarly, the reduced POD space is not sufficient to provide low-cost to LSPG because the residual minimization problem from Eq. \ref{eq:regression_obj} is always grid-size dependent. Hyper-reduction techniques are capable of providing the supplementary approximation needed to obtain reasonable computational savings in the so-called ``reduce-then-project'' approach. Succinctly, spatial modes are reduced using a non-random sampling algorithm before projection. In some cases, pre-computation of strong non-linearities can be made possible by the use of lifting transformations \cite{willcox_05_lifting}. On one hand, the invasiveness of the additional layer of approximation introduced by hyper-reduction is avoided. On the other hand, a non-uniquely defined lifted system has to be derived and supplementary variables are introduced.

The gappy POD method \cite{Sirovich_gappy}, which was originally used in facial image reconstruction with incomplete data sets, provides the framework to the reduce-then-project procedure also used in other studies \cite{Willcox_02_gappy, Kevin01}. Within this approach, given a subset of indices $J = \{ j_1, \dots , j_s \} \subset \{ 1, \dots , N_g \}$, a solution vector $\mathbf{q}$ is approximated by shrinking the spatial modes using a mask projection matrix $\mathbf{P} = [\mathbf{e_{j_1}, \dots , e_{j_{s}}}]^T \in \mathbb{R}^{N_g \times s}$ to construct the estimated solution $\Tilde{\mathbf{q}} \approx \mathbf{P^T \Phi a}$. Here, $s$ is the number of indices retained from the original vector of size $N_g$ and $\mathbf{e_{j_k}}$ denotes the vector with a $1$ in the $j_k$-th coordinate and $0$'s elsewhere. Temporal modes $\mathbf{a}$ are then calculated by minimizing the error $\epsilon$ between the gapless solution $\mathbf{q}$ and $\Tilde{\mathbf{q}}$. Hence, the error is defined as
\begin{equation}
    \epsilon = \|  \mathbf{P^T q} - \mathbf{P^T \Phi a} \|_{\mathbf{\Pi}}^2 \,\,\mbox{ ,}
\end{equation}
\noindent
which is equivalent to the minimization problem
\begin{equation}
    \| \mathbf{I} - \langle \mathbf{\tilde{\Phi}_i,\tilde{\Phi}_j} \rangle_{\mathbf{\Pi}} \|_2^2 \,\,\mbox{ ,}
    \label{eq:gappy_min}
\end{equation}

\noindent
where $\mathbf{\tilde{\Phi}} =  \mathbf{P^T \Phi}$. Random sampling is used when no dynamical insight of the problem being solved is previously available. This approach usually leads to considerably bigger mask matrices and randomness makes reproduction of results impossible. Fortunately, this is not the case when using a POD basis. Hyper-reduction has been a very active research topic in the past two decades and, as a result, several methods have been developed and tested.
The optimal solution of Eq. \ref{eq:gappy_min} for a set of given size is an intractable combinatorial optimization problem even for relatively small problems. However, the missing point estimation (MPE) method \cite{Willcox_03_MPE} eliminates points by evaluating the condition number 
\begin{equation}
    c\left(\mathbf{\langle \Tilde{\Phi}_i, \Tilde{\Phi}_j \rangle}\right) \equiv \frac{\lambda_{max} \left(\mathbf{\langle \Tilde{\Phi}_i, \Tilde{\Phi}_j \rangle}\right)}{\lambda_{min} \left(\mathbf{\langle \Tilde{\Phi}_i, \Tilde{\Phi}_j \rangle}\right)}
\end{equation}
of the approximated identity matrix $\langle \mathbf{\Tilde{\Phi}_i , \Tilde{\Phi}_j} \rangle \approx \mathbf{I}$. This procedure is performed looping over all components until a user specified near-optimal set of points is left. Computational cost of the MPE method may quickly become prohibitive, although the method is cheaper than the optimal solution.
%

The discrete empirical interpolation method (DEIM) \cite{Saifon01} is a popular substitute to MPE \cite{Willcox_04_DEIM,navon_deim,isoz_deim,ghavamiam_deim}. This method was derived from the empirical interpolation method (EIM) \cite{maday_eim} and it first appeared as a hyper-reduction approach for equations containing strong non-linear terms. However, even though it is simple and cost effective, the DEIM has an important shortcoming: the number of interpolation points has to be at most the same as the number of POD modes used in the reconstruction. This is an obvious limiting element, especially for non-trivial evolution equations or problems using few POD modes. Hence, one may require alternative sampling methods capable of circumventing this limitation in order to obtain stable and accurate ROMs.

The accelerated greedy MPE procedure \cite{Willcox01} introduces a massive improvement to the classic MPE and, therefore, considerable time complexity reductions are possible while still using the MPE principle. Briefly, the greedy point selection is equivalent to picking the spatial mode index responsible for the largest growth in the condition number of the modified eigenvalue problem. The costly modified symmetric eigenvalue problem is not actually solved, but the index selection occurs by analyzing properties of the candidate indices. In the present work, we adopt the accelerated greedy MPE technique for hyper-reduction.





\subsection{Non-conservative compressible Navier-Stokes equations}
\label{NS:equations}

Generally, the conservative form of the compressible Navier-Stokes equations is preferred in CFD applications, especially in applications involving discontinuities. However, the non-linear rational functions of the variables impede speed-ups when applying a Galerkin projection framework because pre-computation of the ROM coefficients is impossible in the conservative form. This issue can be avoided either by applying hyper-reduction methods to cut down complexity or choosing a non-conservative formulation as shown by \cite{iollo_01}. The conservative form of the compressible Navier-Stokes equations was recently applied with hyper-reduction in the context of projection-based reduced-order models \cite{grimberg2020stability}. Hyper-reduction is unattractive and to be avoided when possible since conservation properties are lost. Moreover, the second degree polynomial non-linearity of the non-conservative equations allows for pre-computation and cost reduction without any further approximations, thus being the approach used in this study.

In this work, we solve the $2D$ non-conservative compressible Navier-Stokes equations presented by \cite{iollo_01}. They can be written as
\begin{subequations}
    \begin{align}
    \begin{split}
        \zeta_t = & \zeta (u_x + v_y) - u \zeta_x - v \zeta_y   \mbox{ ,}
    \end{split}\\
    \begin{split}
        u_t = & - u u_x - v u_y - \zeta p_x + \frac{Ma}{Re} \zeta \bigg[ \bigg( \frac{4}{3} u_x - \frac{2}{3} v_y \bigg)_x + (v_x + u_y)_y  \bigg]   \mbox{ ,}
    \end{split}\\
    \begin{split}
        v_t = & - u v_x - v v_y - \zeta p_y + \frac{Ma}{Re} \zeta \bigg[ \bigg( \frac{4}{3} v_y - \frac{2}{3} u_x \bigg)_y + (v_x + u_y)_x \bigg]   \mbox{ , and}
    \end{split}\\
    \begin{split}
        p_t = & - u p_x - v p_y - \gamma p (u_x + v_y) + \frac{\gamma Ma}{Re Pr} [(p \zeta)_{xx} + (p \zeta)_{yy}] \\
        & + \frac{(\gamma - 1) Ma}{Re} \bigg[ u_x \bigg( \frac{4}{3} u_x - \frac{2}{3} v_y \bigg) + v_y \bigg( \frac{4}{3} v_y - \frac{2}{3} u_x \bigg) + (v_x + u_y)^2 \bigg]  \mbox{ ,}
    \end{split}
    \end{align}
    \label{eq:nonconsNS}
\end{subequations}
\noindent
where $\zeta = 1 / \rho$ is the specific volume, $u$ and $v$ are the $x$ and $y$ velocity components, respectively, and $p$ is the pressure. In the set of Eqs. \ref{eq:nonconsNS}, $Pr$, $Re$ and $Ma$ denote the reference Prandtl, Reynolds and Mach numbers, respectively. Subscripts denote partial derivatives and $\gamma$ is the specific heat ratio. 

\subsection{Calibration of projection-based reduced-order models}

Calibration methods are mostly developed and applied to Galerkin-type ROMs. They provide additional terms to the generally non-linear system of ordinary differential equations arising from projecting the POD modes in the governing equations
\begin{equation}
    \mathbf{\dot{a}} = \mathbf{f(a)} \mbox{ ,}
    \label{eq:cauchy_problem}
\end{equation}
\noindent
where $\mathbf{f(a)}$ is the right-hand side of Eq. \ref{galerkin_ode_cav}. Calibration could be, in principle, applied to any ROM originating from an initial value problem. For example, calibration of data-driven ROMs \cite{Brunton3932, lui_wolf_2019} is a work in progress.
Usually, only linear terms are adopted in the calibration process, but non-linear terms can also be considered and have been used to some extent \cite{iliescu_03_calibration}. Here, constant $\mathbf{e^c}$ and linear $\mathbf{A^c}$ calibration terms will be added as
\begin{equation}
    \mathbf{\dot{a}} = \mathbf{f(a) + e^c + A^c \, a} \mbox{ ,}
\end{equation}
and details of the numerical procedure are provided in the following sections.



\subsubsection{Linear least-squares calibration}
\label{section_calibration_lls}

In a linear least-squares calibration, the goal may be to minimize the error $E_1$ between temporal modes obtained by solving the system of non-linear equations of the reduced-order model $\mathbf{a^{ROM}}$ and the original POD temporal modes $\mathbf{a^{POD}}$ in a user specified training window $0\leq t \leq T$ as
\begin{equation}
    E_1 = \sum_{i=1}^{M} \int_0^T \mathbf{\left(a_i^{POD}(t) - a_i^{ROM} (t)\right)^2} dt \mbox{ .}
\end{equation}
%
Similarly, temporal modes should also satisfy $\mathbf{\dot{a} = f(a)}$ and, hence, a second error norm $E_2$ can be defined by the ODE system and this is the preferred method used in this work
\begin{equation}
    E_2 = \sum_{i=1}^{M} \int_0^T \mathbf{\left(\dot{a}_i^{POD}(t) - f_i(a^{ROM})\right)^2} dt \mbox{ .}
\end{equation}

Linearization is obtained by enforcing $\mathbf{f(a^{ROM}) = f(a^{POD})}$ in the error norm $E_2$ \cite{couplet2005}, where $ \mathbf{\dot{a}_i^{POD}(t)}$ is computed using a high-order finite difference scheme. Suppression of the non-linear constraint to obtain an affine operator may seem extreme at first, but this can be understood as a measure of how the non-calibrated ROM alters the solution in each time step. If successful, calibration should systematically compensate for any deviations from the POD temporal modes
%
\begin{equation}
    E_2 = \sum_{i=1}^{M} \int_0^T \mathbf{\left(\dot{a}_i^{POD}(t) - f_i(a^{POD})\right)^2} dt \mbox{ .}
\end{equation}
Here, $E_2$ will be used to indicate the non-calibrated affine approximation error while $E_2^c$ will indicate the same error with constant $\mathbf{e^c}$ and linear $\mathbf{A^c}$ calibration terms included
\begin{equation}
    E_2^c (\mathbf{e^c,A^c})
    = \sum_{i=1}^{M} \int_0^T \mathbf{\left(\dot{a}_i^{POD}(t) - f_i(e^c, A^c, a^{POD})\right)^2} dt \mbox{ .}
    \label{eq:error_E2c}
\end{equation}

Calibrating the ODE system by directly minimizing the error $E_2^c$ inside the training window could fail to generalize the knowledge to longer time periods. Overfitting can be overcome by controlling the relation between the calibrated and original ROM terms. In this case, a functional $J(\mathbf{e^c, A^c,} \theta)$ can be defined as
\begin{equation}
    J(\mathbf{e^c, A^c,} \theta) = \theta \frac{E_2^c (\mathbf{e^c,A^c})}{E_2} + (1 - \theta) \frac{\mathbf{ \| e^c \|^2 + \| A^c \|^2}}{\mathbf{ \| e \|^2 + \| A \|^2}} \mbox{ ,} 
    \label{eq:func_J1}
\end{equation}
%
where the first term in the right hand side corresponds to the normalized prediction error and the second one provides the relative weight of the calibration coefficients compared to the coefficients of the original ROM. The parameter $\theta \in [0 , 1]$, and values of $\theta$ close to $0$ add more importance to the original ROM while $\theta$ close to $1$ add a higher weight to the prediction quality of the calibrated model along the training window. Other functional choices, favoring for example constant terms, could be explored. In this work, we use the following functional form 
\begin{equation}
    \tilde{J} (\mathbf{e^c,A^c},\theta) = E_2^c(\mathbf{e^c, A^c}) + \tilde{\theta} (\mathbf{\| e^c \|^2 + \| A^c \|^2})
    \label{eq:calibration}
\end{equation}
\noindent
with
\begin{equation}
    \tilde{\theta} = \frac{1 - \theta}{\theta} \frac{E_2}{\mathbf{\| e \|^2 + \| A \|^2}} \mbox{ .}
\end{equation}

Minimizing the functional $\tilde{J}$ gives rise to a linear system. For clarity, the different calibration coefficients are grouped such that $\mathbf{K^c = [e^c \; A^c]}$ and the enriched temporal mode vector $\mathbf{a^{*}} = [1 \; a_1 \; \dots a_{M}] = [1 \;  \mathbf{a}]$, so that the functional to be minimized is written as
\begin{equation}
    \tilde{J} (\mathbf{K^c}, \theta) = \sum_{i=1}^{M} \int_{0}^{T} \left(\dot{a}_i (t) - f_i (a^{POD}) - \sum_{j=1}^{M + 1} K_{ij}^{c} a_j^{*} (t) \right)^2 dt + \tilde{\theta} \mathbf{\| K^c \|^2} \mbox{ ,}
\end{equation}
\noindent
where $f_i$ is the general non-linear system of ordinary differential equations defining the reduced-order model being calibrated.

For a given parameter $\theta$, the optimality condition $\partial \tilde{J} / \partial K_{ij}^c = 0$ leads to solving $M$ linear systems of size $M + 1$ or, for $i = 1, \dots , M$
\begin{equation}
    \mathbf{D^T  K_i^c = b^i} \mbox{ ,}
\end{equation}
\noindent
where $\mathbf{K_i^c}$ is the $i^{th}$ row of $\mathbf{K^c}$,
\begin{equation}
    D_{ij} = \int_{0}^{T} a_i^{*} (t) \, a_j^{*} (t) dt + \tilde{\theta} \delta_{ij}
\end{equation}
\noindent
and
\begin{equation}
    b_i^j = \int_{0}^{T} \left(\dot{a}_j (t) - f_j (t)\right) a_i^{*} (t) dt \mbox{ .}
\end{equation}
Matrix $\mathbf{D}$ is calculated only once while vector $\mathbf{b^i}$ must be evaluated for each mode during the calibration phase. Further details and explanations can be found in \cite{bourguet_these}.

\subsubsection{Calibration of least-squares Petrov-Galerkin}

The LSPG method should provide better stability properties compared to Galerkin projection. However, the method still lacks a priori stability and accuracy guarantees, especially when combined with hyper-reduction. Thus, the technique could possibly benefit from calibration but it has to be tailored to fit the approach developed in Section \ref{section_calibration_lls} because it solves the fully discrete residual minimization problem. In other words, the problem must be applicable in the error norm given by Eq. \ref{eq:error_E2c}.

This inconvenience can be worked around by extracting a function $\mathbf{f(a)}$ equivalent to the right-hand side of Eq. \ref{eq:cauchy_problem} from the LSPG formulation. Instead of solving the fully discrete residual minimization problem, the trick is to solve the spatial-discrete residual minimization version of the problem for the time derivative of the temporal mode $\mathbf{\dot{a} = f(a)}$ at each $n$-th time-step using the POD trial basis projection with hyper-reduction 
\begin{equation}
     \mathbf{f (a_{POD}^{n})} = \ arg \ min \ \| \mathbf{R (a_{POD}^{n})} \|_{\mathbf{\Pi}}^{2} \mbox{ .}
     \label{eq:regression_argmin_cali}
\end{equation}

Although hyper-reduction is non-compulsory in Eq. \ref{eq:regression_argmin_cali} when evaluating $\mathbf{f (a_{POD}^{n})}$, it is strongly encouraged for a cheaper calibration procedure. After all values of $\mathbf{f (a_{POD}^{n})}$ are determined, calibration terms can be easily calculated according to Section \ref{section_calibration_lls}. Finally, constant $\mathbf{e^c}$ and linear $\mathbf{A^c}$ calibration terms are integrated systematically afterwards to the model at each time-step in a predictor-corrector fashion as
\begin{subequations}
    \begin{align}
        \begin{split}
            \mathbf{\tilde{a}} = & \ arg \ min \ \| \mathbf{R (a^{n})} \|_{\mathbf{\Pi}}^{2} \mbox{ ,}
        \end{split}\\
        \begin{split}
            \mathbf{a^{n+1}} = & \ \mathbf{\tilde{a}} + \int_{\Delta t} (\mathbf{e^c + A^c a^n}) dt \mbox{ .}
        \end{split}
    \end{align}
    \label{eq:calibrationLSPG}
\end{subequations}

Calibration could be particularly beneficial to systematically accounting for the additional errors introduced by hyper-reduction. Moreover, smaller mask matrices (i.e., a more aggressive hyper-reduction) could be envisioned when combined with calibration, what would reduce both the computational costs and model errors.

\section{Results and Discussion}

In this section, we analyze the performance of Galerkin and LSPG methods with and without calibration for generating reduced-order models of compressible flows. Both ROM approaches were previously tested in \cite{zucatti_01} for the solution of incompressible flows involving convective heat transfer. In this previous reference, the LSPG method was able to present accurate solutions with an aggressive hyper-reduction that used only $0.05 \%$ of the grid nodes. In this work, ROM calibration is tested on compressible flows involving  wave propagation. Firstly, 
we study the flow past a NACA0012 airfoil at freestream Mach number $M_{\infty}=0.3$ and Reynolds number $Re_c=100{,}000$. In this case, boundary layer hydrodynamic instabilities lead to trailing-edge noise generation that, in turn, excites a feedback loop mechanism. Then, an assessment of ROM calibration is presented for a supersonic flow over a NACA0012 airfoil at $M_{\infty}=1.2$ and $Re_c=80{,}000$. This test case involves the transient motion of a bow shock formed at the airfoil leading edge and a shock-vortex interaction occurring at the trailing edge. 


The full order models considered in this work are obtained by numerical simulation of the two-dimensional compressible Navier-Stokes equations. Length scales and flow quantities are non-dimensionalized by the airfoil chord, freestream density and velocity. Numerical results are obtained by a sixth-order accurate compact finite difference scheme \cite{Nagarajan2003} for spatial discretization. The method employs a staggered grid formulation and is able to capture shock waves for low Mach number supersonic flows ($M_{\infty}<1.3$) without the explicit addition of a shock capturing scheme. A hybrid implicit-explicit framework is employed for time marching of the equations through a combination of a third-order Runge Kutta scheme with a modified Beam-Warming implementation \cite{Nagarajan2004}. All flow simulations are conducted with O-type grids with $N_x \times N_y$ grid points in the streamwise and wall-normal directions, respectively. The present grids have $\left(N_x \times N_y \right) = \left(800 \times 600 \right)$ and $\left(768 \times 400 \right)$ points for the present subsonic and supersonic airfoil flows, respectively.

\subsection{Subsonic flow past NACA0012 airfoil}

\begin{figure}
        \centering 
        \subfigure[]{\includegraphics[width=.4\textwidth,trim={0mm -80mm 0mm 20mm},clip]{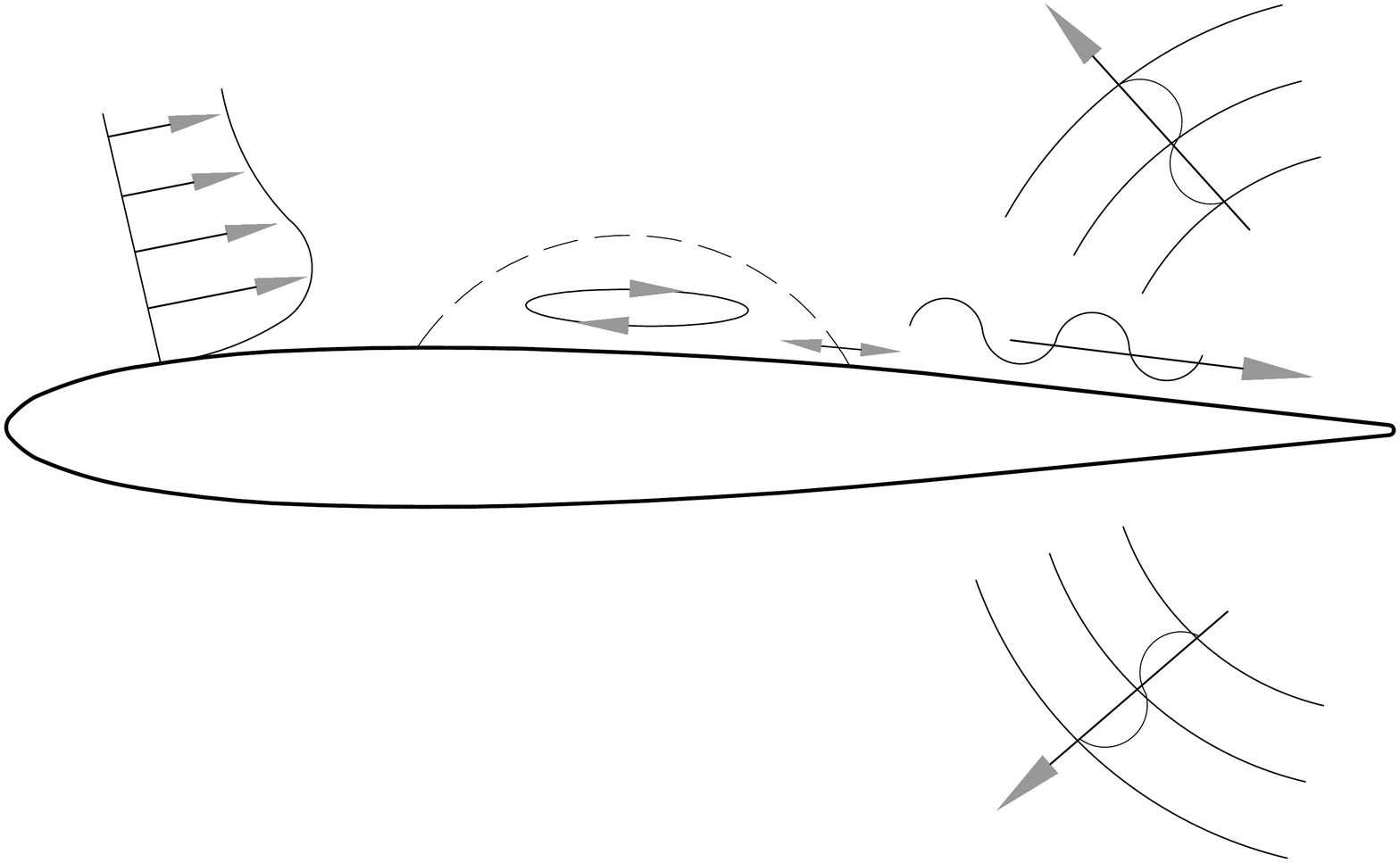}}
        \subfigure[]{\includegraphics[width=.5\textwidth,trim={0mm 0mm 0mm 00mm},clip]{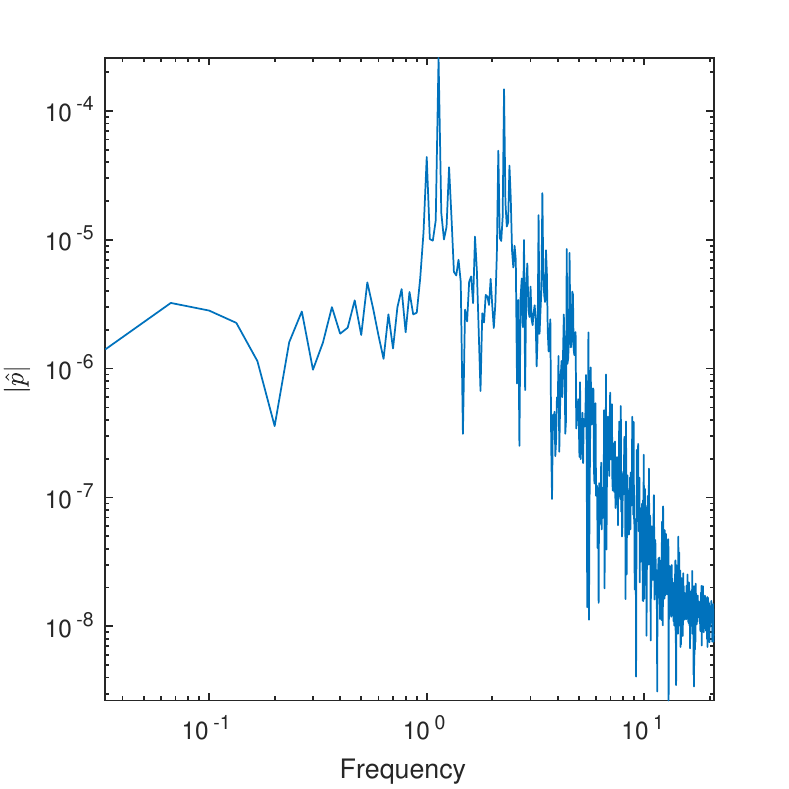}}
    \caption{Sketch of 2D compressible flow at moderate Reynolds number over NACA0012 airfoil (left). Flow instabilities developing over suction side boundary layer are advected along the trailing edge leading to acoustic scattering. These flow instabilities are modulated by a low-frequency motion of the separation bubble, what induces the appearance of equidistant secondary tones in the noise radiation as shown in \cite{RICCIARDI202054}. Pressure spectrum computed one chord above the trailing edge showing a main tone plus secondary tones (right).}
    \label{fig:Airfoil_flow_1}
\end{figure}

In the present analysis, a NACA0012 airfoil with a rounded trailing edge is immersed in a $M_{\infty}=0.3$ flow at 3 deg. angle of incidence with $Re_c=100{,}000$. At this moderate Reynolds number, flow instabilities develop along the suction side boundary layer as can be observed in the sketch shown in Fig. \ref{fig:Airfoil_flow_1}(a). These instabilities are advected past the trailing edge, generating acoustic waves that propagate upstream forming a feedback loop mechanism \cite{Arbey1983, Desquesnes2007}. For the present flow, Ricciardi {\em et al.} \cite{RICCIARDI202054} show that a thin recirculation bubble forms on the airfoil suction side. This bubble has a low frequency flapping that induces a frequency modulation of flow structures that are transported along the boundary layer until reaching the trailing edge. This modulation causes lead and lag of flow instabilities that, in turn, affect the acoustic scattering mechanism, leading to multiple equidistant secondary tones in the acoustic spectrum as shown Fig. \ref{fig:Airfoil_flow_1}(b). Therefore, despite the simple geometrical configuration, this airfoil flow at moderate Reynolds number is a suitable candidate to evaluate the performance of ROMs since it offers rich dynamics. More details about this flow can be found in \cite{RICCIARDI202054}.







Results obtained from the full order model are recorded for $10{,}000$ snapshots with a constant non-dimensional time step $\Delta t_{snapshot} = 3 \times 10^{-3}$. Half of the snapshots are used to construct a reduced-order basis by the snapshot-POD method introduced in section \ref{subsection:POD}. This is equivalent to 15 non-dimensional time units. The length of the training window and the number of sampling snapshots need to be chosen wisely to avoid an ill-resolved POD basis. In this case, the low-frequency motion of the separation bubble  requires a wide training window. On the other hand, high sampling frequency rates are essential for resolving the finer flow scales appearing along the boundary layer. This combination of features is crucial for stability and accuracy of the present non-linear ROM.





\begin{figure}
    \centering 
    \subfigure[]{\includegraphics[width=.45\textwidth,trim={0mm 0mm 0mm 0mm},clip]{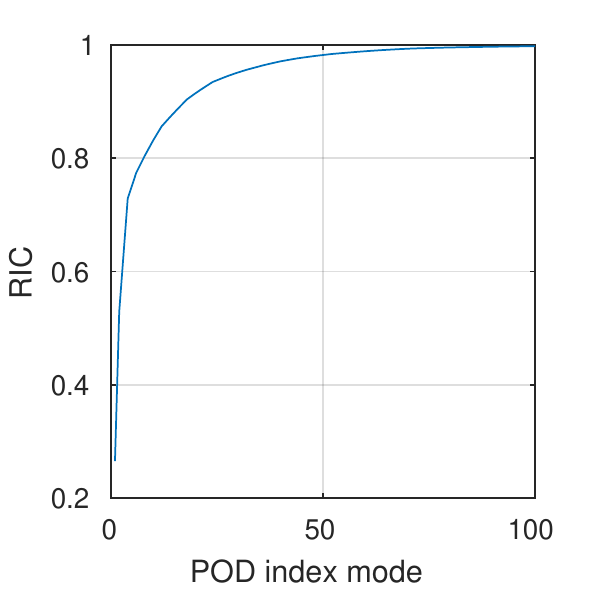}}
    \subfigure[]{\includegraphics[width=.45\textwidth,trim={0mm 0mm 0mm 0mm},clip]{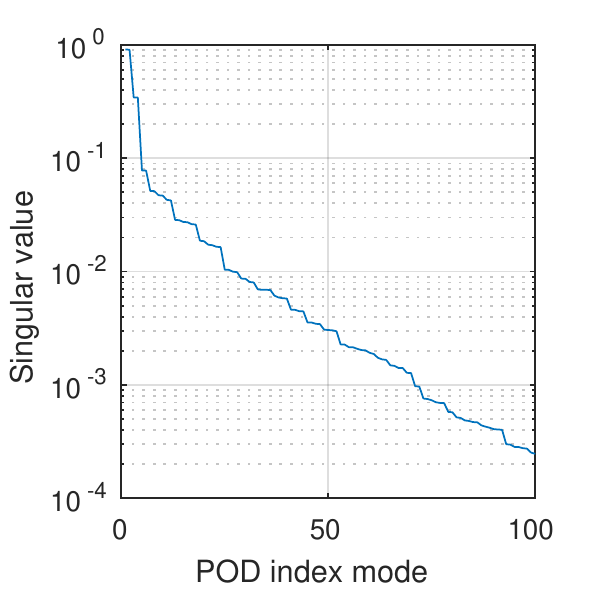}}
    \caption{Relative information content (left). Spetrum of singular values (right)}
    \label{fig:ric}
\end{figure}

The optimality property of the POD method is expected to produce a basis where only a small number of modes should be necessary to reconstruct the input data and, thus, the benefit of including additional modes is expected to rapidly decay. Usually, the number of POD modes used in the ROM is chosen according to the relative information content $\mbox{RIC} (M) = \sum_{i = 1}^{M} \lambda_i / \sum_{i = 1}^{N} \lambda_i$ and should satisfy a predefined threshold. The evolution of the RIC for this case is presented in Fig. \ref{fig:ric}(a) for a basis composed of the first 100 modes (out of 5,000). This basis represents $99.75\%$ of the model RIC and, therefore, should be sufficient to lead to an accurate flow representation. Additionally, the corresponding spectrum of singular values is presented in Fig. \ref{fig:ric}(b), where it is possible to see the fast magnitude decay of the first modes. As can be also seen in this figure, the magnitudes are similar for mode pairs, what indicates that such modes contain similar frequency information, differing only with respect to phase. 
%

%

\begin{figure}
    \begin{subfigure}
        \centering
        \includegraphics[scale=.32,trim={20mm 30mm 10mm 30mm},clip]{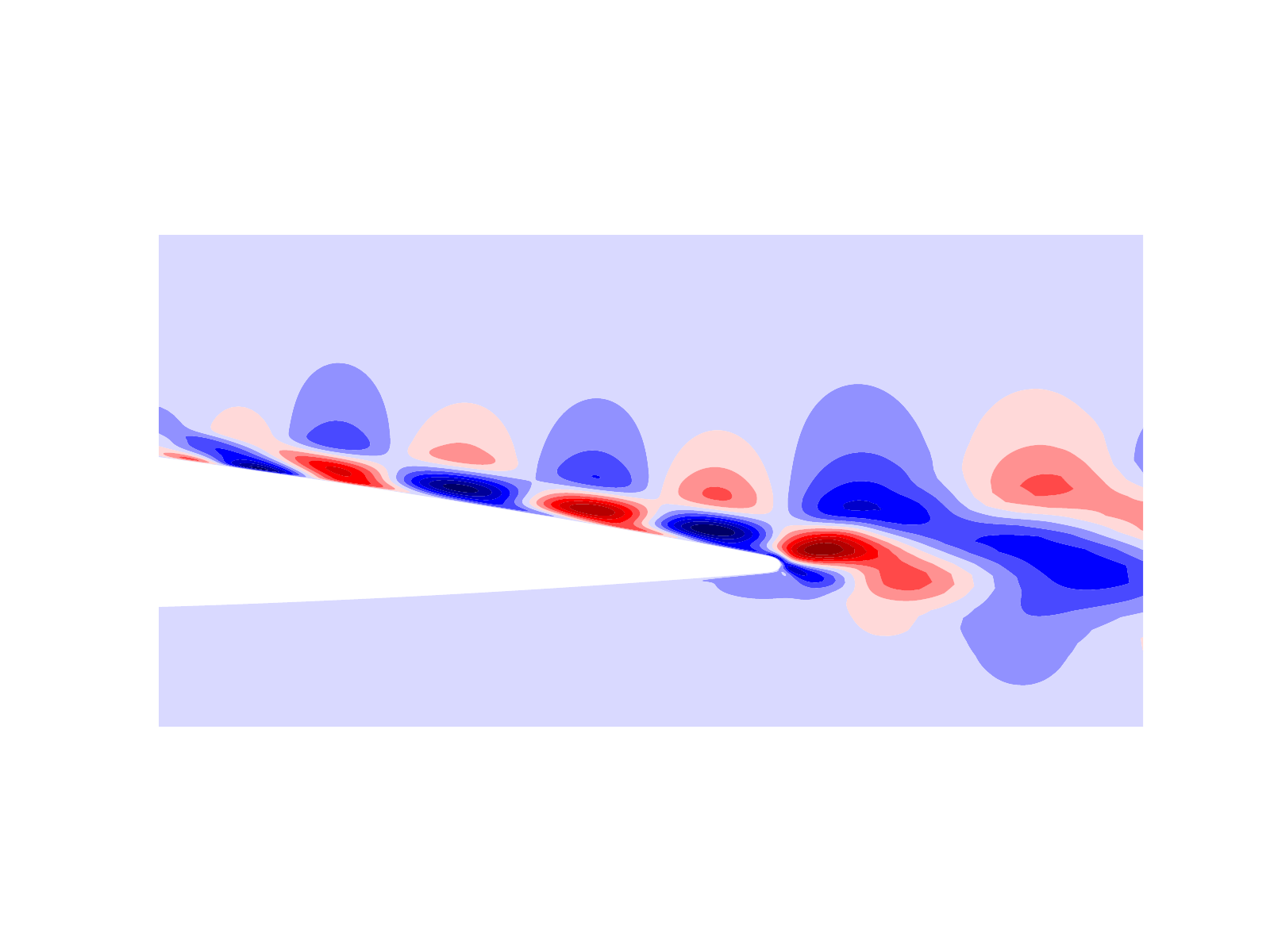}
    \end{subfigure}
    \begin{subfigure}
        \centering
        \includegraphics[scale=.32,trim={20mm 30mm 10mm 30mm},clip]{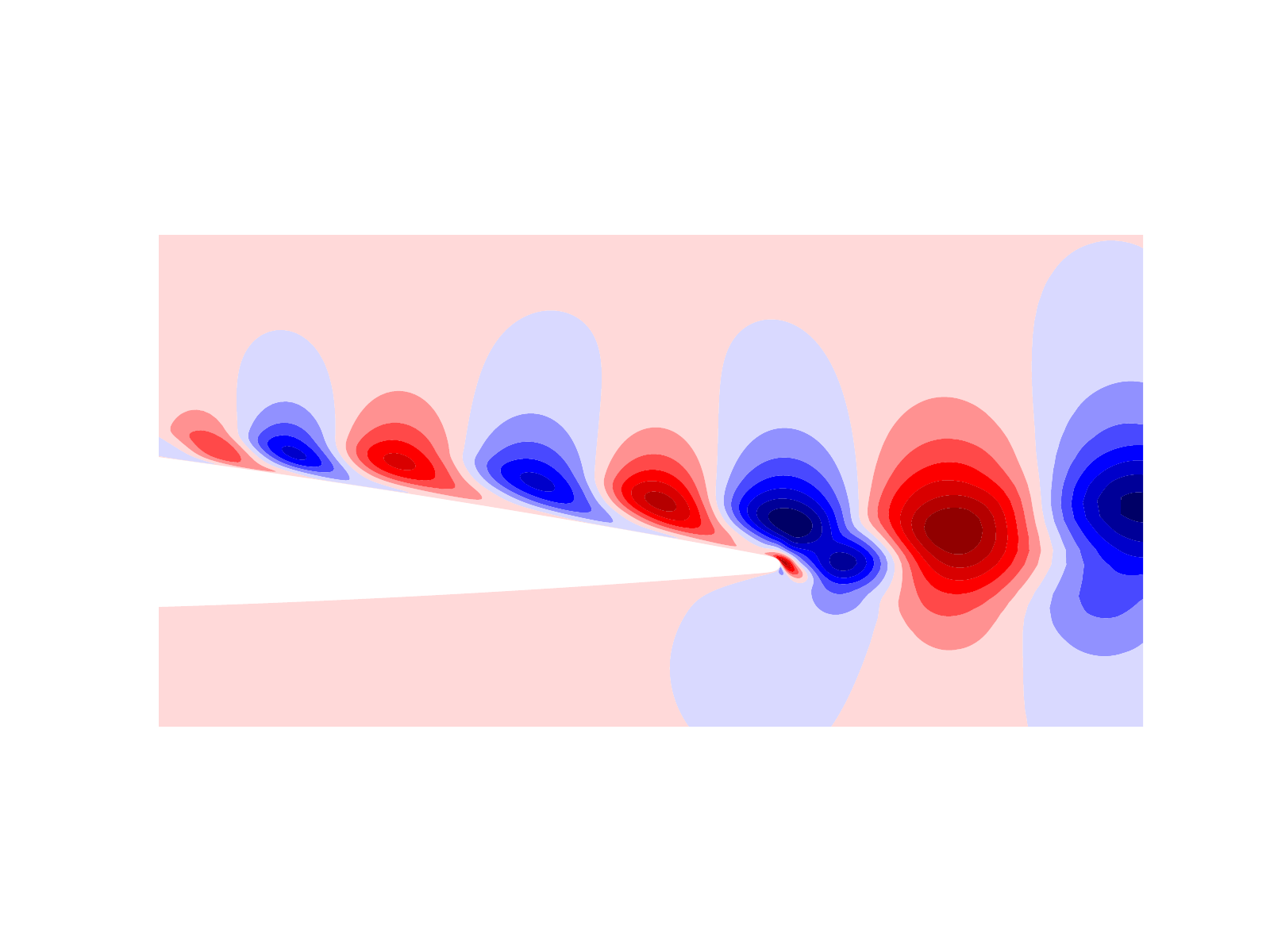}
    \end{subfigure}
    \begin{subfigure}
        \centering
        \includegraphics[scale=.32,trim={20mm 30mm 10mm 30mm},clip]{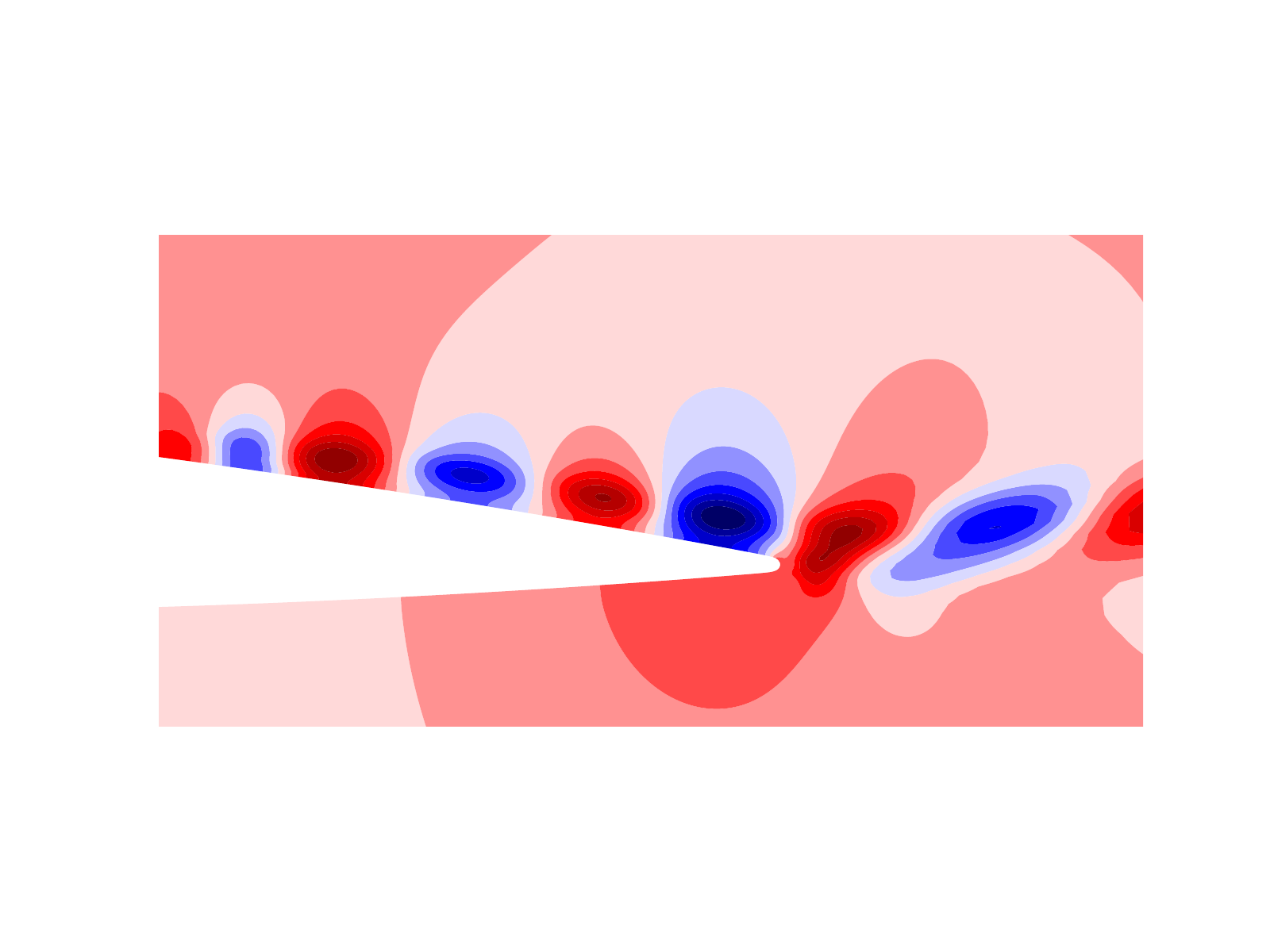}
    \end{subfigure}\\
    \begin{subfigure}
        \centering
        \includegraphics[scale=.32,trim={20mm 30mm 10mm 30mm},clip]{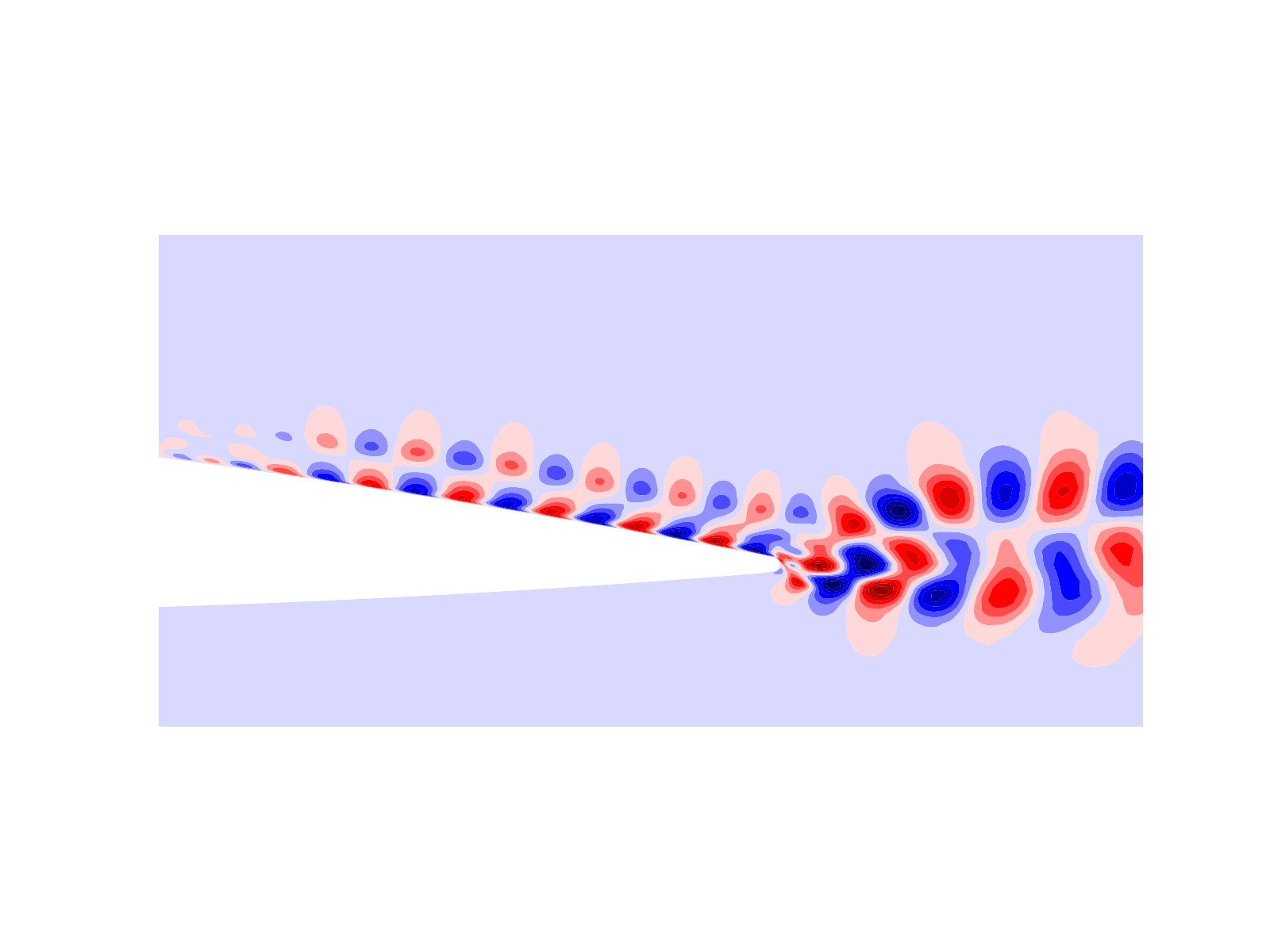}
    \end{subfigure}
    \begin{subfigure}
        \centering
        \includegraphics[scale=.32,trim={20mm 30mm 10mm 30mm},clip]{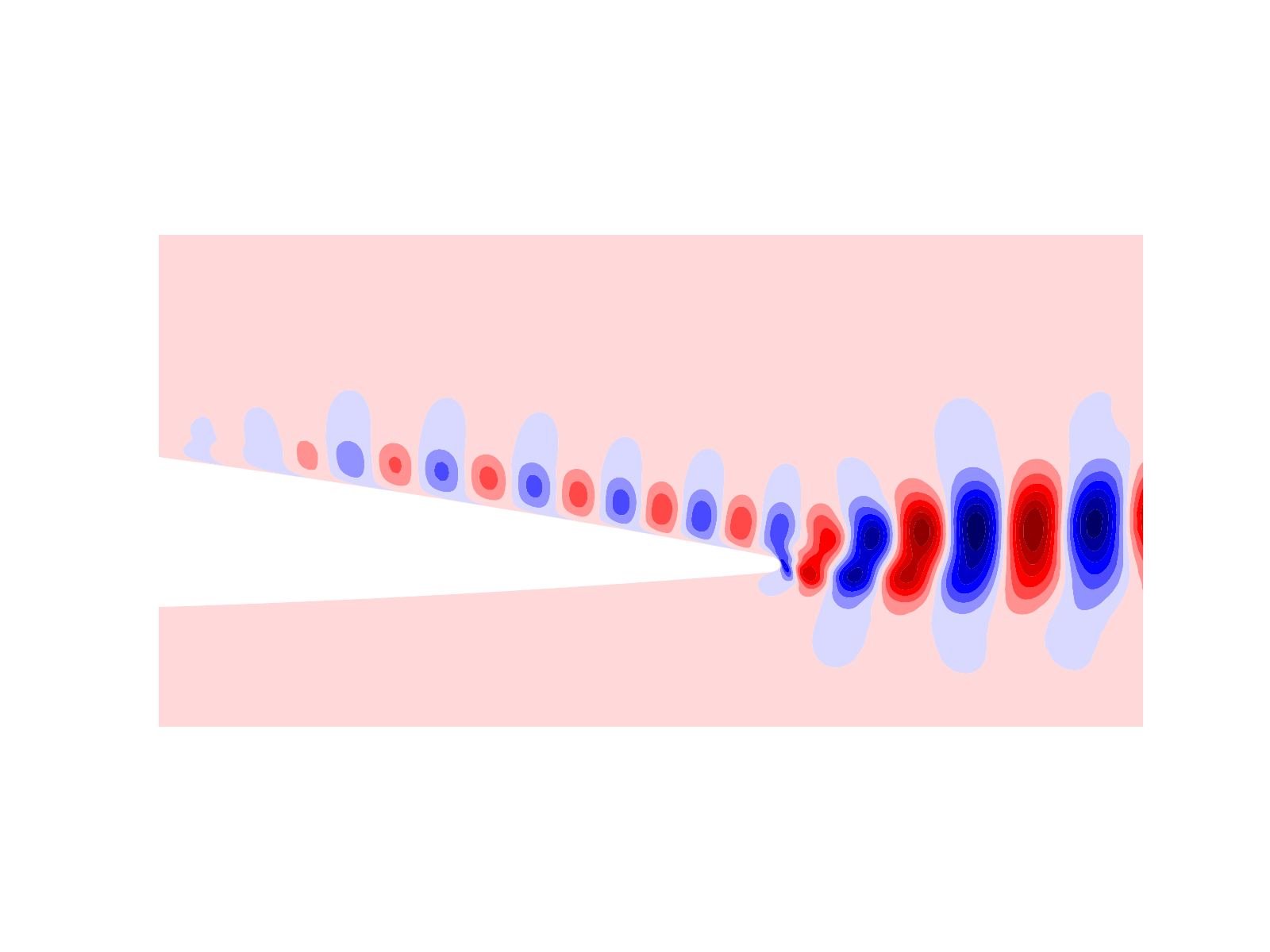}
    \end{subfigure}
    \begin{subfigure}
        \centering
        \includegraphics[scale=.32,trim={20mm 30mm 10mm 30mm},clip]{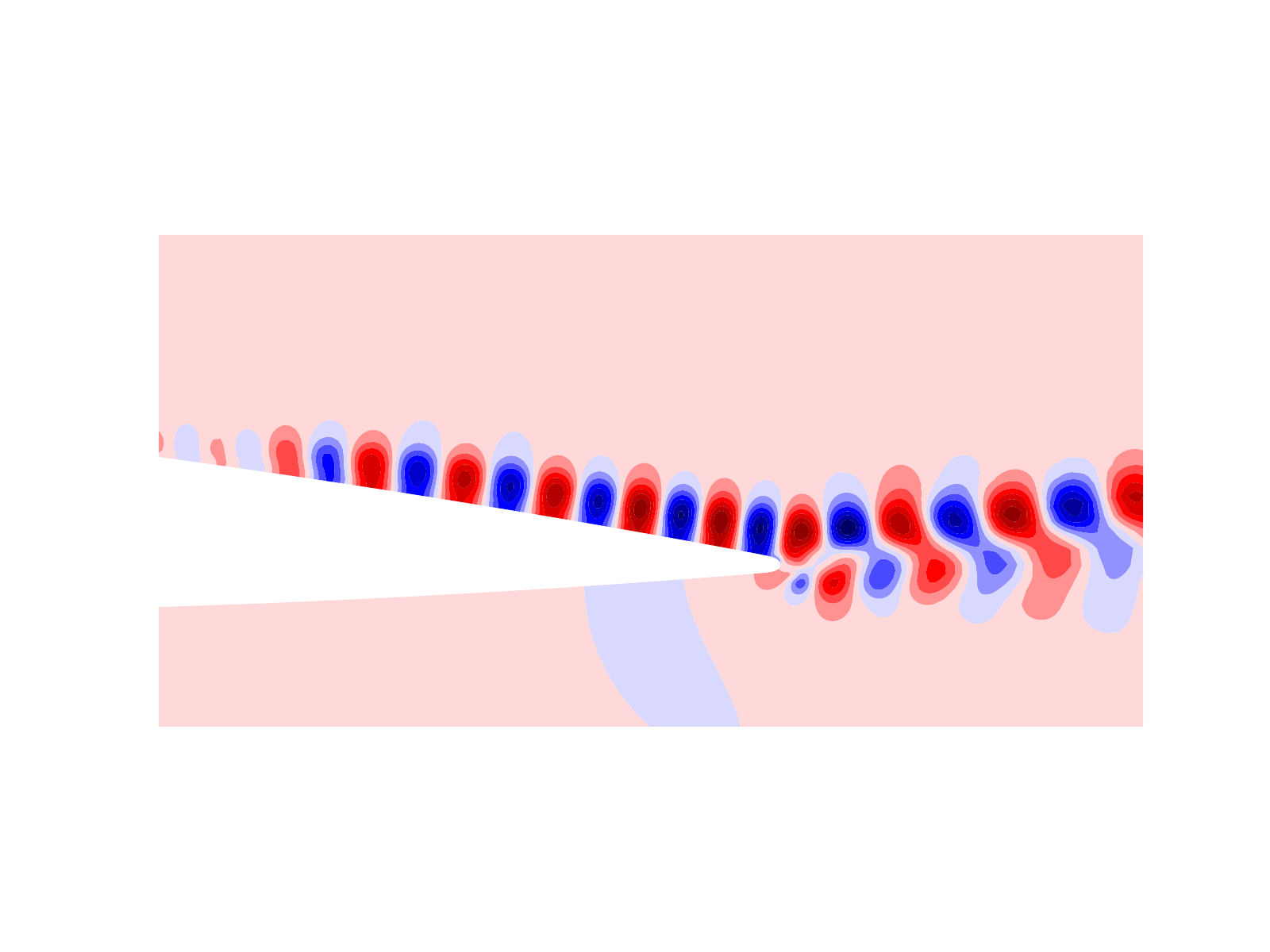}
    \end{subfigure}\\
    \begin{subfigure}
        \centering
        \includegraphics[scale=.32,trim={20mm 30mm 10mm 30mm},clip]{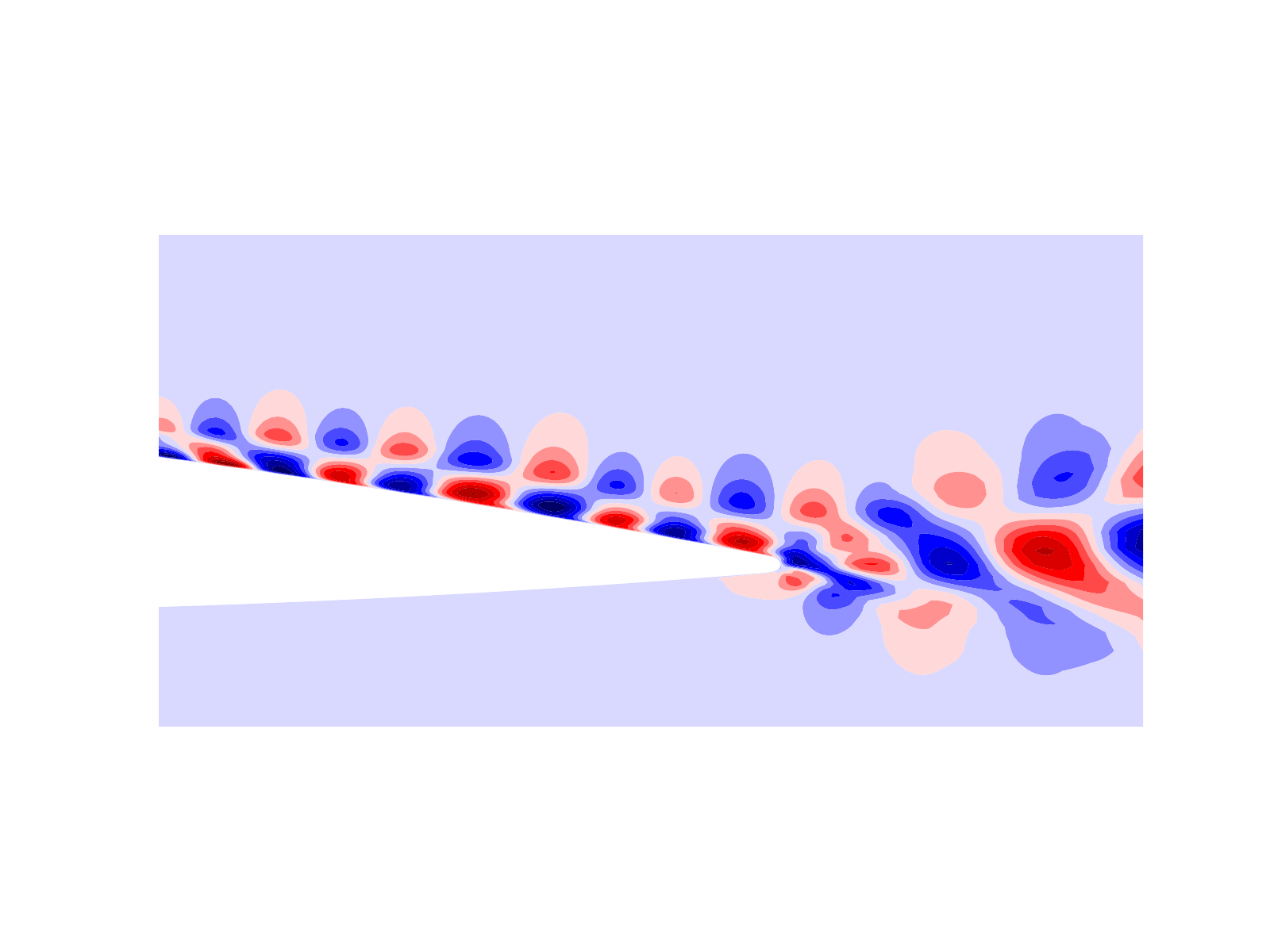}
    \end{subfigure}
    \begin{subfigure}
        \centering
        \includegraphics[scale=.32,trim={20mm 30mm 10mm 30mm},clip]{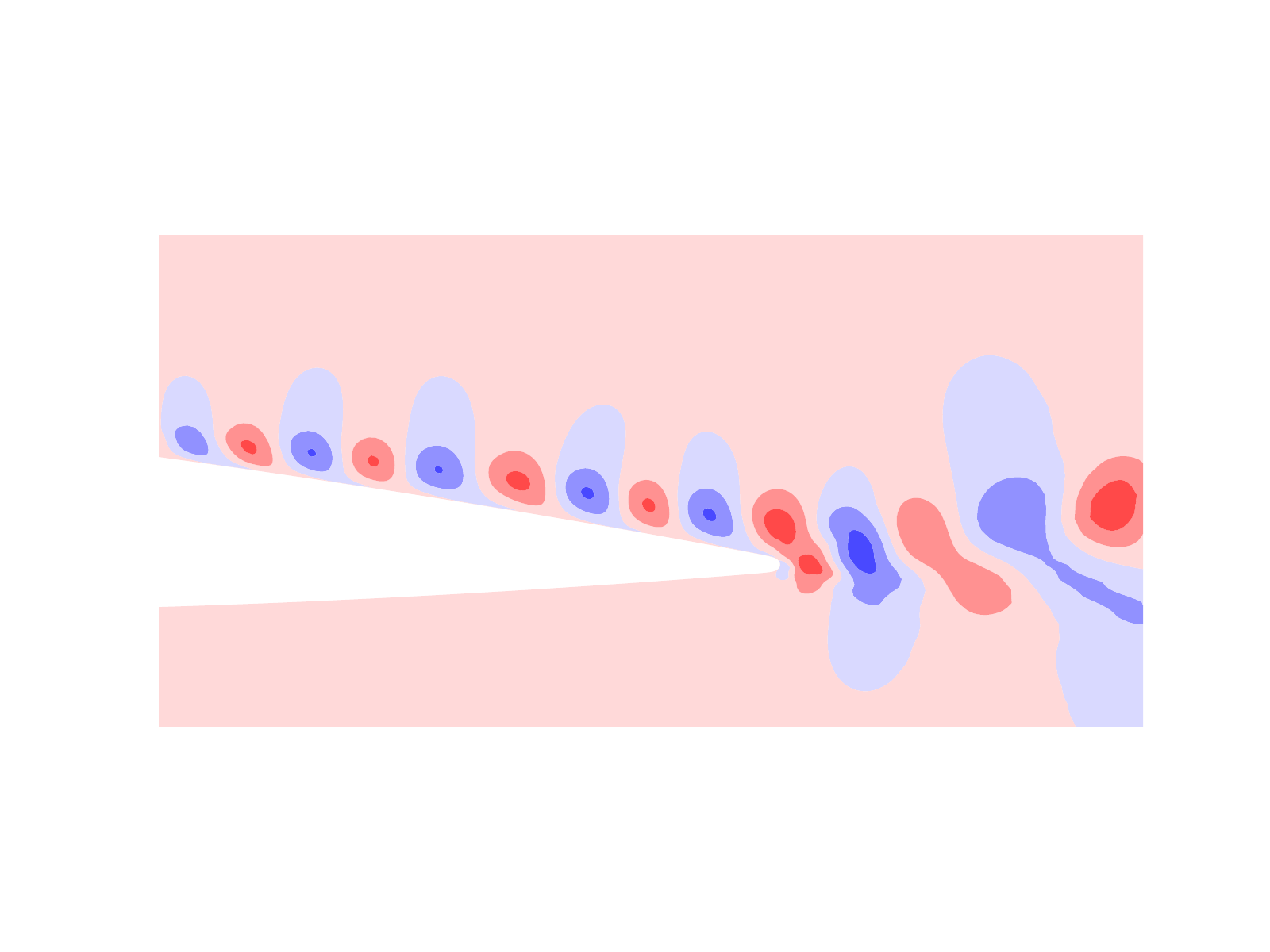}
    \end{subfigure}
    \begin{subfigure}
        \centering
        \includegraphics[scale=.32,trim={20mm 30mm 10mm 30mm},clip]{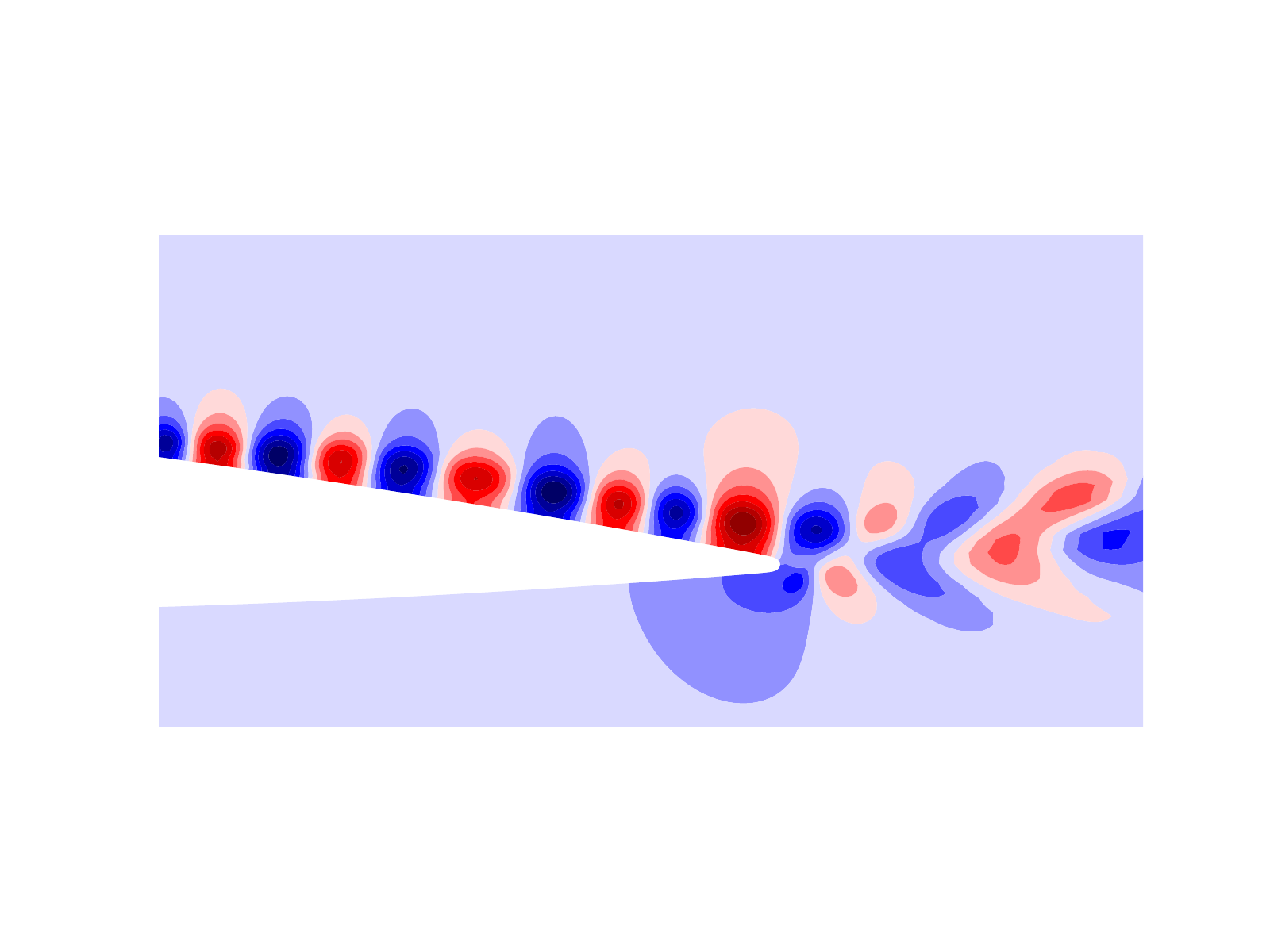}
    \end{subfigure}\\
    \begin{subfigure}
        \centering
        \includegraphics[scale=.32,trim={20mm 30mm 10mm 30mm},clip]{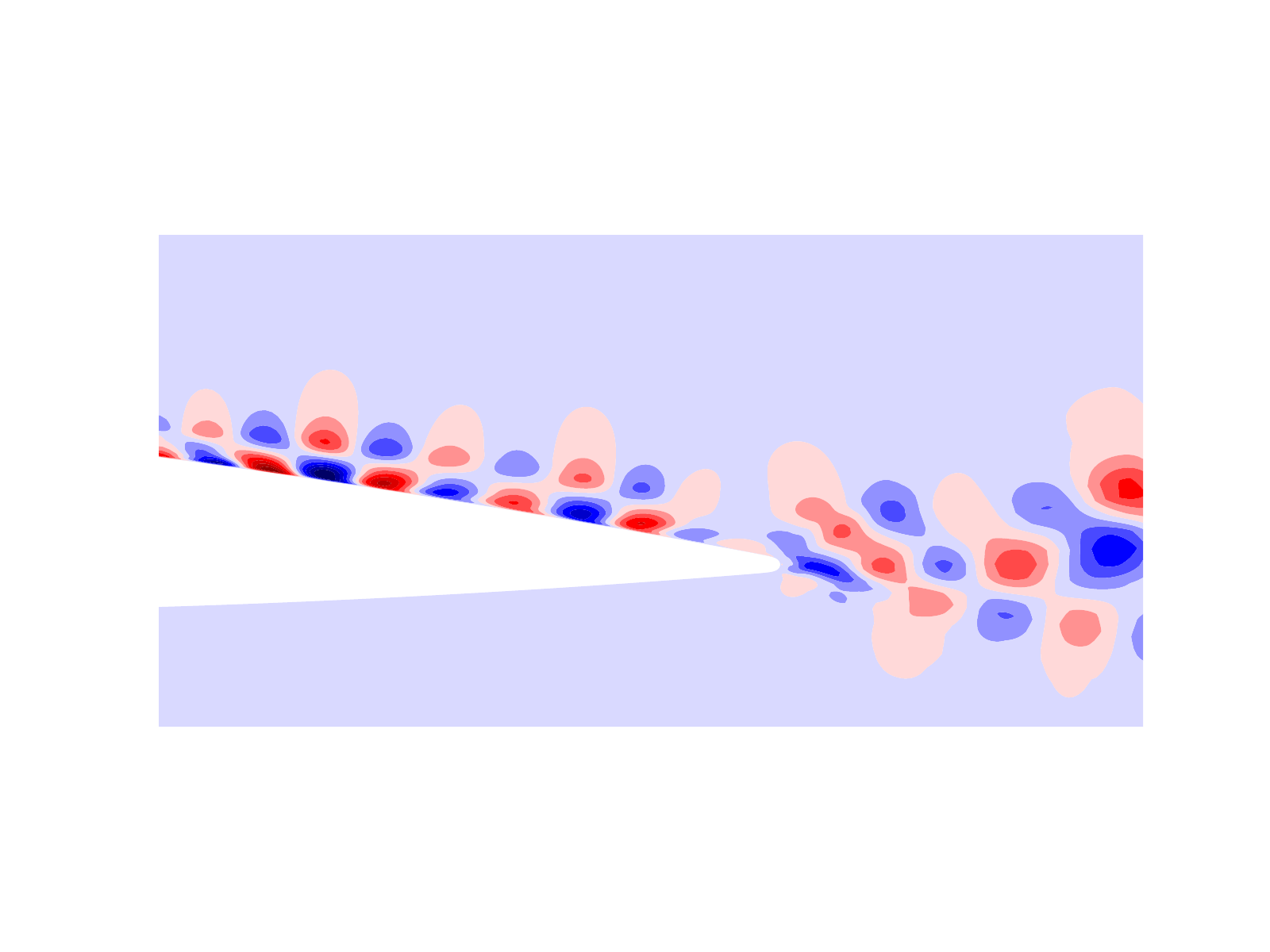}
    \end{subfigure}
    \begin{subfigure}
        \centering
        \includegraphics[scale=.32,trim={20mm 30mm 10mm 30mm},clip]{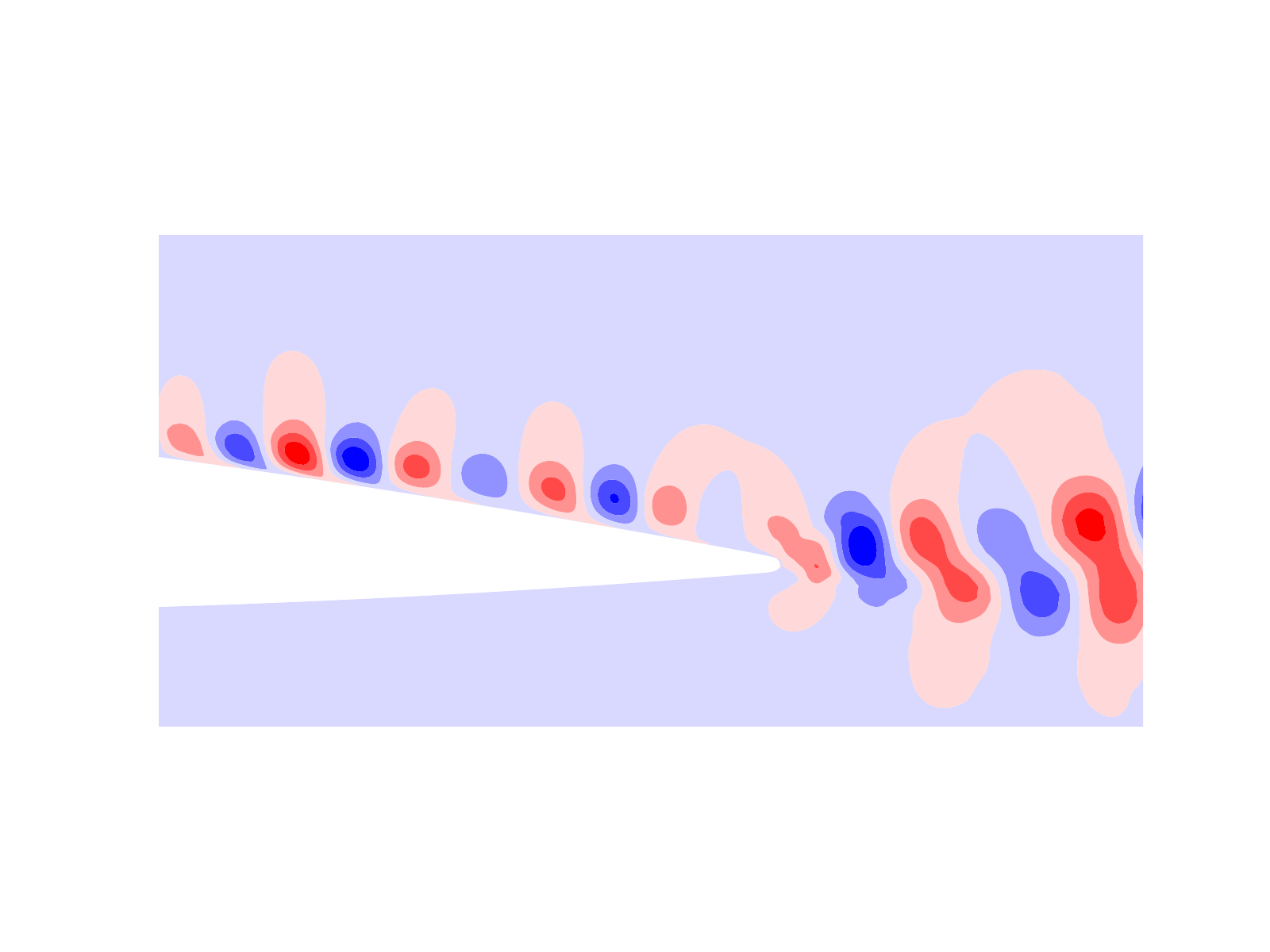}
    \end{subfigure}
    \begin{subfigure}
        \centering
        \includegraphics[scale=.32,trim={20mm 30mm 10mm 30mm},clip]{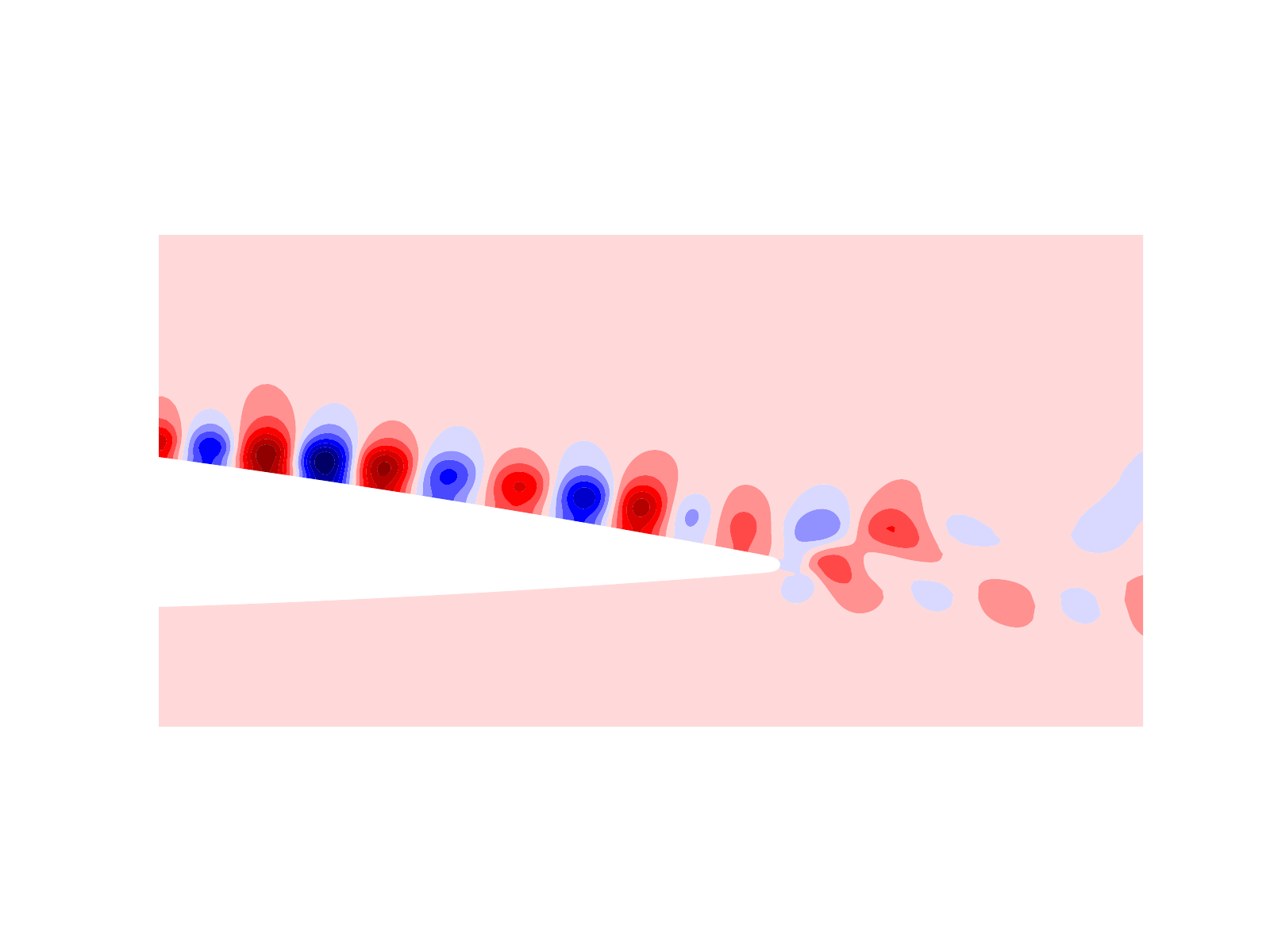}
    \end{subfigure}\\
    \begin{subfigure}
        \centering
        \includegraphics[scale=.32,trim={20mm 30mm 10mm 30mm},clip]{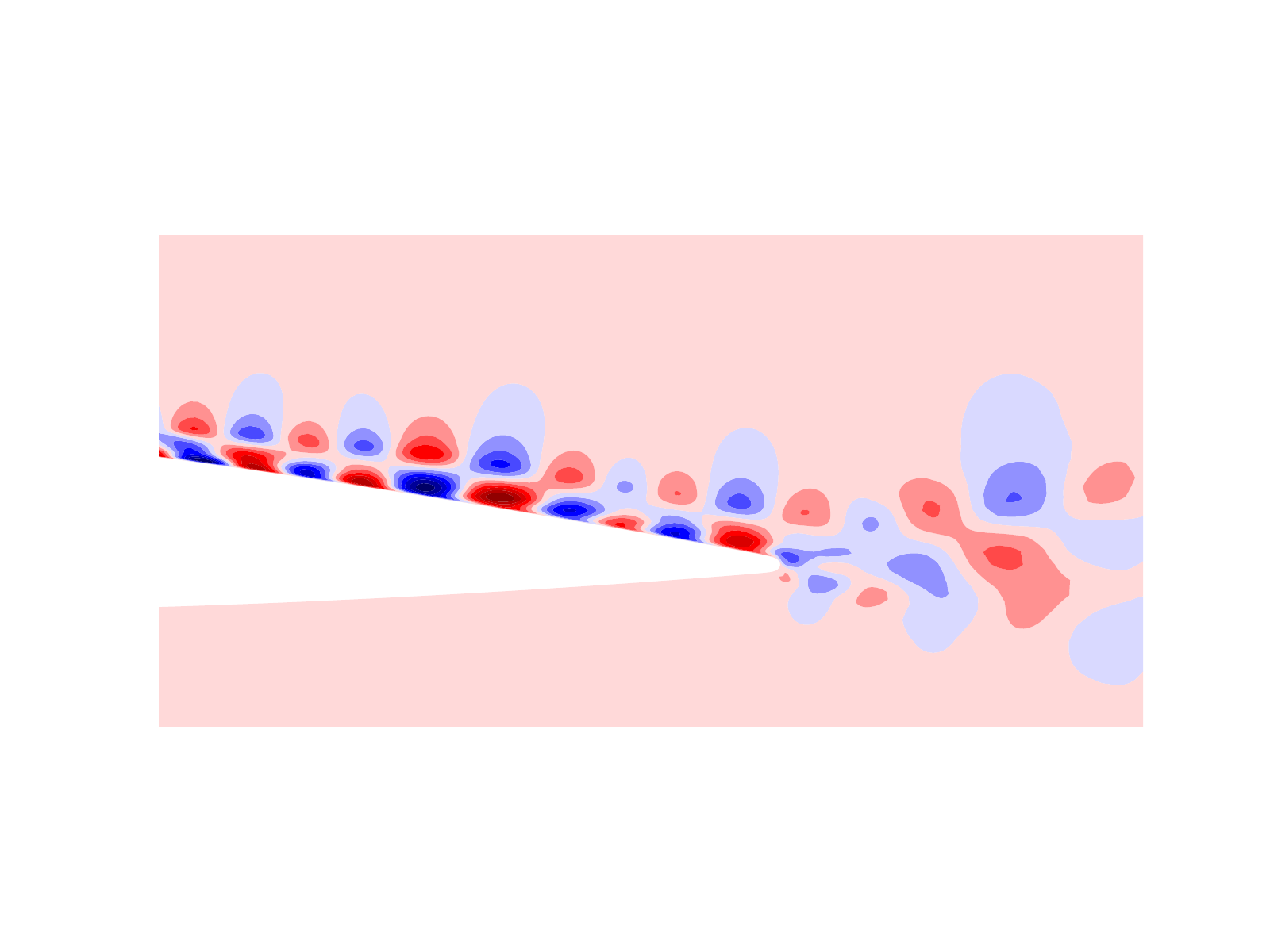}
    \end{subfigure}
    \begin{subfigure}
        \centering
        \includegraphics[scale=.32,trim={20mm 30mm 10mm 30mm},clip]{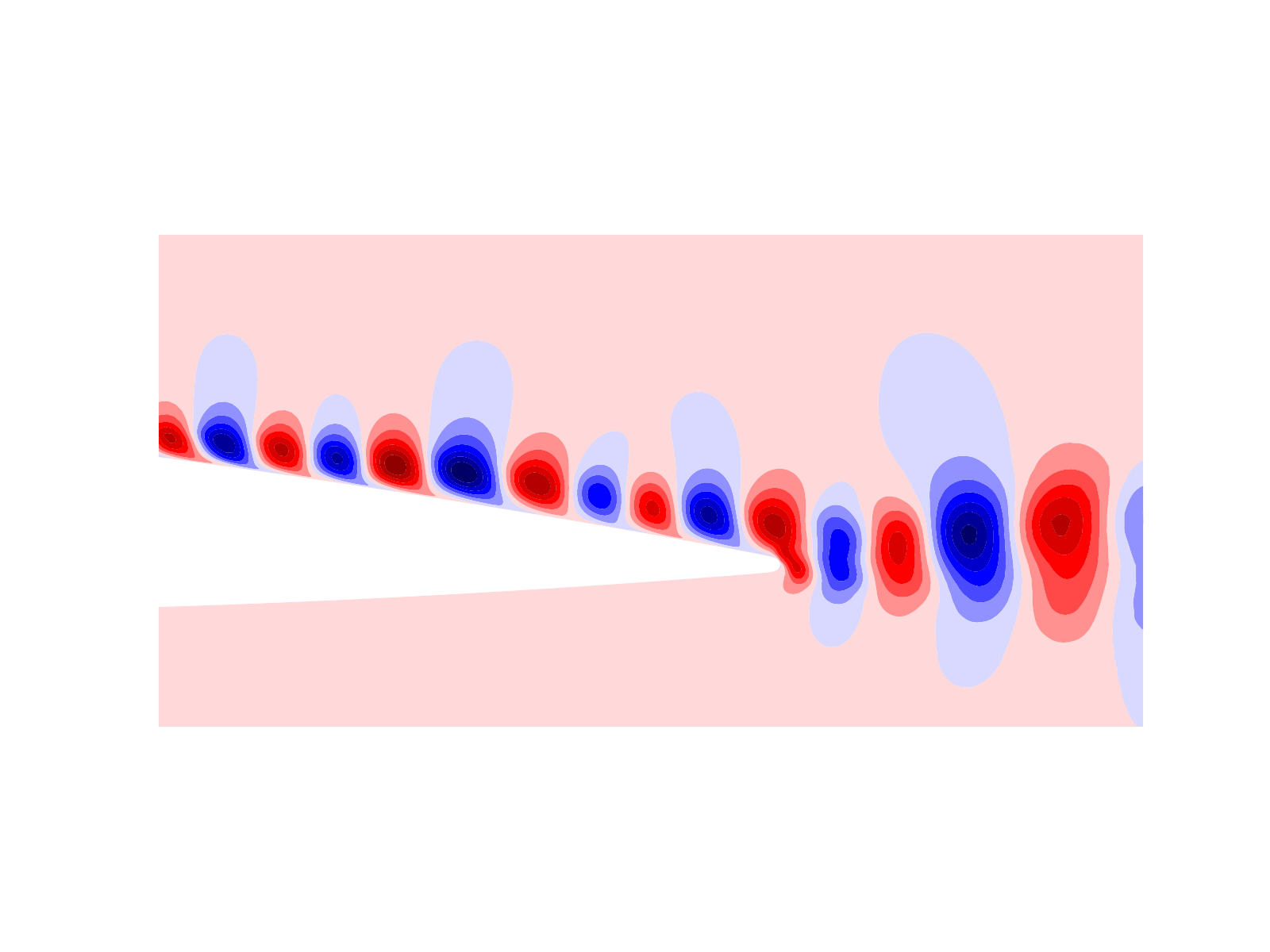}
    \end{subfigure}
    \begin{subfigure}
        \centering
        \includegraphics[scale=.32,trim={20mm 30mm 10mm 30mm},clip]{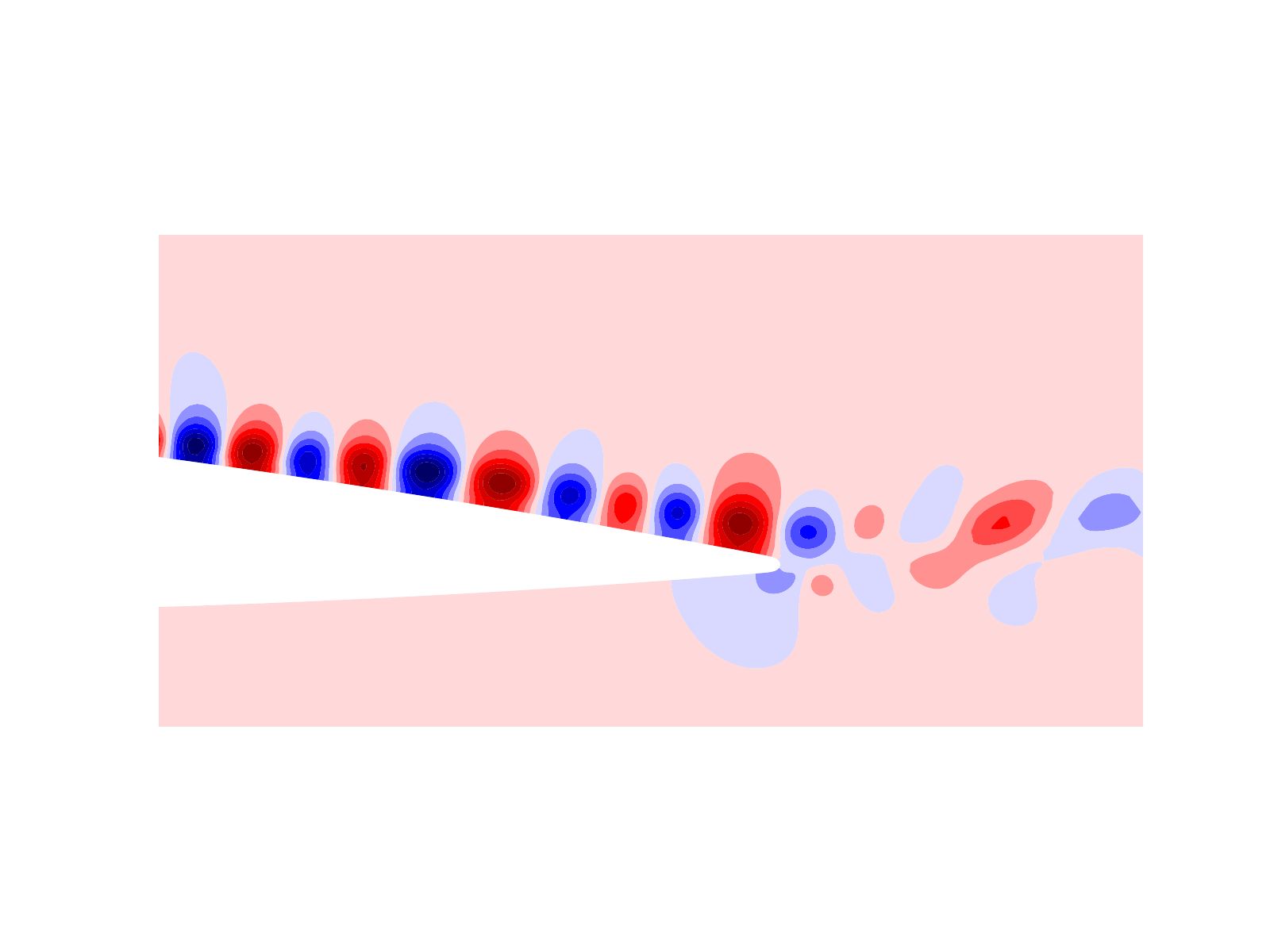}
    \end{subfigure}
    \caption{Contours of POD eigenfunctions for modes $1$, $5$, $9$, $13$ and $17$ (top to bottom) for $u$-velocity (left), $v$-velocity (center) and pressure (right).}
    \label{fig:pod_modes}
\end{figure}

\begin{figure}
    \begin{subfigure}
        \centering
        \includegraphics[width=.8\textwidth,trim={-20mm 5mm 20mm 5mm},clip]{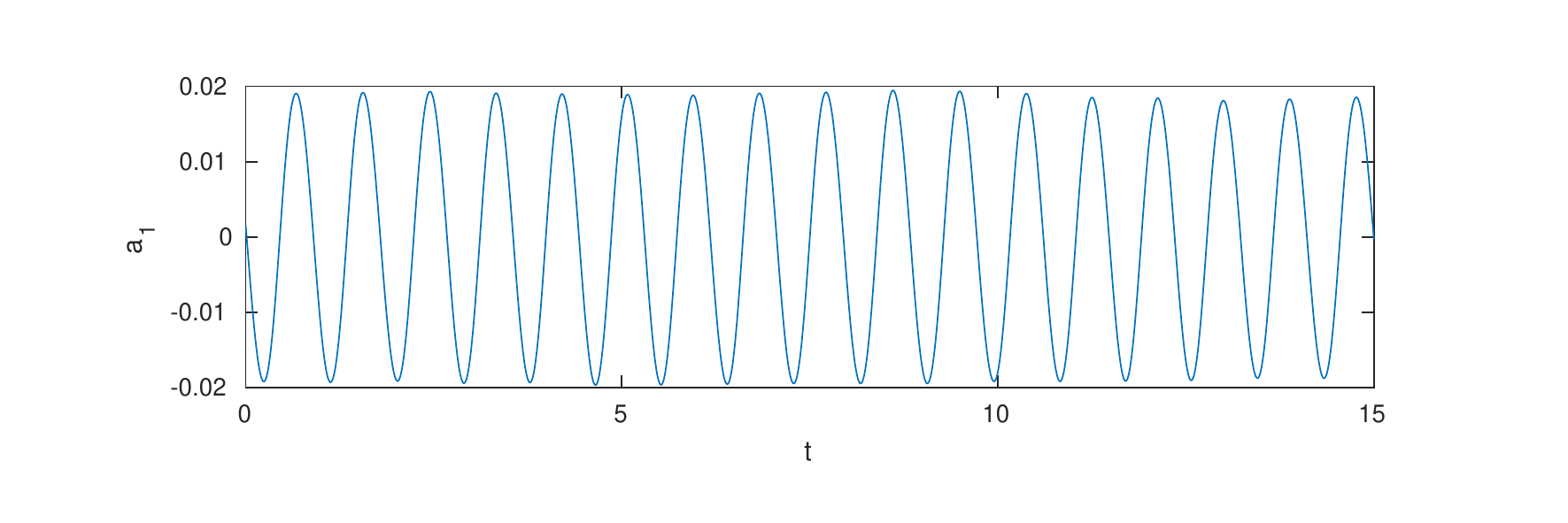}
    \end{subfigure}
    \begin{subfigure}
        \centering
        \includegraphics[width=.8\textwidth,trim={-20mm 5mm 20mm 5mm},clip]{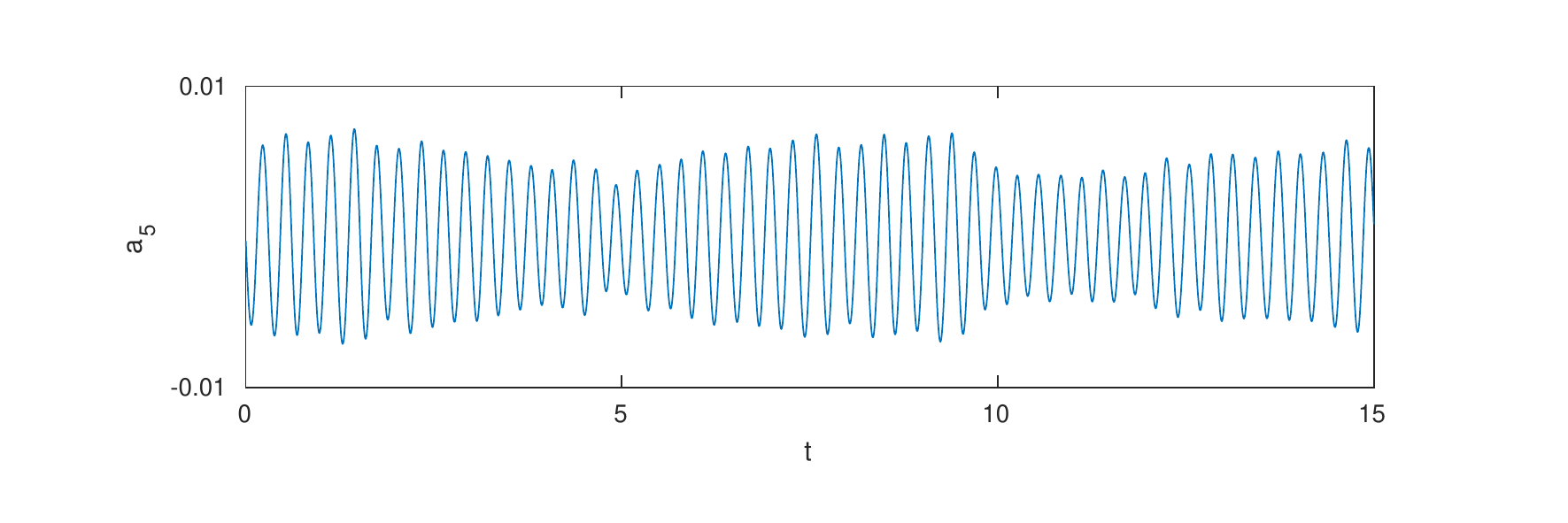}
    \end{subfigure}
    \begin{subfigure}
        \centering
        \includegraphics[width=.8\textwidth,trim={-20mm 5mm 20mm 5mm},clip]{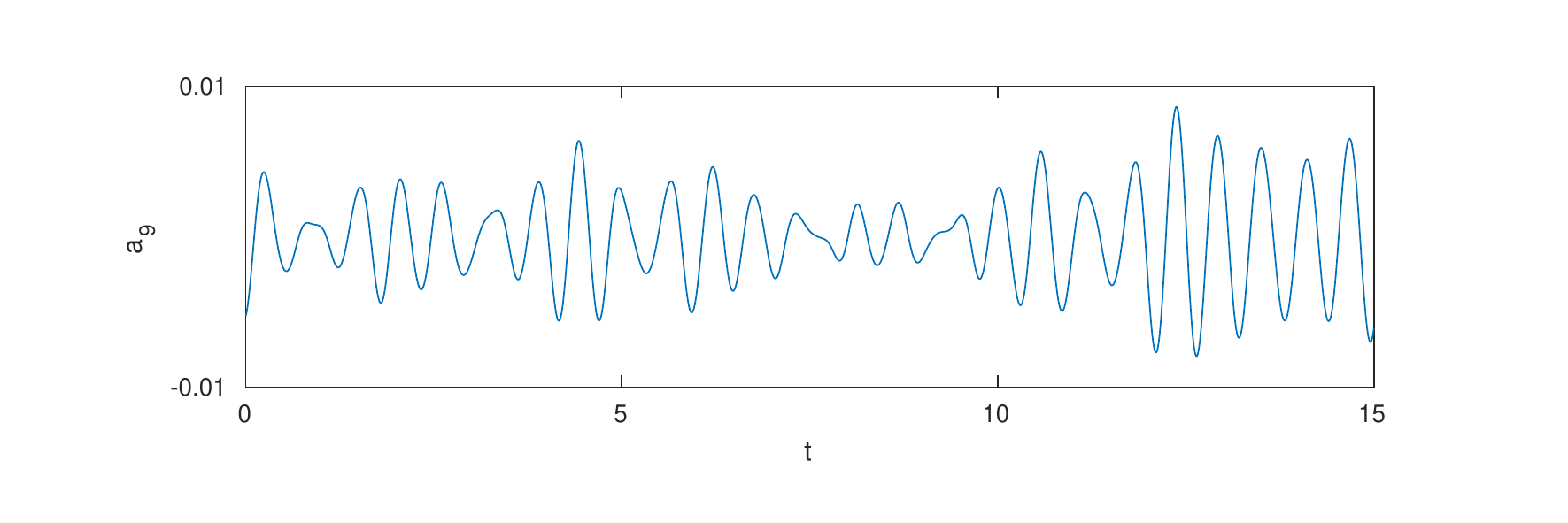}
    \end{subfigure}
    \begin{subfigure}
        \centering
        \includegraphics[width=.8\textwidth,trim={-20mm 5mm 20mm 5mm},clip]{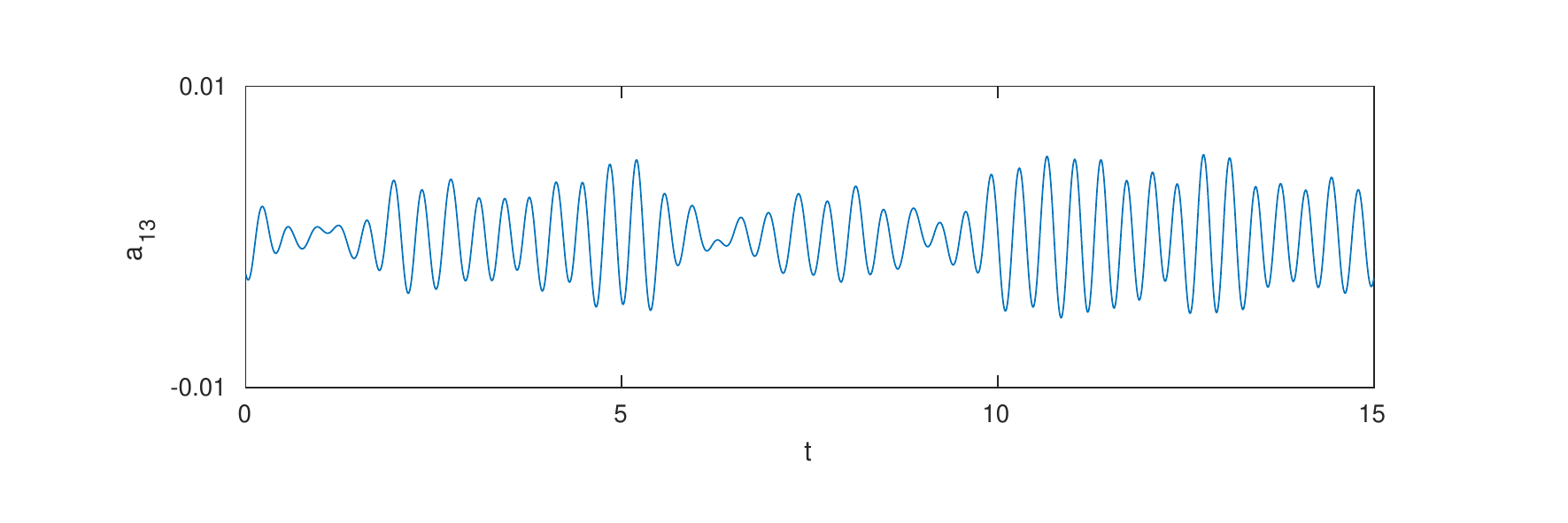}
    \end{subfigure}
    \begin{subfigure}
        \centering
        \includegraphics[width=.8\textwidth,trim={-20mm 5mm 20mm 5mm},clip]{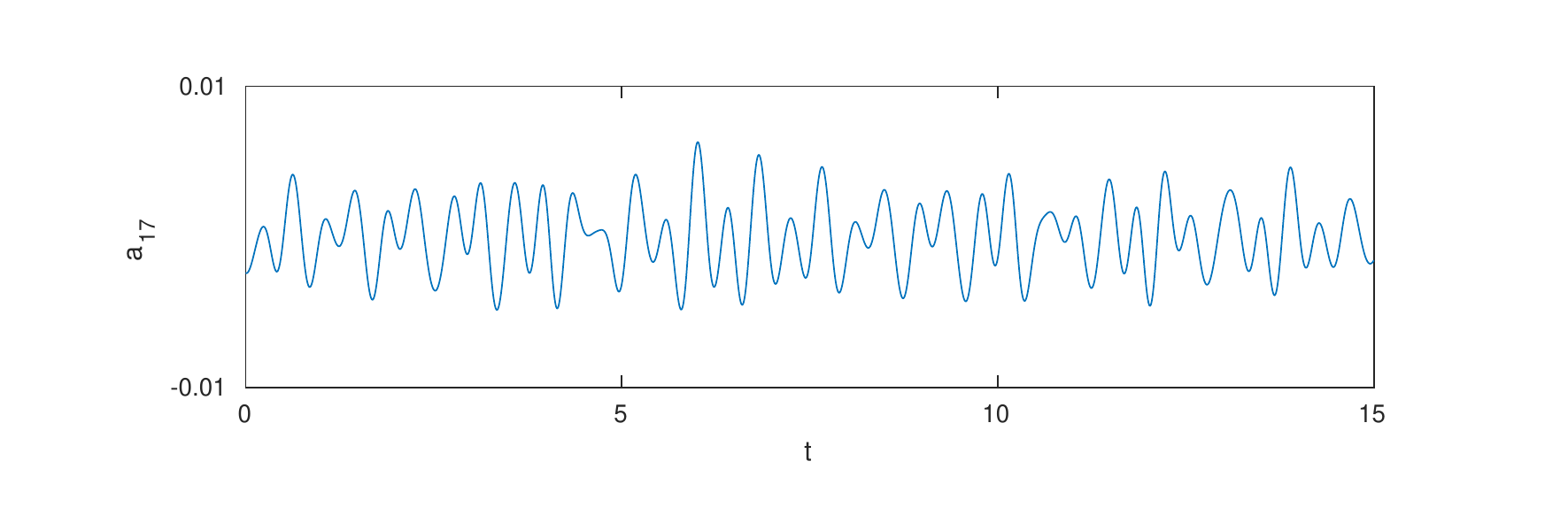}
    \end{subfigure}
    \caption{Temporal dynamics of POD modes $1$, $5$, $9$, $13$ and $17$.}
    \label{fig:time_modes}
\end{figure}

It is expected that the first mode pairs represent the most energetic coherent structures responsible for the flow dynamics. In order to have a better understanding of the present flow, the POD spatial eigenfunctions are shown in Fig. \ref{fig:pod_modes} for modes $1$, $5$, $9$, $13$ and $17$. These modes are presented for $u$ and $v$ velocity components, and pressure fluctuations ($\zeta'$ has the same spatial distribution as $p'$). They are chosen according to their dynamical content, depicted in Fig. \ref{fig:time_modes}. For all modes, it is possible to notice that the flow structures appear along the suction side boundary layer and wake regions. Analyzing both figures together, it is clear that mode $1$ contains the most energetic large-scale structures on the boundary layer, which are associated with the main tone of the spectrum shown in Fig. \ref{fig:Airfoil_flow_1}(b). Mode $5$ is associated with finer scales and displays a modulated higher frequency content. Modes $9$, $13$ and $17$ share similarities in terms of flow structures as can be observed from Fig \ref{fig:pod_modes}. The temporal dynamics of the former two modes are similar and include multiple frequencies strongly modulated. However, the main frequency of mode $9$ is lower than that of mode $13$. On the other hand, the temporal dynamics of mode $17$ appears to have a more clear pattern composed of sinusoids with weaker modulation.

For this case, the ROMs are constructed using the first 100 POD modes. For the Galerkin scheme, the constant, linear and non-linear coefficients are computed and stored in a pre-processing stage following the equations shown in \ref{appendix_galerkin}. The spatial derivatives of the POD modes are computed using a $10th$-order accuracy compact finite difference scheme for both Galerkin and LSPG techniques. Zucatti {\em et al.} \cite{zucatti_01} show that the error evolution of projection-based ROMs is sensitive to the computation of spatial derivatives, being considerably reduced when high-resolution schemes are applied.

\begin{figure}
    \centering
    \includegraphics[scale=.6,trim={10mm 5mm 25mm 10mm},clip]{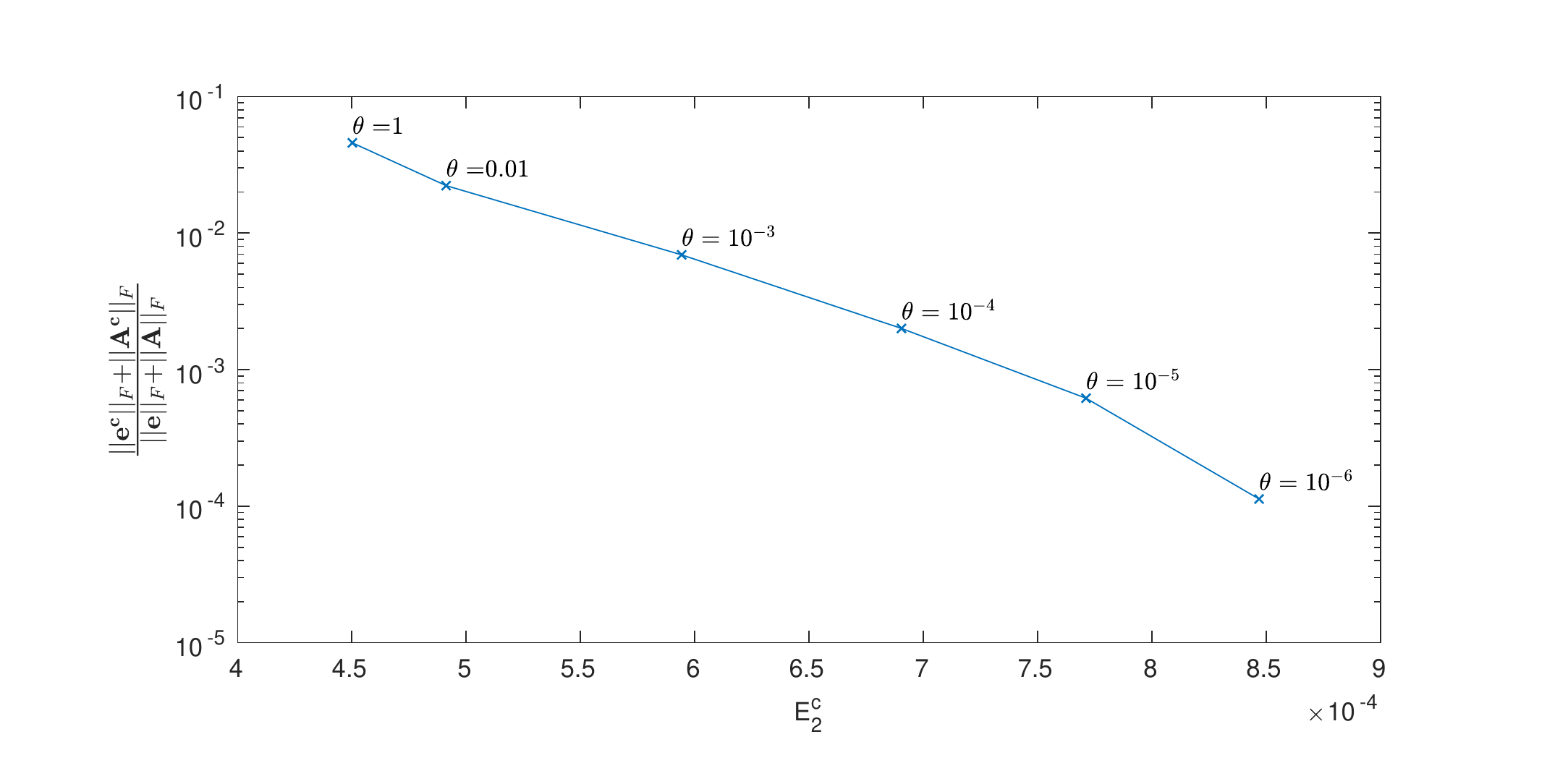}
    \caption{Ratio of Frobenius norms computed for calibrated and Galerkin coefficients as a function of approximation error for different values of  $\theta$ (see Eq. \ref{eq:calibration} for details).}
    \label{fig:theta_tulio}
\end{figure}

The present Galerkin model is computed using a maximum calibration parameter ($\theta = 1$) chosen after an assessment of different values. For this particular flow, the solution error is not very sensitive with respect to the calibration parameter, as can be observed in Fig. \ref{fig:theta_tulio}. This figure shows that for $\theta = 1$ the sum of the Frobenius norms $\| \cdot \|_F$ of the constant and linear calibrated terms is an order of magnitude smaller than that computed for the original Galerkin operators. Thus, relatively low-intrusive terms are obtained in the calibration procedure even though the prediction quality along the training window is favored during calibration. However, this is not always the case and, for other flows, it is important to choose $\theta$ aiming to improve the model quality and, at the same time, avoid overfitting. The calibrated LSPG models do not depend explicitly on $\theta$ since they are computed using Eq. \ref{eq:calibrationLSPG}.
%
%
%
%
%
%
%

\begin{figure}
    \centering
    \includegraphics[width=.5\textwidth,trim={0mm 0mm 0mm 0mm},clip]{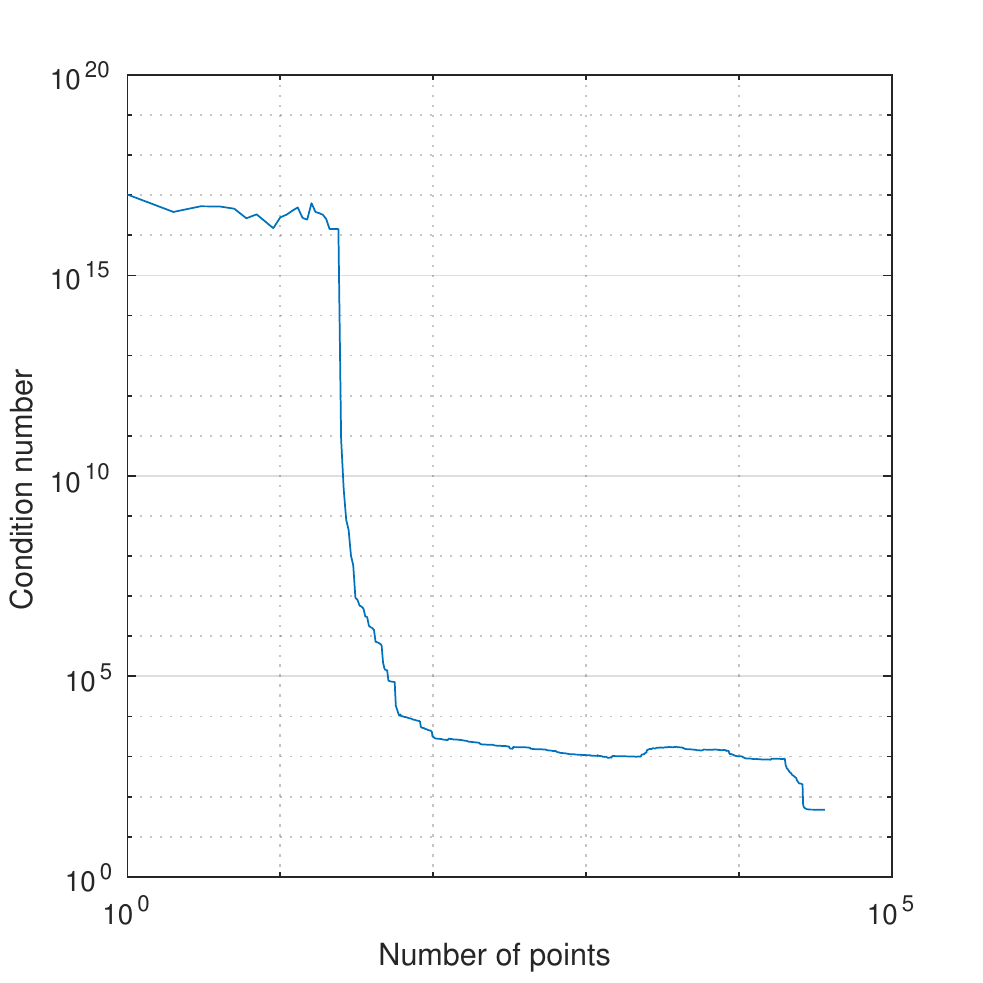}
    \caption{Condition number of approximation mode matrix $\mathbf{\langle \Tilde{\Phi}_i, \Tilde{\Phi}_j \rangle}$ obtained from hyper-reduction.}
    \label{fig:condition_number}
\end{figure}

As previously discussed, the LSPG method should provide better stability properties than the Galerkin projection. However, as stated by \cite{Kevin01}, its computational cost can scale with that of the FOM if all degrees of freedom are used on the model reconstruction. In order to reduce the model assembly cost, we use the hyper-reduction technique previously described. Before a particular number of grid points is employed in the hyper-reduction, it is important to assess the condition number behavior of the system. This is a good way to have an {\it a priori} estimate of the number of points to be used in the ROM. On one hand, we desire to use as few as possible grid points in the model reconstruction to reduce its computational cost. On the other hand, the condition number should be kept as small as possible to reduce the errors associated with the procedure and to be able to represent the relevant flow dynamics. Figure \ref{fig:condition_number} shows the condition number as a function of the number of grid points used in the hyper-reduction. 

\begin{figure}
    \centering
        \subfigure[]{\includegraphics[width=.5\textwidth,trim={20mm 45mm 10mm 45mm},clip]{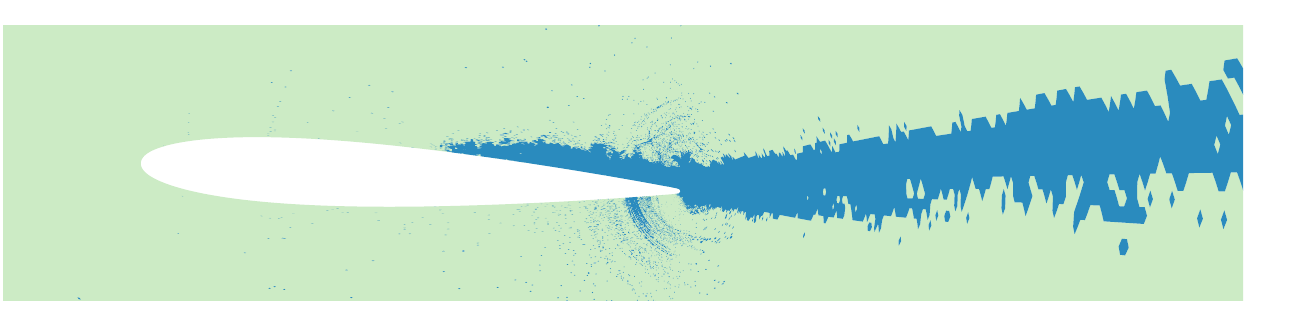}}
        \subfigure[]{\includegraphics[width=.5\textwidth,trim={20mm 20mm 10mm 20mm},clip]{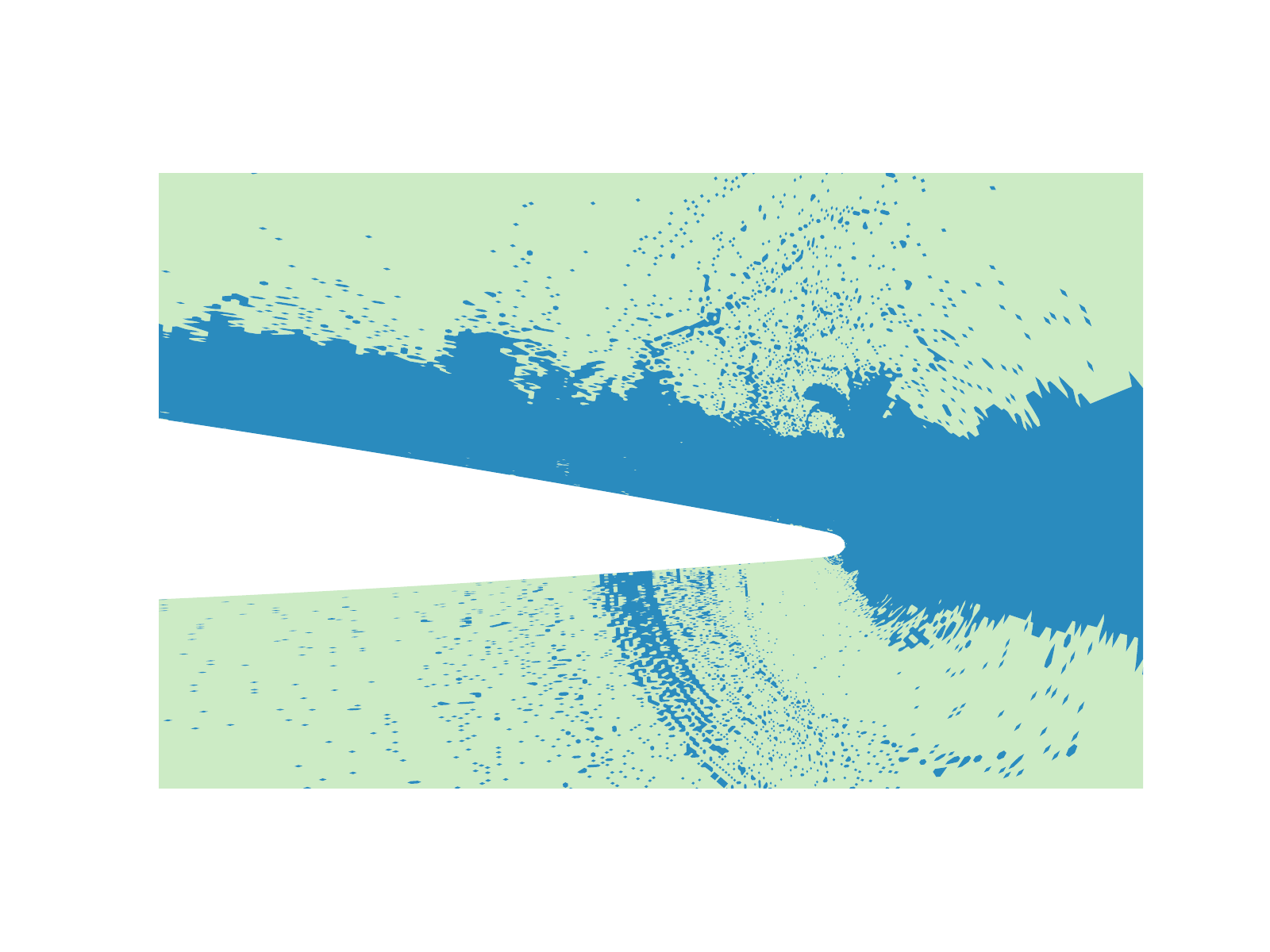}}
    \caption{Sample points (in blue) chosen by the accelerated greedy-MPE hyper-reduction algorithm.}
    \label{fig:sample_points}
\end{figure}

One can see that the condition number does not fall monotonically, what is expected since the present hyper-reduction algorithm is locally optimal but not globally optimal, as previously discussed. Eventually, as the number of grid points is increased, the condition number will become unity once the approximation mode matrix recovers the original full POD spatial mode matrix. As can be observed from the figure, there is a first considerable drop in the condition number when approximately 50 grid points are used in the mask matrix. However, the LSPG ROMs built for this range of condition number were either unstable or inaccurate. Increasing the number of grid points resulted in a second drop in the condition number which provided stable models. For the present problem, hyper-reduction is then applied using $29{,}000$ grid points ($6.04\%$ of the total) which resulted in a condition number $c(\mathbf{\langle \Tilde{\Phi}_i, \Tilde{\Phi}_j \rangle}) = 47.65$. Figure \ref{fig:sample_points} shows the sample points computed by the accelerated greedy-MPE algorithm. In Fig. \ref{fig:sample_points}(a), one can see that the hyper-reduction approach picks the grid points along the wake and suction-side boundary layer since these regions contain the most relevant flow dynamics. A detail view of the trailing-edge region is presented in Fig. \ref{fig:sample_points}(b) and it is possible to notice that some grid points are also selected in the near acoustic field region, where cylindrical sound waves radiate from the trailing edge.

Reduced-order models obtained either by Galerkin or LSPG schemes consist in a set of coupled non-linear ordinary differential equations. In order to describe the dynamics of the system at hand, these equations need to be temporally integrated using a time-marching scheme. A suit of explicit and implicit techniques is available and, here, some are tested to assess the accuracy and stability properties of the ROMs. All non-linear least squares problems emerging from implicit time integration are solved using the Levenberg–Marquardt algorithm previously described. The model time step is selected the same as that between snapshots $\Delta t_{ROM} = 3 \times 10^{-3}$ and this value is $16$ times larger compared to the time step used in the FOM simulation. For this time step the highest frequencies observed in the POD temporal modes are still resolved with at least $20$ points per wavelength. Carlberg {\em et al.} present a detailed theoretical assessment of time integrators for Galerkin and LSPG schemes in \cite{Kevin02_comp}. Here, we analyze the performance of implicit and explicit methods together with calibration techniques. Results are computed for a probe located on the boundary layer at $(x,y) = (0.6919, 0.0083)$. At this location, flow instabilities are advected towards the trailing edge while acoustic waves are propagated upward through the boundary layer. 
\begin{figure}
     \centering
        \subfigure[Galerkin model]{\includegraphics[width=.65\textwidth,trim={15mm 5mm 20mm 10mm},clip]{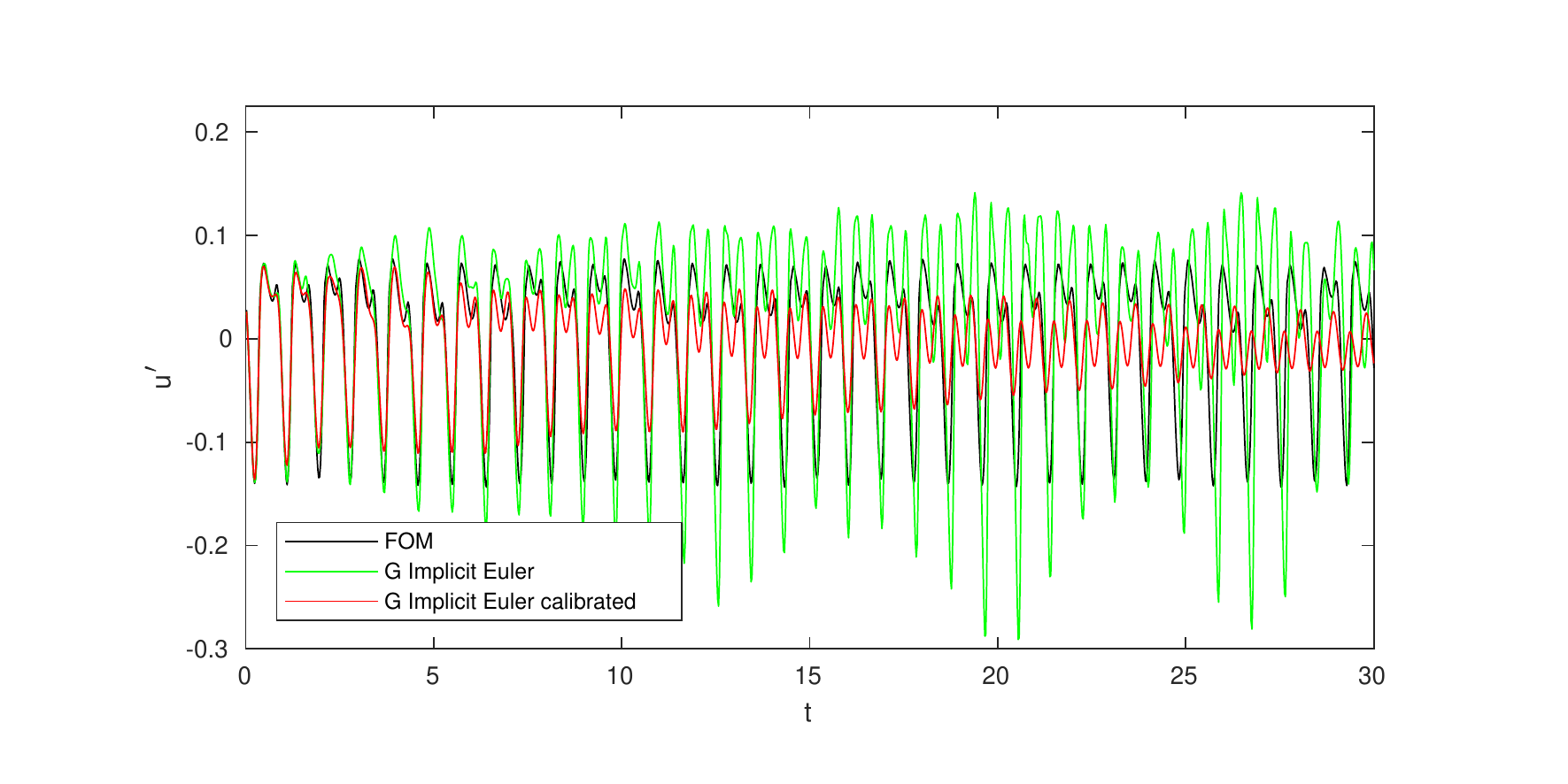}}
        \subfigure[LSPG model]{\includegraphics[width=.65\textwidth,trim={15mm 5mm 20mm 10mm},clip]{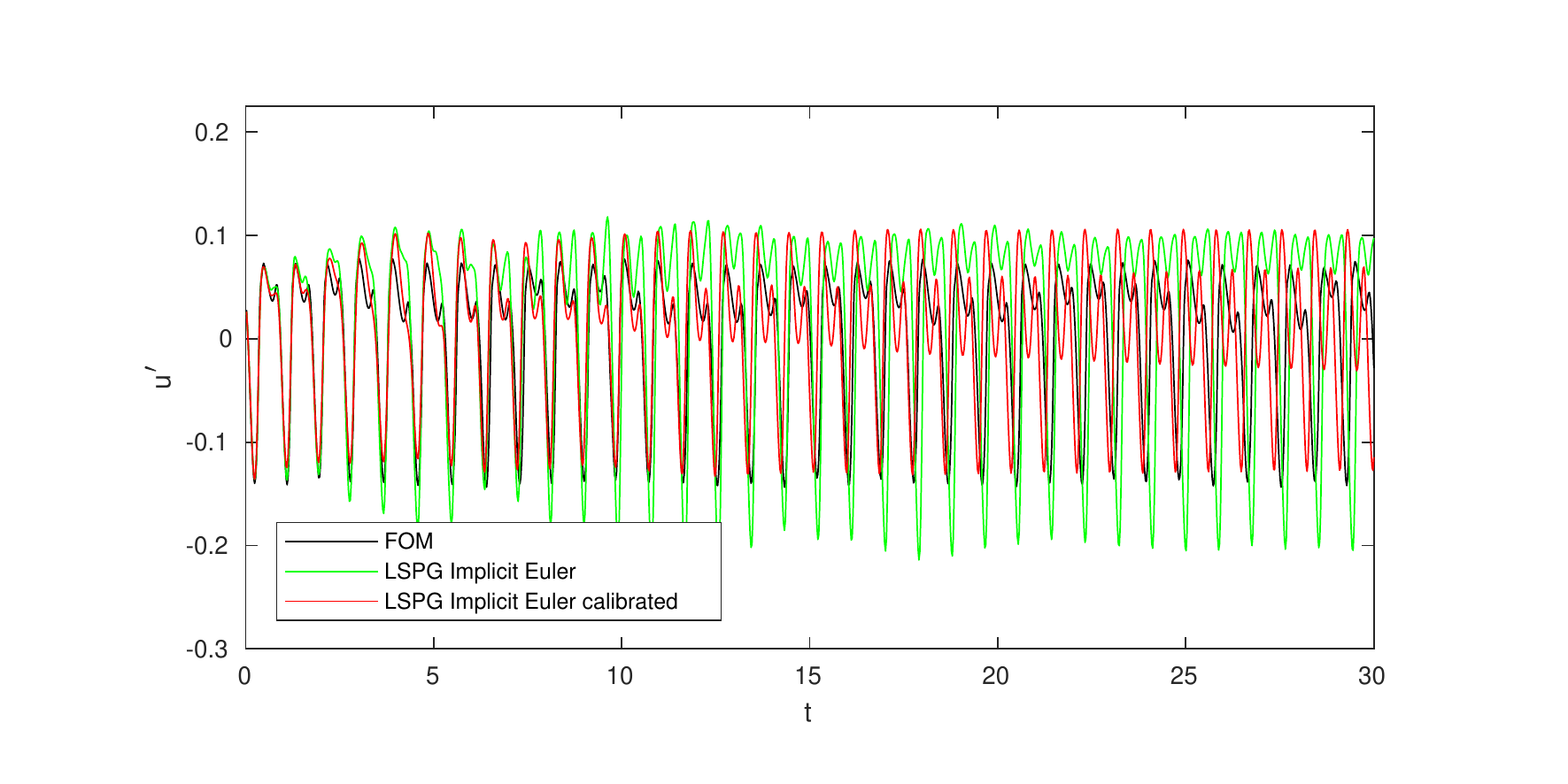}}
    \caption{Time histories of $u$-velocity fluctuations computed by implicit Euler time marching scheme at $(x,y) = (0.6919, 0.0083)$.}
    \label{fig:probe_IE}
\end{figure}

\begin{figure}
        \centering
        \subfigure[Galerkin model]{\includegraphics[width=.65\textwidth,trim={15mm 5mm 20mm 10mm},clip]{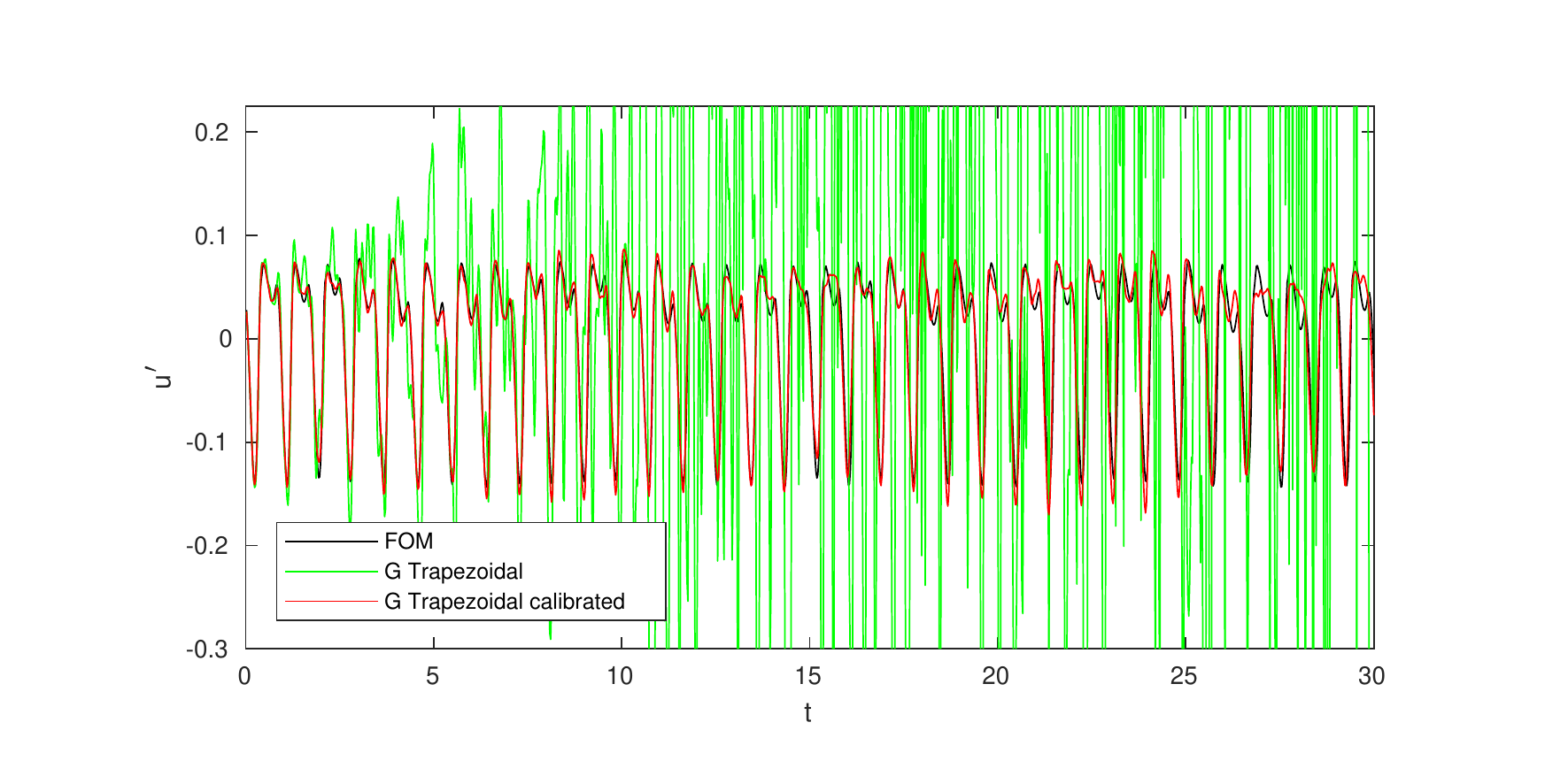}}
         \subfigure[LSPG model]{\includegraphics[width=.65\textwidth,trim={15mm 5mm 20mm 10mm},clip]{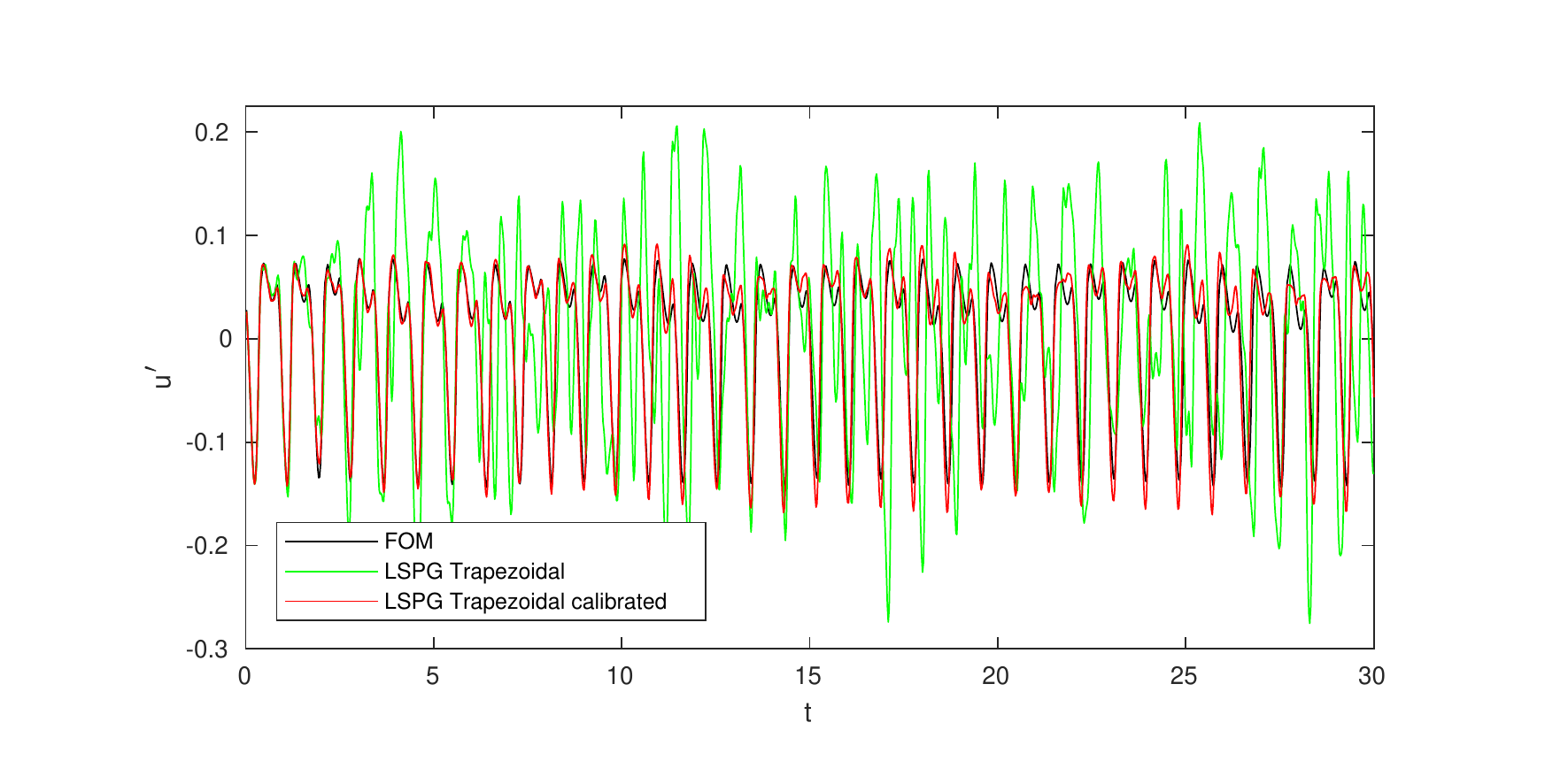}}
    \caption{Time histories of $u$-velocity fluctuations computed by trapezoidal time marching scheme at $(x,y) = (0.6919, 0.0083)$.}
    \label{fig:probe_trap}
\end{figure}

\begin{figure}
        \centering
         \subfigure[Galerkin model]{\includegraphics[width=.65\textwidth,trim={15mm 5mm 20mm 10mm},clip]{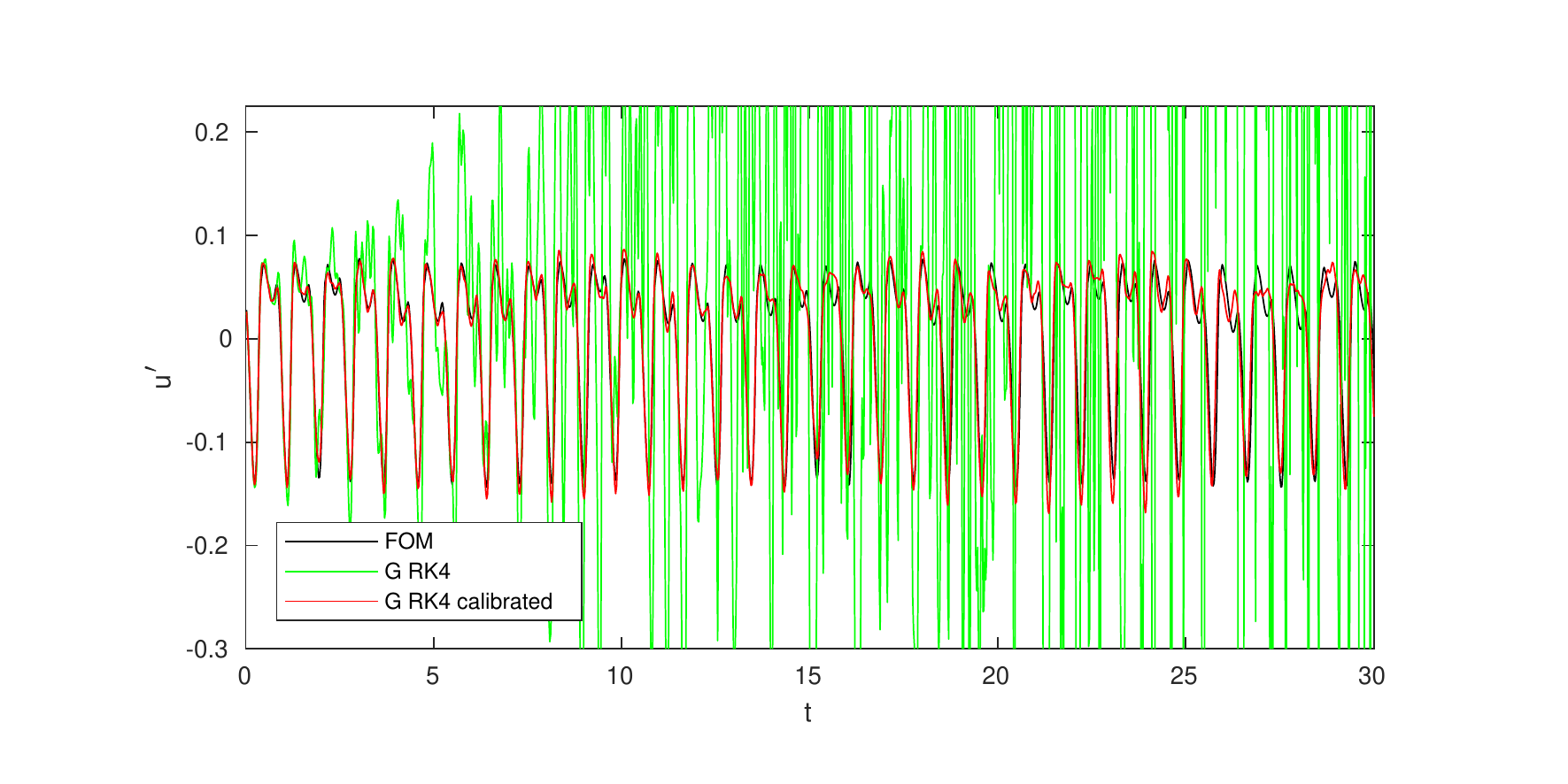}}
         \subfigure[LSPG model]{\includegraphics[width=.65\textwidth,trim={15mm 5mm 20mm 10mm},clip]{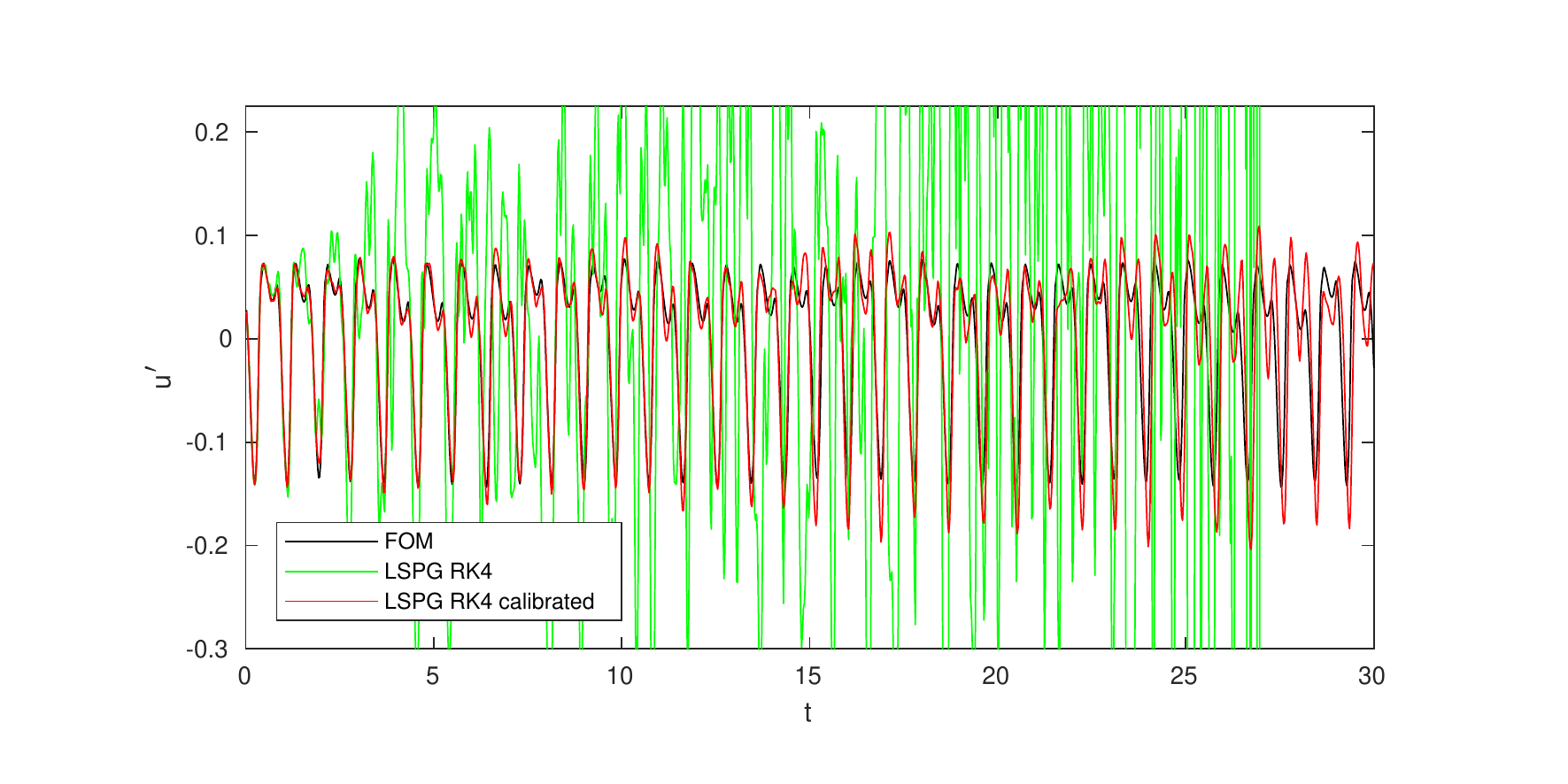}}
    \caption{Time histories of $u$-velocity fluctuations computed by RK4 time marching scheme at $(x,y) = (0.6919, 0.0083)$.}
    \label{fig:probe_RK4}
\end{figure}

Figures \ref{fig:probe_IE}--\ref{fig:probe_RK4} show the time histories of $u$-velocity fluctuations computed by the FOM and the Galerkin and LSPG models. Solutions are presented for the training region ($0 \leq t \leq 15$) and for an extrapolation period for which FOM solutions are available ($15 < t \leq 30$). The impact of model calibration is analyzed for different time marching schemes and, in Fig. \ref{fig:probe_IE}, the implicit Euler method is evaluated. This 1st-order scheme is tested due to its stability properties. As can be observed in the figure, both Galerkin and LSPG models are stable along the training and extrapolation periods even without calibration, but they show a large amplitude error. Calibration leads to an excessive damping for the Galerkin method. On the other hand, the LSPG solution shows a reasonable agreement with the FOM, displaying a small error in amplitude and phase.

In Fig. \ref{fig:probe_trap}, the ROMs are integrated using the trapezoidal method. One should expect an improvement in terms of amplitude and phase errors for this 2nd-order implicit scheme, while retaining stability. For this case, the Galerkin scheme becomes unstable without calibration while the LSPG maintains stability, but presents an inaccurate solution. When calibration is employed, both ROMs become stable and relatively accurate, with small discrepancies in the capture of high-frequency oscillations. It is also worth mentioning that calibration render implicit models much cheaper because of the faster convergence rate towards the solution despite imposing additional terms.
%
%

Solutions obtained for the 
explicit fourth-order Runge Kutta (RK4) scheme are presented in Fig. 
\ref{fig:probe_RK4}. Both Galerkin and LSPG solutions become unstable when this explicit time integration scheme is applied without calibration. However, calibration provides stable Galerkin and LSPG models, with the Galerkin scheme being visually more accurate in terms of amplitude. 

Table \ref{tab:error} shows the computed absolute and root mean-squared errors for the different time integration schemes. The mean-squared error is computed for the fluctuation field which is integrated along the entire mesh from $0\leq t \leq 30$ as
\begin{equation}
	\mbox{Error} = \frac{\| \mathbf{u_{FOM}'} (\mathbf{x}, t) - \mathbf{u_{ROM}'} (\mathbf{x}, t) \|_{L_2}}{\| \mathbf{u_{FOM}'} (\mathbf{x}, t) \|_{L_2}} \mbox{ }
\end{equation}
and, therefore, is prone to any small phase and amplitude differences with respect to the FOM solution. 
Results obtained for diagonally-implicit second and third-order Runge Kutta schemes (DIRK2 and DIRK3) are also included in the table. As can be noticed, calibration not only stabilizes the solutions but leads to a considerable error reduction. Except for the implicit Euler scheme employed with the Galerkin model, the mean-squared error is always reduced when calibration is applied. However, LSPG combined with implicit Euler performed reasonably well and would be the preferred method if calibration was not to be used. The trapezoidal method obtains the most accurate solutions for both the calibrated Galerkin and LSPG models. For the present period of integration, even without calibration, this scheme also obtained stable solutions for the integration period tested, but with large errors. The explicit Runge-Kutta scheme also presents accurate solutions and is a viable option due to its reduced computational cost compared to the implicit methods. Models obtained by the LSPG display larger errors than those computed by the Galerkin in all calibrated cases with the exception of the trapezoidal absolute error. Additionally, in order to evaluate hyper-reduction effects, a calibrated LSPG model using the RK4 without hyper-reduction was built. Although better than its low-cost counterpart, the gapless model was not only incapable of outperforming the Galerkin results but it was also very expensive due to the large number of degrees of freedom in the optimization problem. The root mean-squared and absolute errors for the gapless LSPG-RK4 model are $0.2087$ and $0.5175$, respectively.

\begin{table}[h]
\centering
\begin{tabular}{| c | c | c | c | c | c | c | c | c |}
    \hline
    & \multicolumn{4}{c|} {Root mean-squared error} & \multicolumn{4}{c|} {Absolute error} \\
    \cline{2-9}
    Time  integrator & \multicolumn{2}{c|} {Calibrated} & \multicolumn{2}{c|} {Non-calibrated} & \multicolumn{2}{c|} {Calibrated} & \multicolumn{2}{c|} {Non-calibrated} \\
    \cline{2-9}
    &  Galerkin & LSPG & Galerkin & LSPG &  Galerkin & LSPG & Galerkin & LSPG\\
    \hline
    IE          & 0.6948 & 0.8547 & 0.6135 & 0.8569 
                & 1.0144 & 1.4716 & 1.4120 & 1.8011 \\
    Trapezoidal & 0.1679 & 0.1846 & 5.6267 & 1.1706 
                & 0.4132 & 0.4127 & 22.060 & 1.8693 \\
    DIRK2       & 0.1704 & 0.3788 & 5.3730 & 388.21 
                & 0.4215 & 1.1302 & 15.070 & 4138.8 \\
    DIRK3       & 0.1735 & 0.3672 & 4.3519 & 4.3538 
                & 0.4303 & 1.1074 & 14.334 & 18.721 \\
    RK4         & 0.1733 & 0.3703 & 5.1665 & NaN    
                & 0.4299 & 1.0844 & 17.550 & NaN \\
    \hline
\end{tabular}
\caption{Mean-square and absolute errors for the different time integration schemes.}
\label{tab:error}
\end{table}

\begin{figure}
        \subfigure[FOM]{\includegraphics[width=.32\textwidth,trim={10mm 5mm 38mm 14mm},clip]{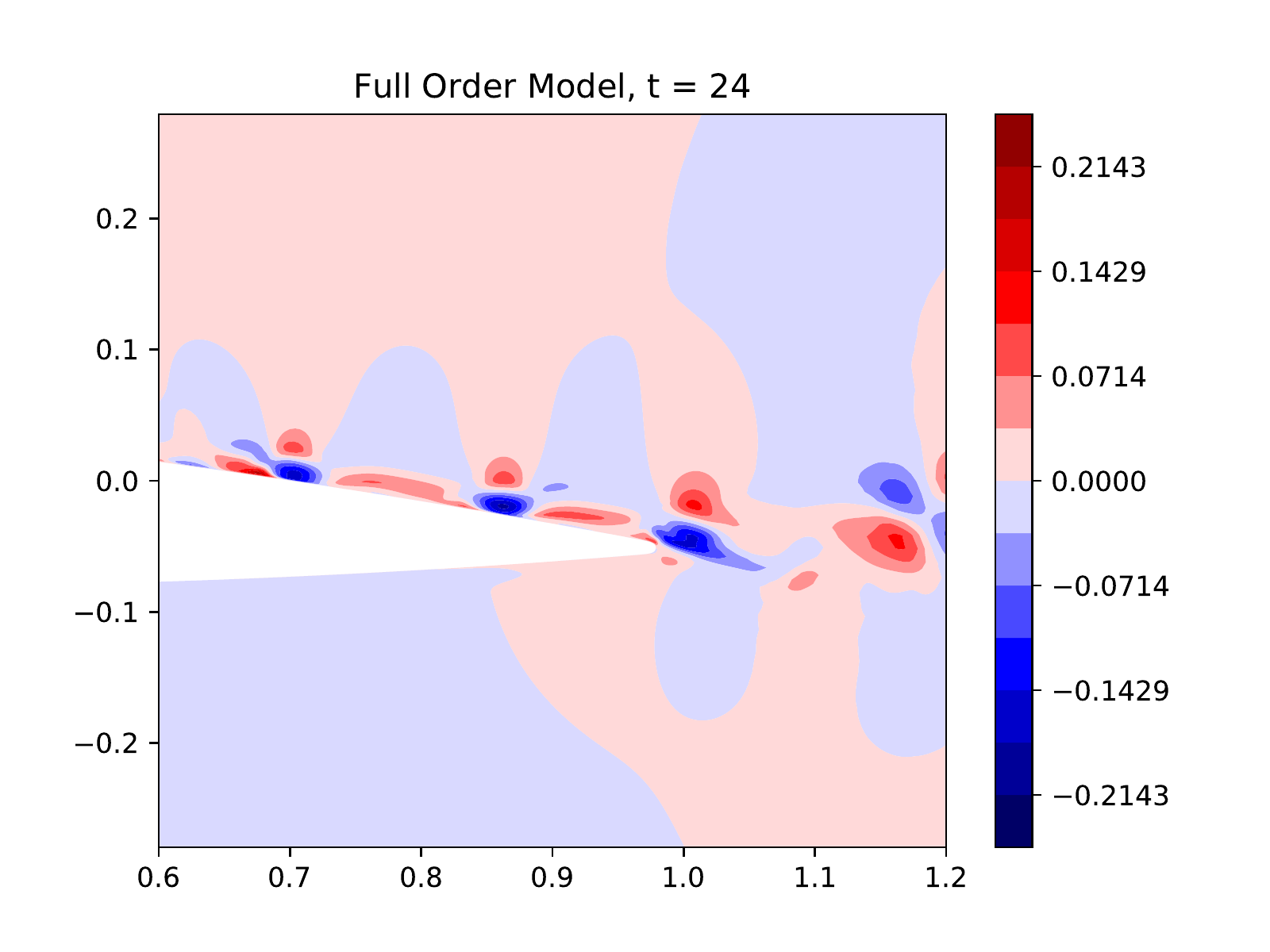}}
        \subfigure[Galerkin]{\includegraphics[width=.32\textwidth,trim={10mm 5mm 38mm 14mm},clip]{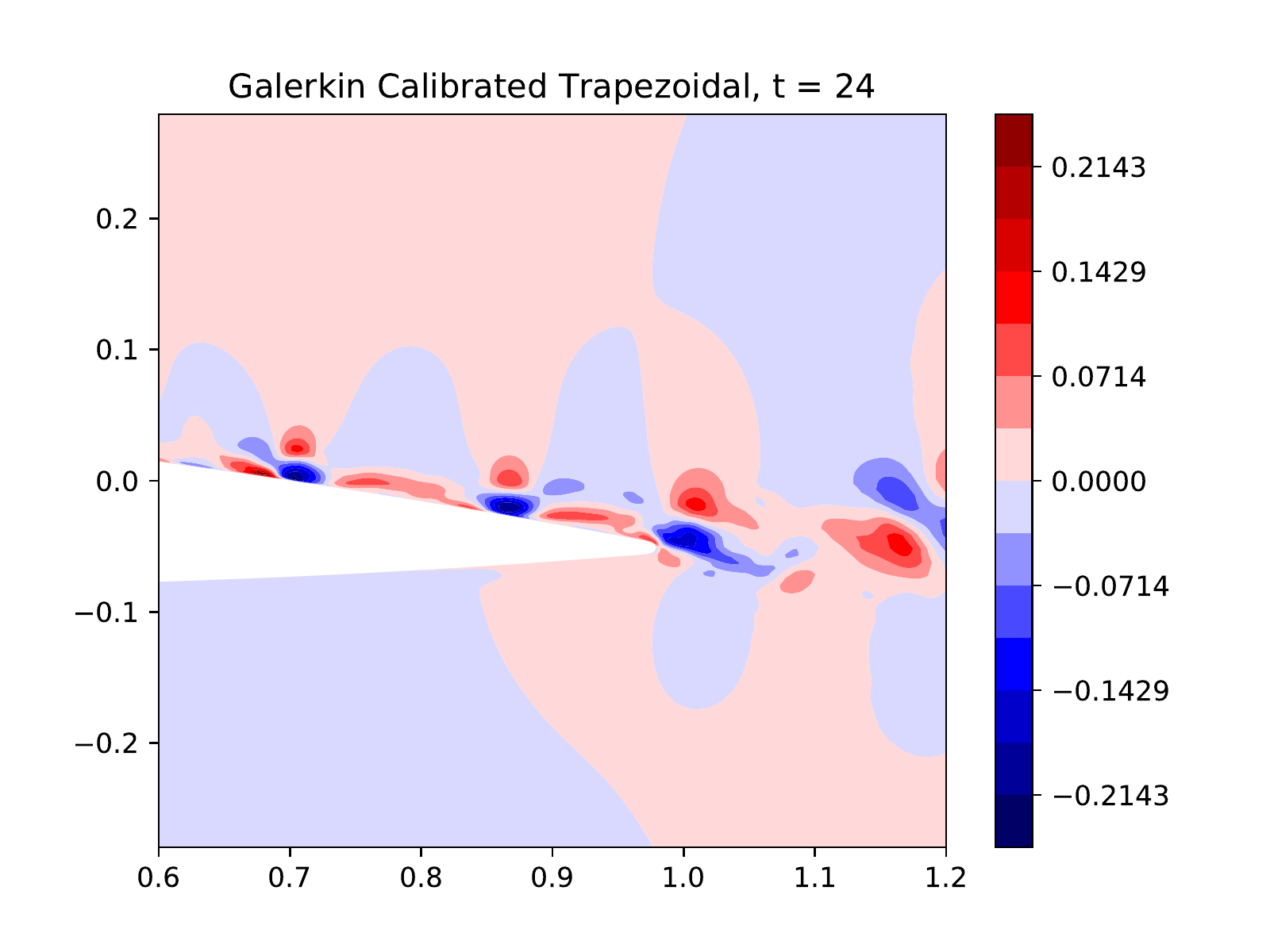}}
        \subfigure[LSPG]{\includegraphics[width=.32\textwidth,trim={10mm 5mm 38mm 14mm},clip]{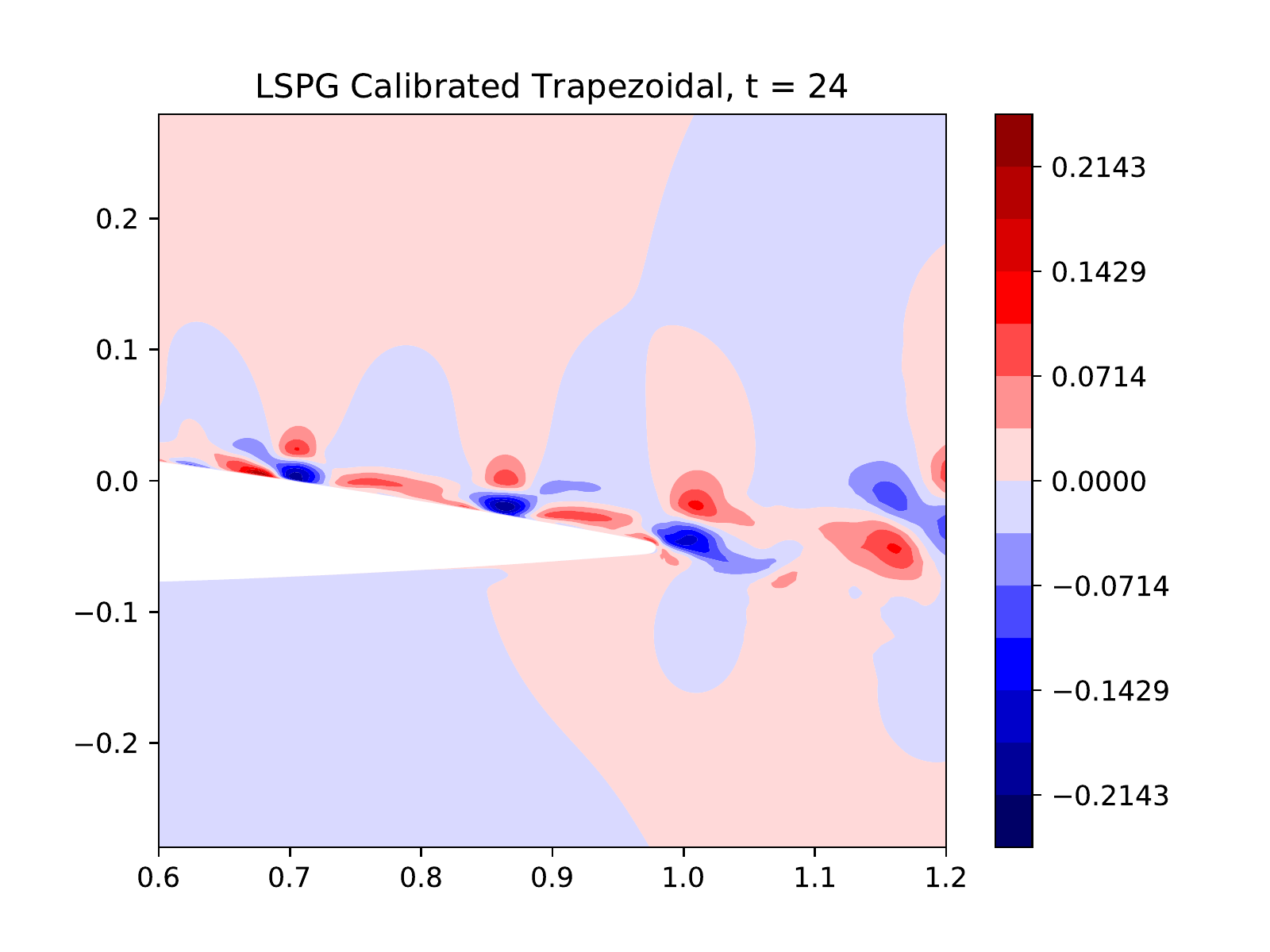}}
    \caption{Contours of $u$-velocity computed at $t = 24$ using calibrated models with trapezoidal integration.}
    \label{fig:u_contours}
\end{figure}

\begin{figure}
        \subfigure[FOM]{\includegraphics[width=.32\textwidth,trim={10mm 5mm 38mm 14mm},clip]{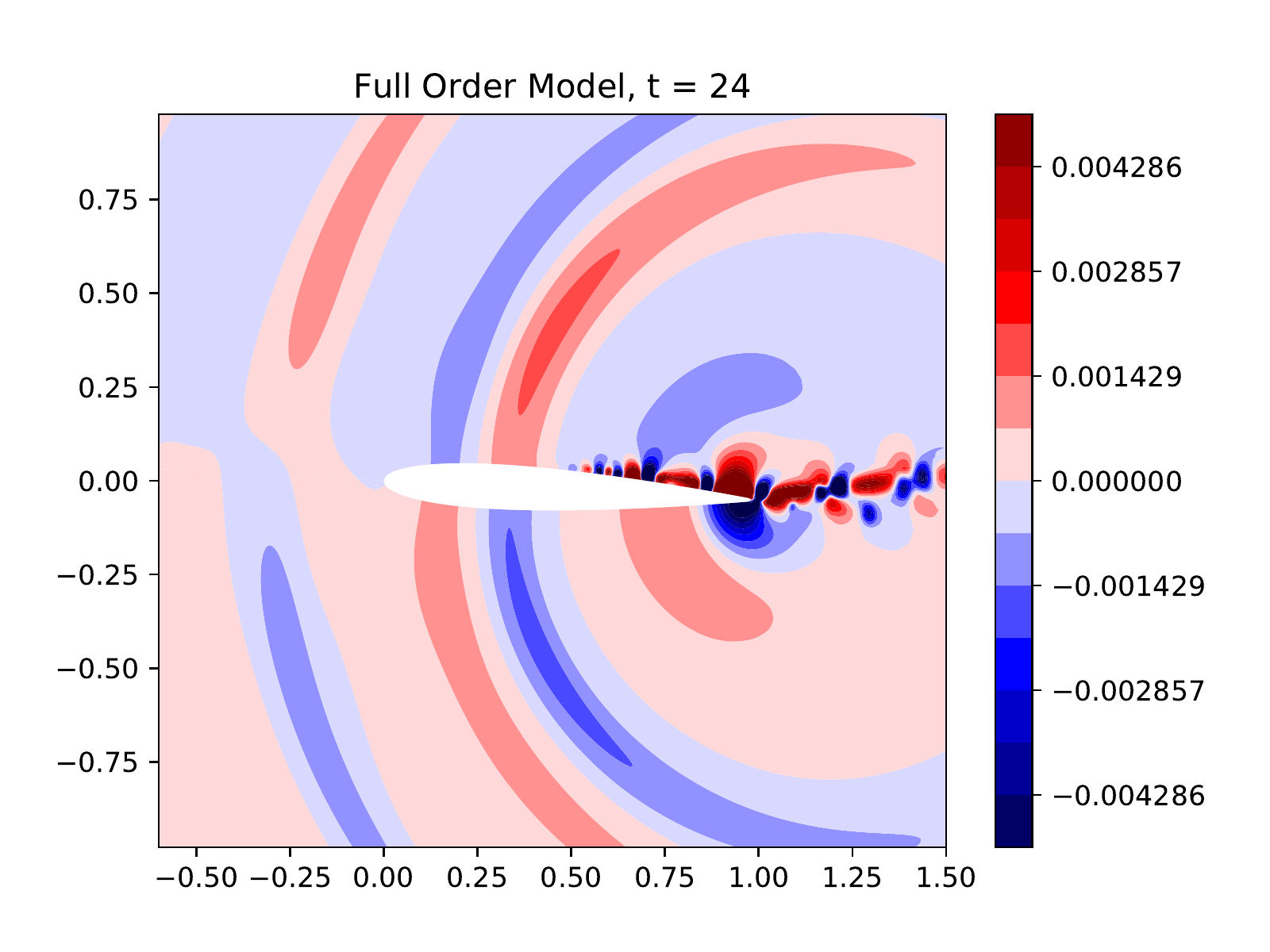}}
        \subfigure[Galerkin]{\includegraphics[width=.32\textwidth,trim={10mm 5mm 38mm 14mm},clip]{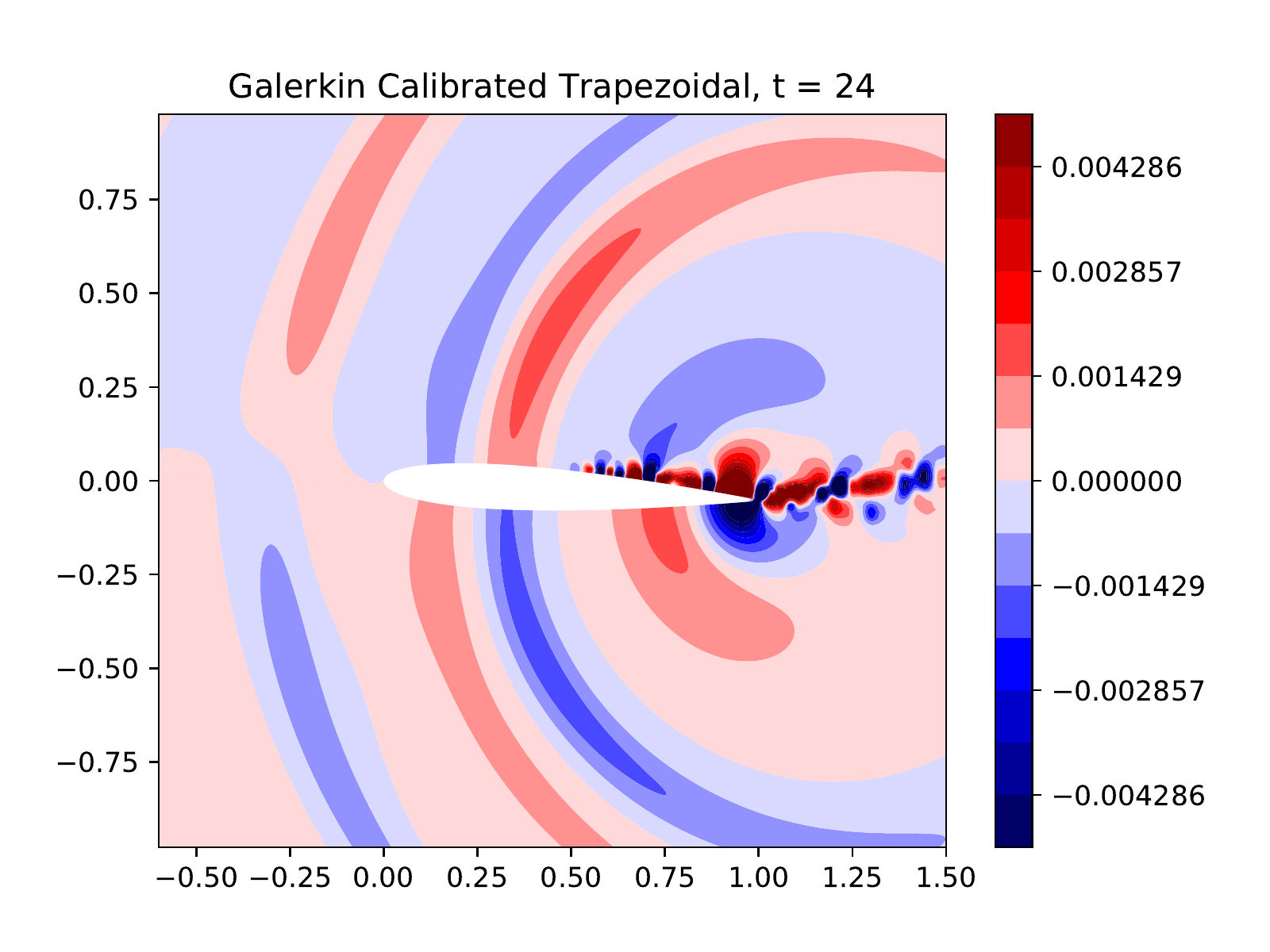}}
        \subfigure[LSPG]{\includegraphics[width=.32\textwidth,trim={10mm 5mm 38mm 14mm},clip]{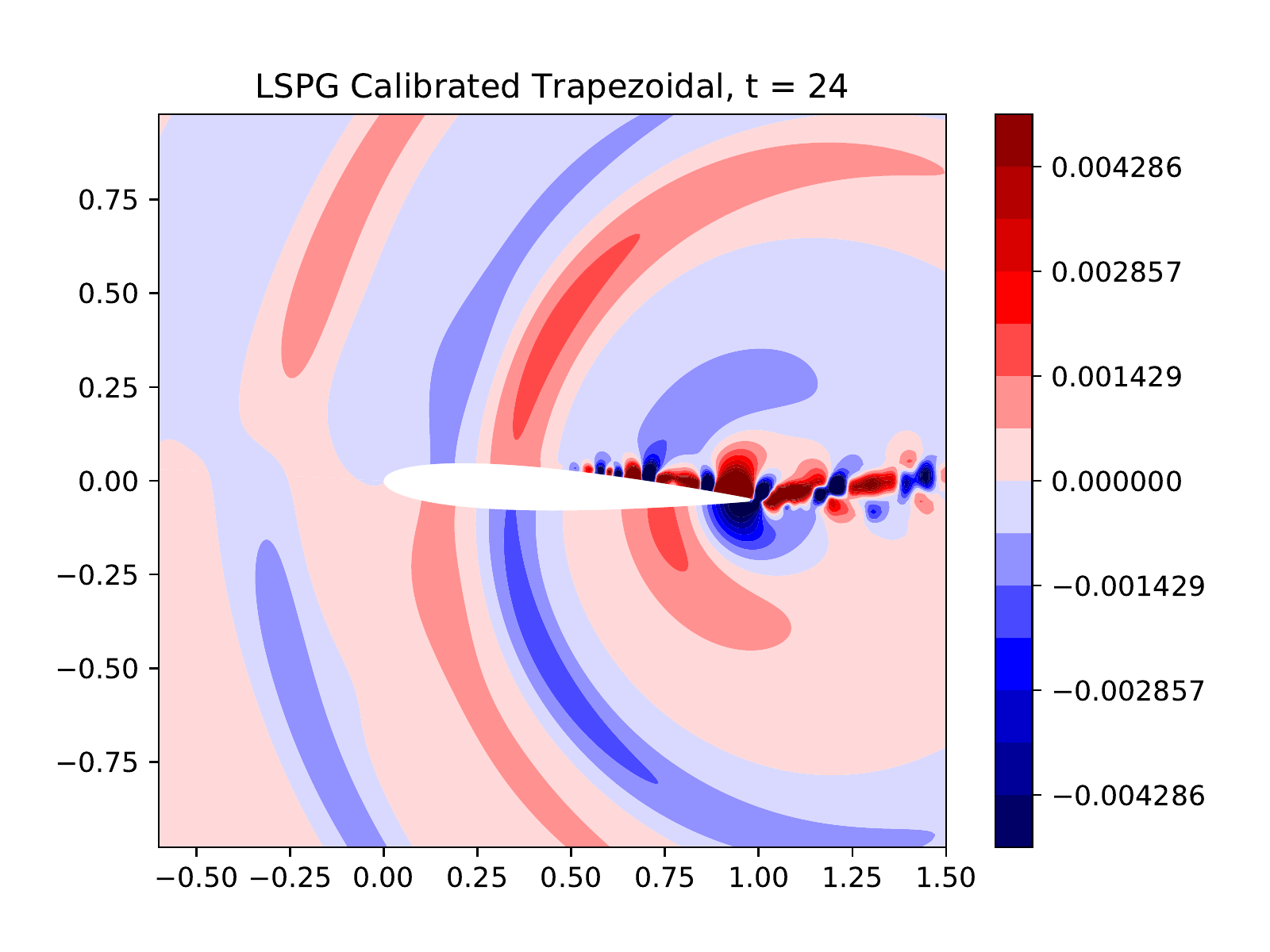}}
    \caption{Pressure contours computed at $t = 24$ using calibrated models with trapezoidal integration.}
    \label{fig:p_contours}
\end{figure}

Snapshots of $u$-velocity and pressure fluctuations obtained at $t = 24$ are presented in Figs. \ref{fig:u_contours} and \ref{fig:p_contours}, respectively. These figures allow a comparison of results between the FOM and calibrated ROMs using the trapezoidal method for time integration. Although small discrepancies between the ROM and FOM solutions can be observed, especially for small-scale flow structures, the main features of the flow are recovered. In the FOM, vortex merging taking place at the mid-chord location leads to flow instabilities that develop along the boundary layer. Figure \ref{fig:u_contours} shows the velocity fluctuations resulting from these instabilities along the trailing-edge region. As one can see, both the Galerkin and LSPG models are able to capture the relevant flow features. Acoustic scattering occurs due to the advection of the flow instabilities along the trailing edge, what leads to sound waves that propagate upstream closing a feedback loop mechanism \cite{RICCIARDI202054}. Figure \ref{fig:p_contours} shows that the ROMs accurately capture the acoustic waves generated at the trailing edge. In this figure, one can also observe a hydrodynamic wavepacket that is generated at the airfoil mid-chord being transported along the boundary layer towards the trailing edge. 


%








\subsection{Supersonic flow past NACA0012 airfoil}

The second test case investigated is a supersonic flow over a NACA0012 airfoil. The freestream Mach number is set as $M_{\infty}=1.2$, the Reynolds number is $Re_c =$ 80,000 and the airfoil is at 6 deg. angle of attack. The FOM includes a transient solution where a detached bow shock wave forms at the airfoil leading edge while a fish-tail shock forms on the trailing edge. The bow shock propagates upstream until it settles in front of the airfoil while the oblique tail shock remains almost stationary. A large-scale starting vortex appears at the early stages of the flow and, later, a separation bubble forms along the airfoil suction side leading to vortex shedding which interacts with the oblique shock at the trailing edge.

\begin{figure}
    \centering
    \subfigure[Mode 1.]{\includegraphics[width=.45\textwidth,trim={20mm 20mm 20mm 20mm},clip]{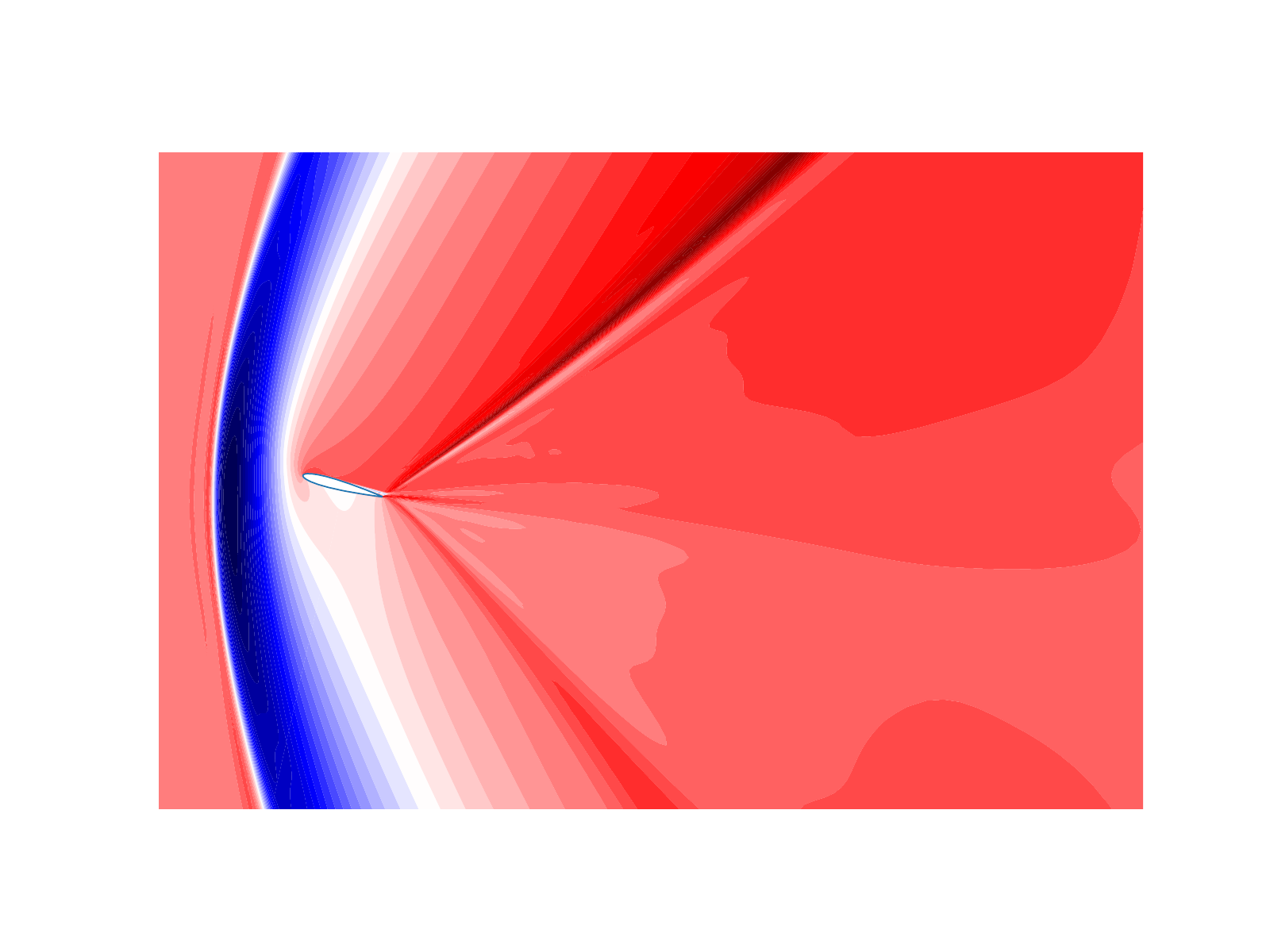}}
    \subfigure[Mode 5.]{\includegraphics[width=.45\textwidth,trim={20mm 20mm 20mm 20mm},clip]{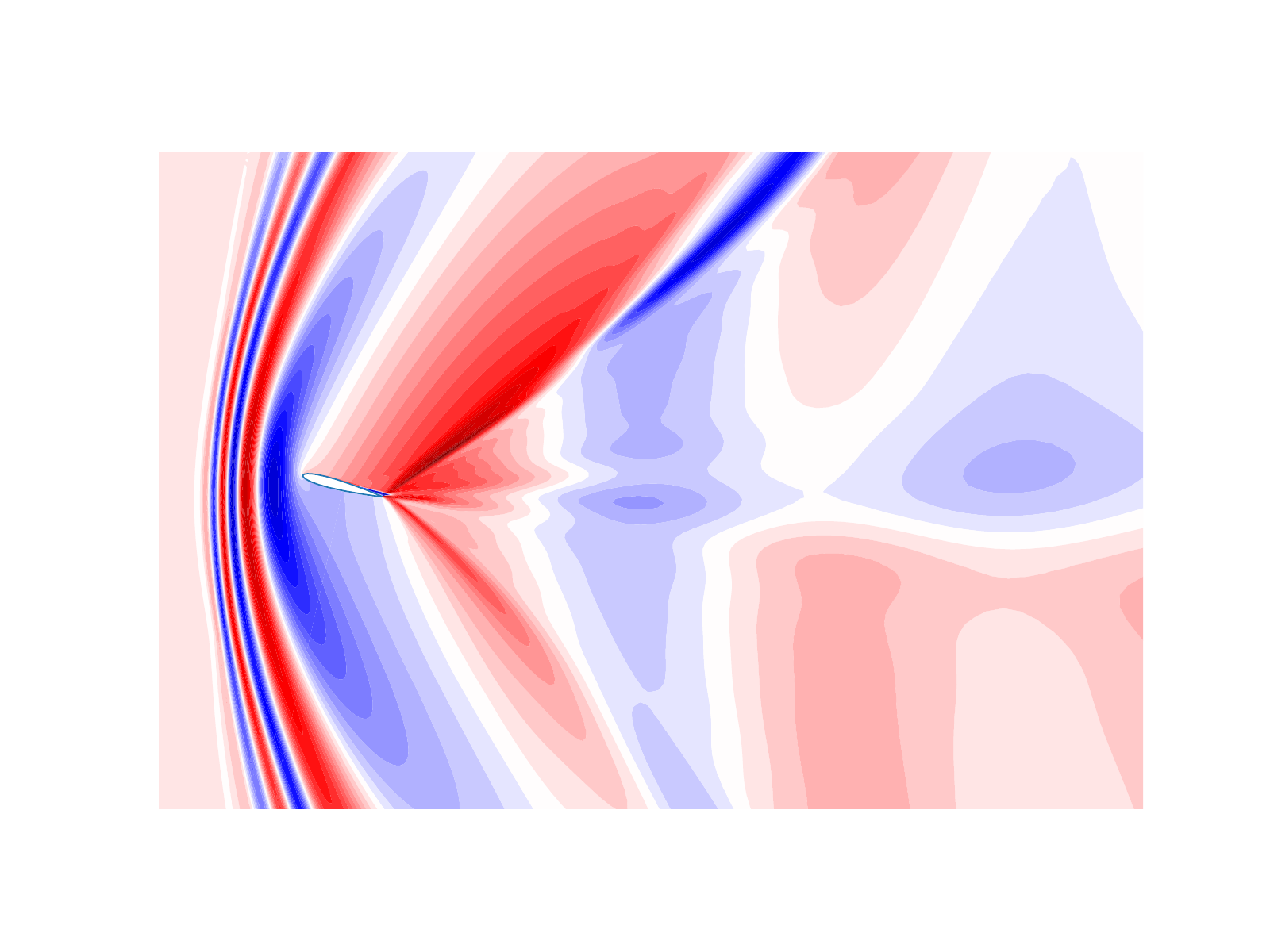}}
    \subfigure[Mode 10.]{\includegraphics[width=.45\textwidth,trim={20mm 20mm 20mm 20mm},clip]{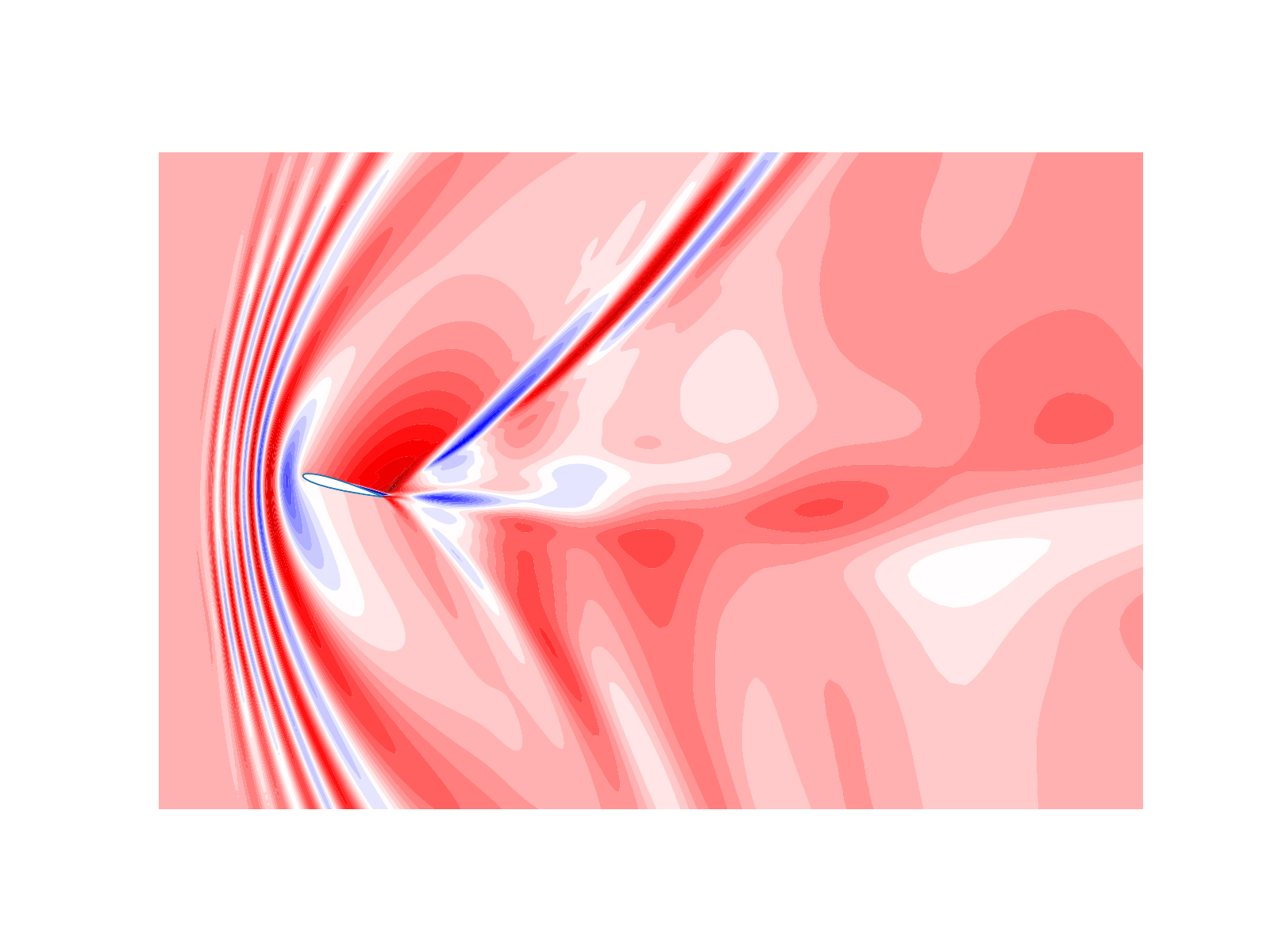}}
    \subfigure[Mode 15.]{\includegraphics[width=.45\textwidth,trim={20mm 20mm 20mm 20mm},clip]{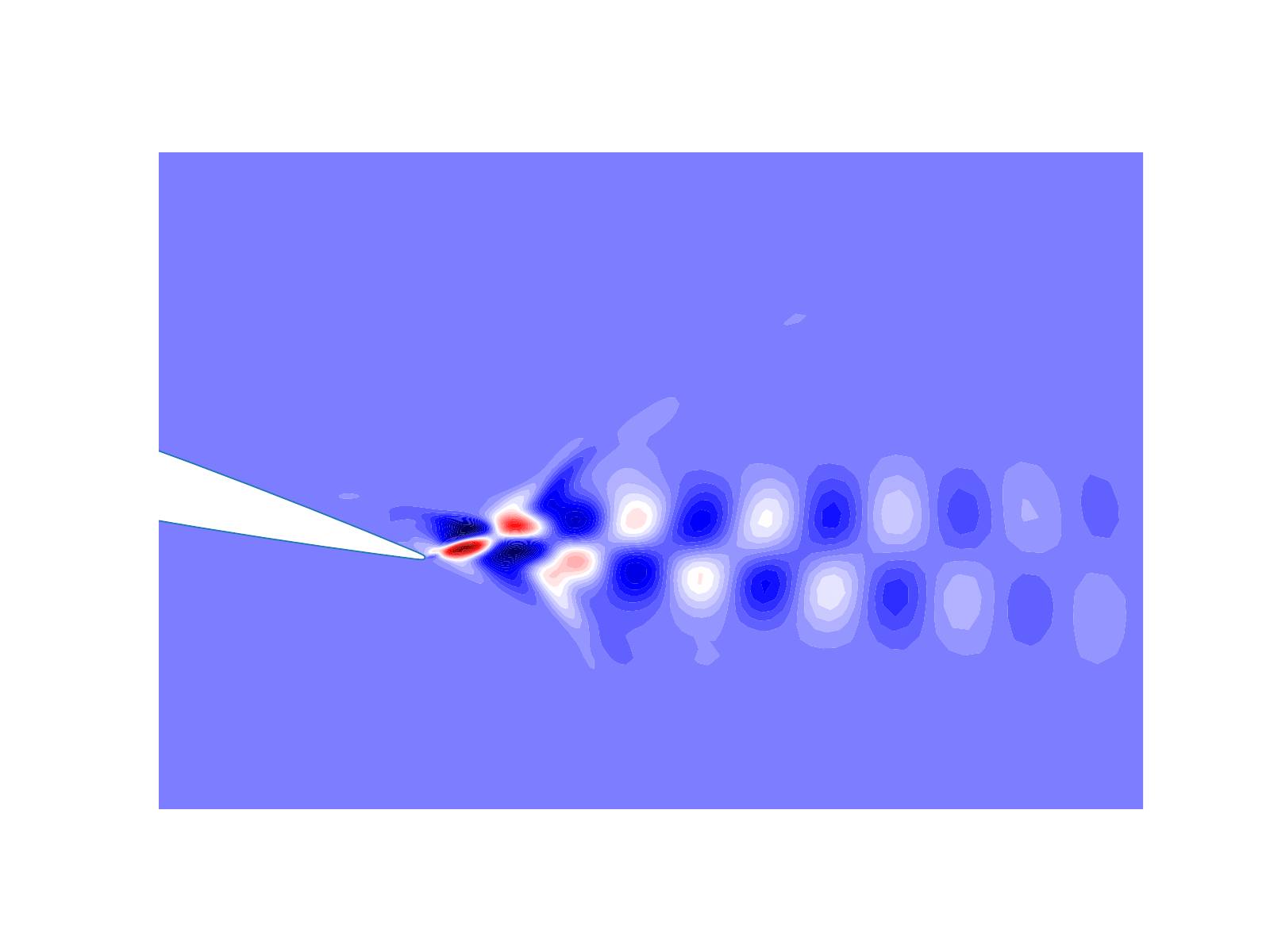}}
    \caption{Contours of POD eigenfunctions for $u$-velocity.}
    \label{fig:space_modes_miotto}
\end{figure}

Some of the flow features described above can be observed in the POD spatial modes computed for the $u$-velocity component in Fig. \ref{fig:space_modes_miotto}. The first mode shown in this figure is solely related to the shock waves and is similar to a mean flow where the detached shock appears stationary upstream the airfoil. Modes 5 and 10 exhibit an oscillatory behavior upstream the airfoil to represent the motion of the bow shock. In these modes, oblique shock waves are observed on the trailing edge and they also have an oscillatory pattern to model the initial transient of the fish-tail shock. Downstream the airfoil, one can also see some oscillations that represent the advection of the starting vortex. Mode 15 shows the oscillatory pattern of vortex shedding which is formed at the trailing edge after the initial transient.
%

\begin{figure}
    \centering
    \includegraphics[scale=.6,trim={10mm 5mm 20mm 10mm},clip]{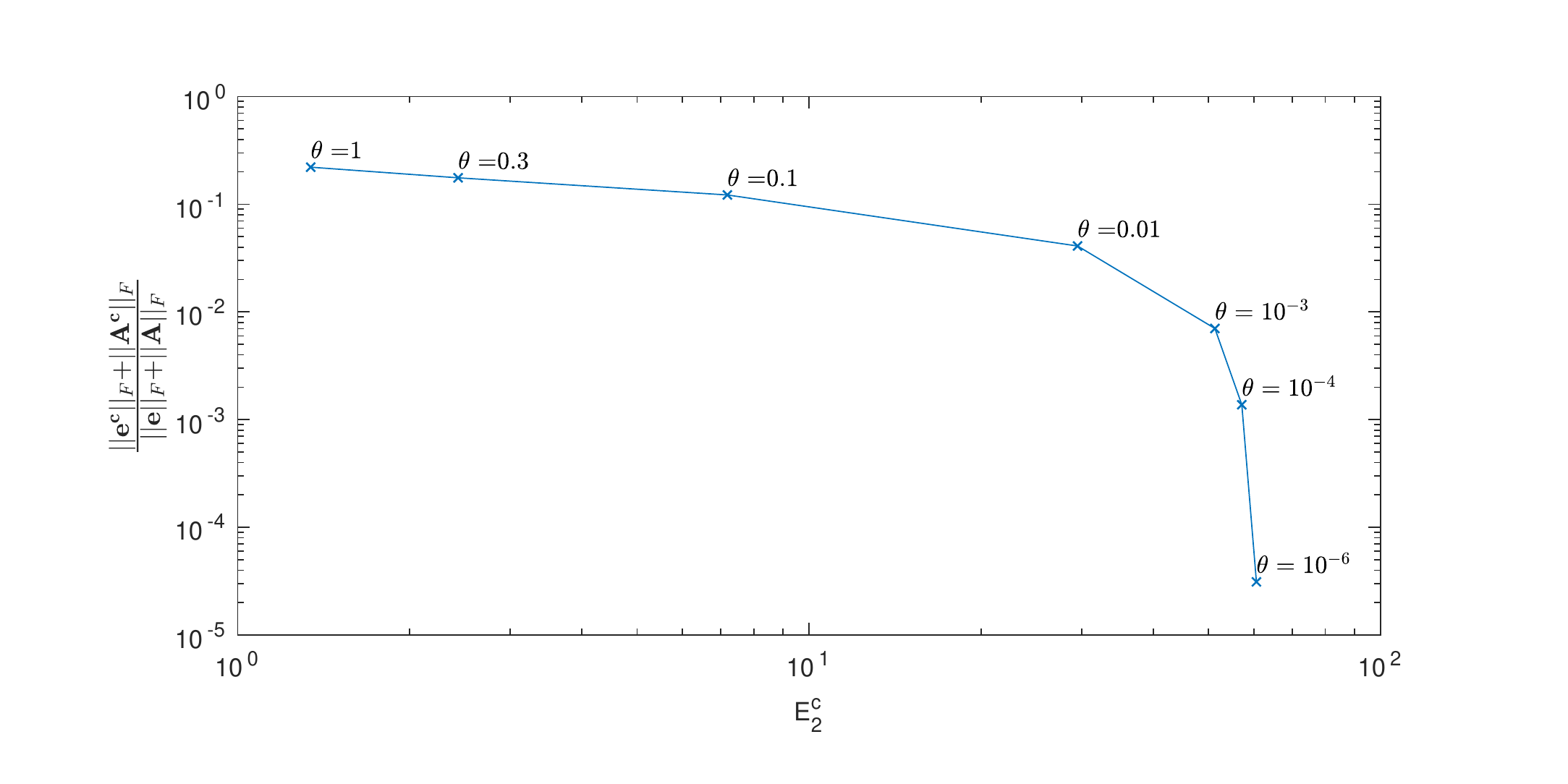}
    \caption{Ratio of Frobenius norms computed for calibrated and Galerkin coefficients as a function of approximation error for different values of $\theta$ (see Eq. \ref{eq:calibration} for details).} 
    \label{fig:theta_miotto}
\end{figure}

Results obtained by the FOM are sampled for $10{,}000$ snapshots with a constant non-dimensional time step $\Delta t_{snapshot} = 0.01$. The first $7{,}000$ snapshots, which represent 70 non-dimensional time units, are used to construct a reduced-order basis by the snapshot-POD method. This period is sufficient to capture the transient motion of the detached shock wave. The ROMs are built using the first $30$ POD modes which contain $99.91 \%$ of the model RIC. The Galerkin models are built using the maximum calibration parameter ($\theta = 1$) and this choice is made after an analysis of Fig. \ref{fig:theta_miotto} that shows the low intrusiveness of the calibration terms  relatively to the original Galerkin operators. For $\theta=1$, the error $E_2^c$ is minimized and the ratio of the Frobenius norms from the calibrated and Galerkin terms is still lower than unity.

\begin{figure}
    \centering
    \subfigure[20,000 points.]{\includegraphics[width=.3\textwidth,trim={20mm 20mm 20mm 20mm},clip]{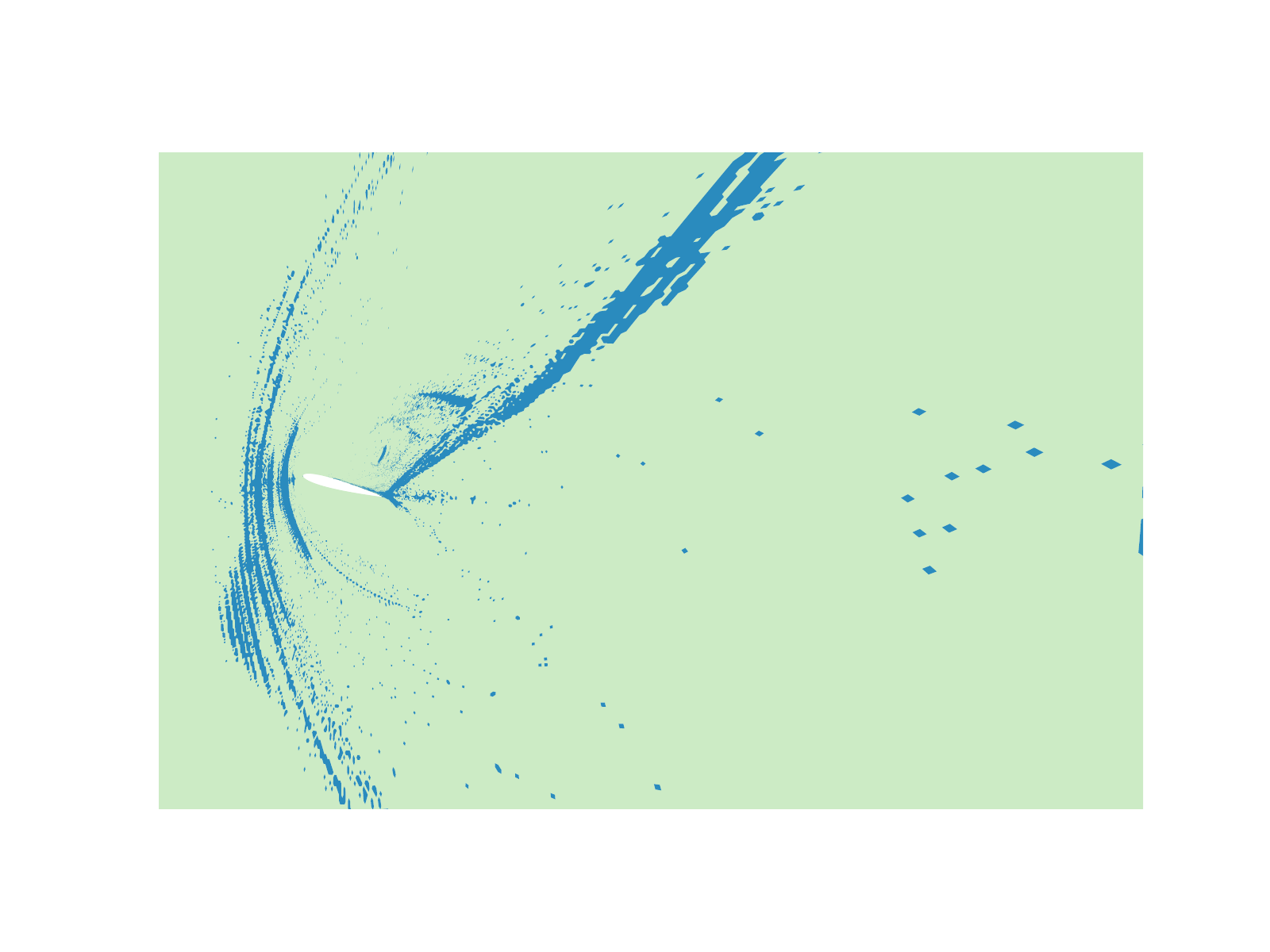}}
    \subfigure[50,000 points.]{\includegraphics[width=.3\textwidth,trim={20mm 20mm 20mm 20mm},clip]{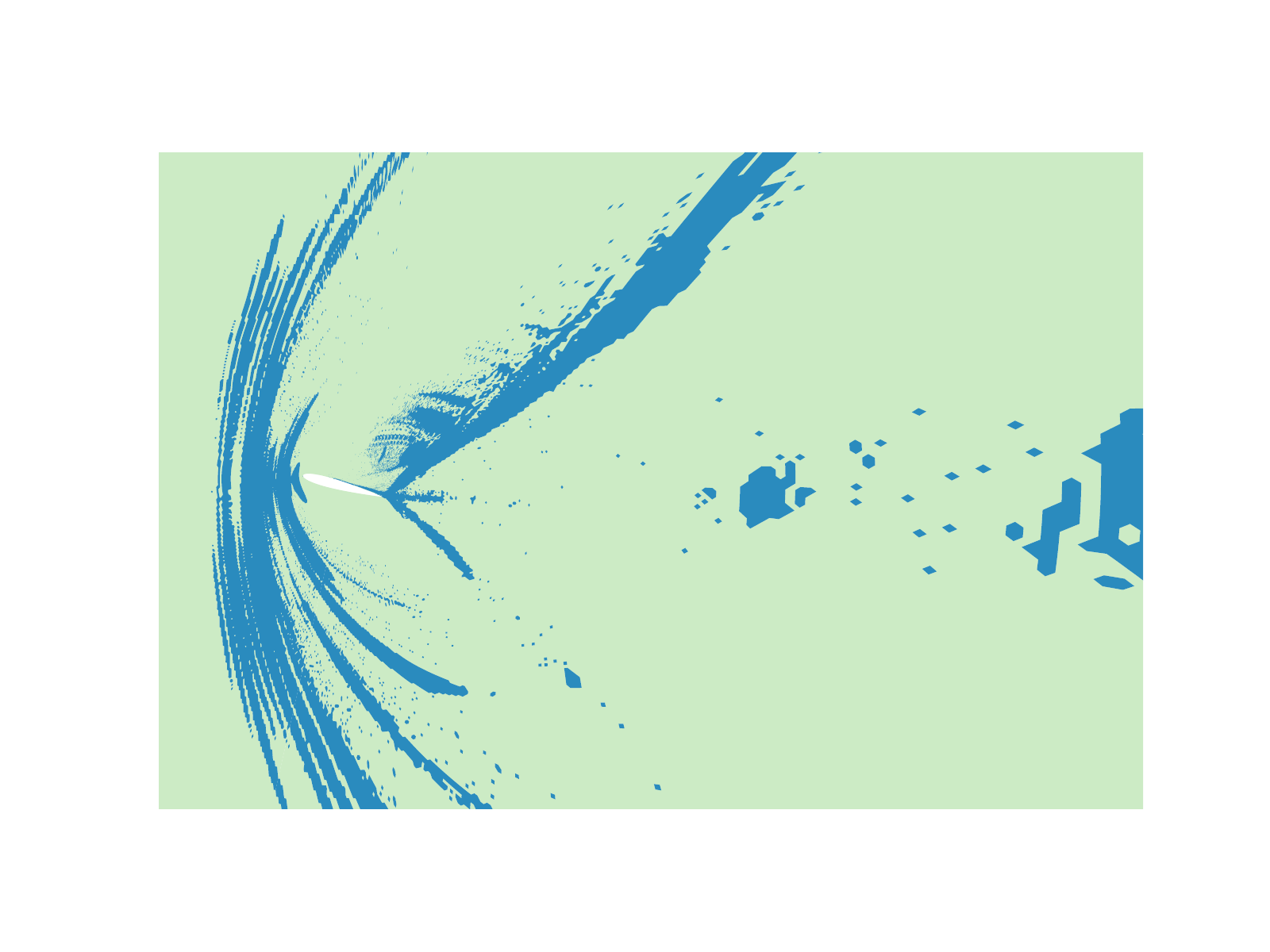}}
    \subfigure[100,000 points.]{\includegraphics[width=.3\textwidth,trim={20mm 20mm 20mm 20mm},clip]{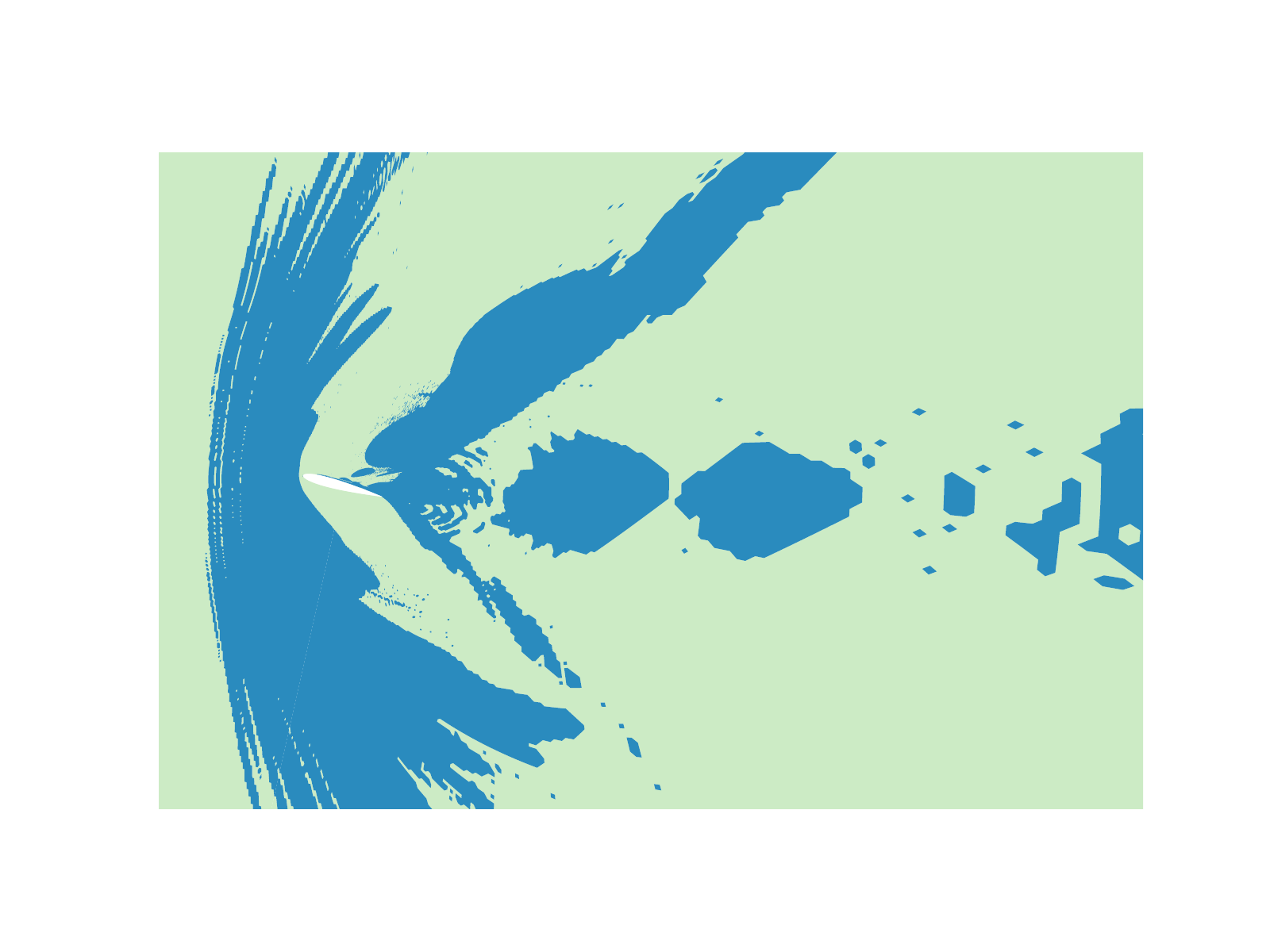}}
    \caption{Sample points (in blue) chosen by the accelerated greedy-MPE hyper-reduction algorithm.}
    \label{fig:hyper_shock}
\end{figure}

The application of hyper-reduction for the present transient flow is a challenge since the entire pathway of motion of the detached shock wave has to be captured by the sampled points. Moreover, the initial transient also includes the starting vortex that is advected along the airfoil wake. Figure \ref{fig:hyper_shock} shows the sampled points chosen by different levels of the accelerated greedy-MPE hyper-reduction. As shown in Fig. \ref{fig:hyper_shock}(a), only the initial pathway of the bow shock appears in first 20,000 points selected by the hyper-reduction. At the same time, the stronger oblique shock on the airfoil suction side is better represented by the sampled points. Figure \ref{fig:hyper_shock}(b) shows that a less aggressive hyper-reduction with 50,000 points captures a wider pathway of the detached shock and adds sample points to both oblique shocks at the trailing edge. Finally, when 100,000 points (out of $\approx$300,000) are employed in the hyper-reduction, the entire pathway of the bow shock is captured besides the tail shocks and near wake. Moreover, some sample points are chosen along the wake to include the motion of the starting vortex.
For the present transient flow, the application of hyper-reduction is not as effective as for the previous subsonic flow since the detached shock wave moves upstream. Hence, in order to capture the entire motion of the shock, the hyper-reduction needs to include its entire pathway ahead of the airfoil, increasing the overall cost of the LSPG ROM.

\begin{figure}
    \begin{subfigure}
        \centering
        \includegraphics[width=.8\textwidth,trim={-20mm 5mm 20mm 5mm},clip]{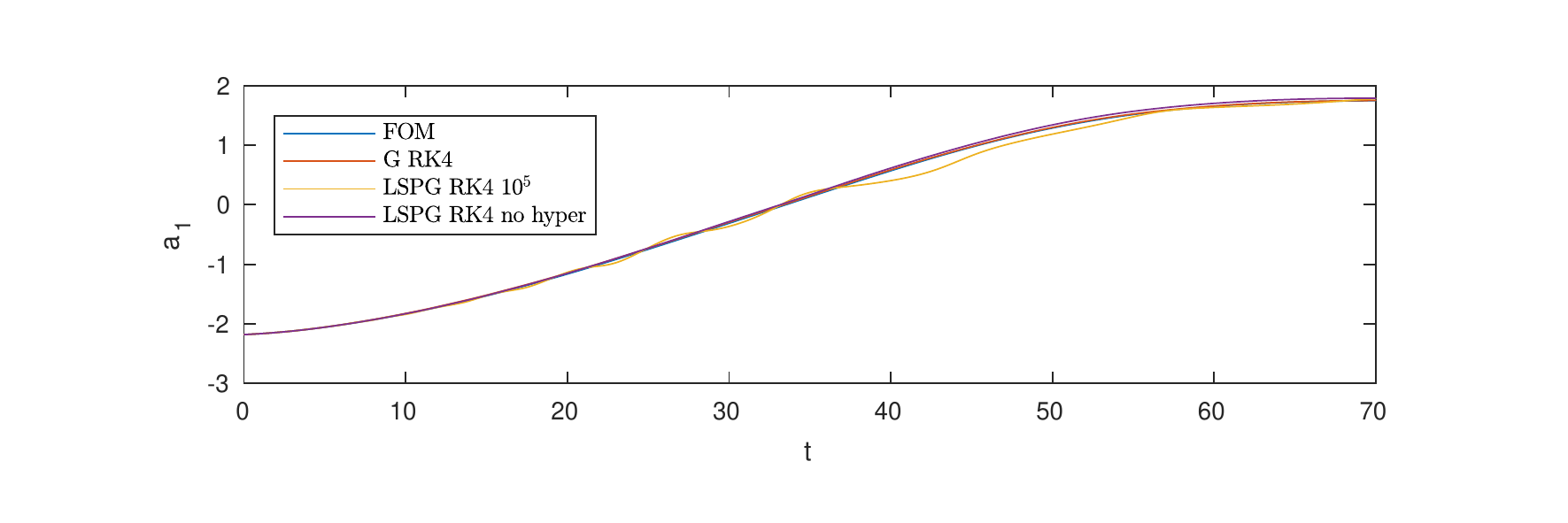}
    \end{subfigure}
    \begin{subfigure}
        \centering
        \includegraphics[width=.8\textwidth,trim={-20mm 5mm 20mm 5mm},clip]{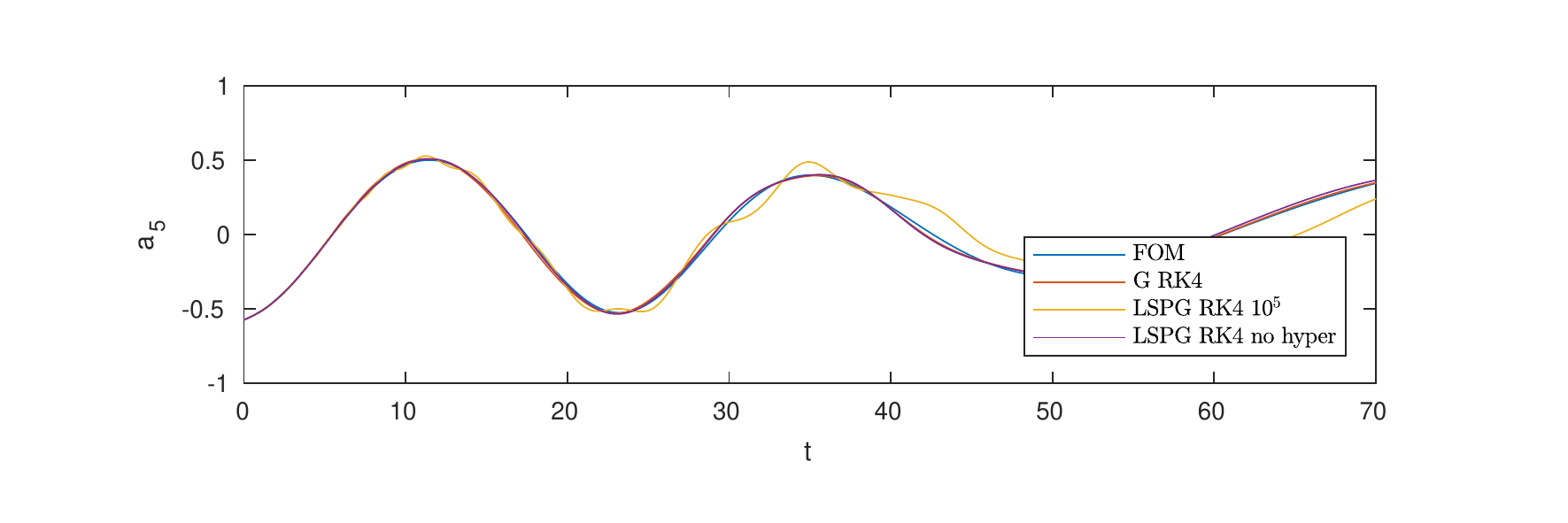}
    \end{subfigure}
    \begin{subfigure}
        \centering
        \includegraphics[width=.8\textwidth,trim={-20mm 5mm 20mm 5mm},clip]{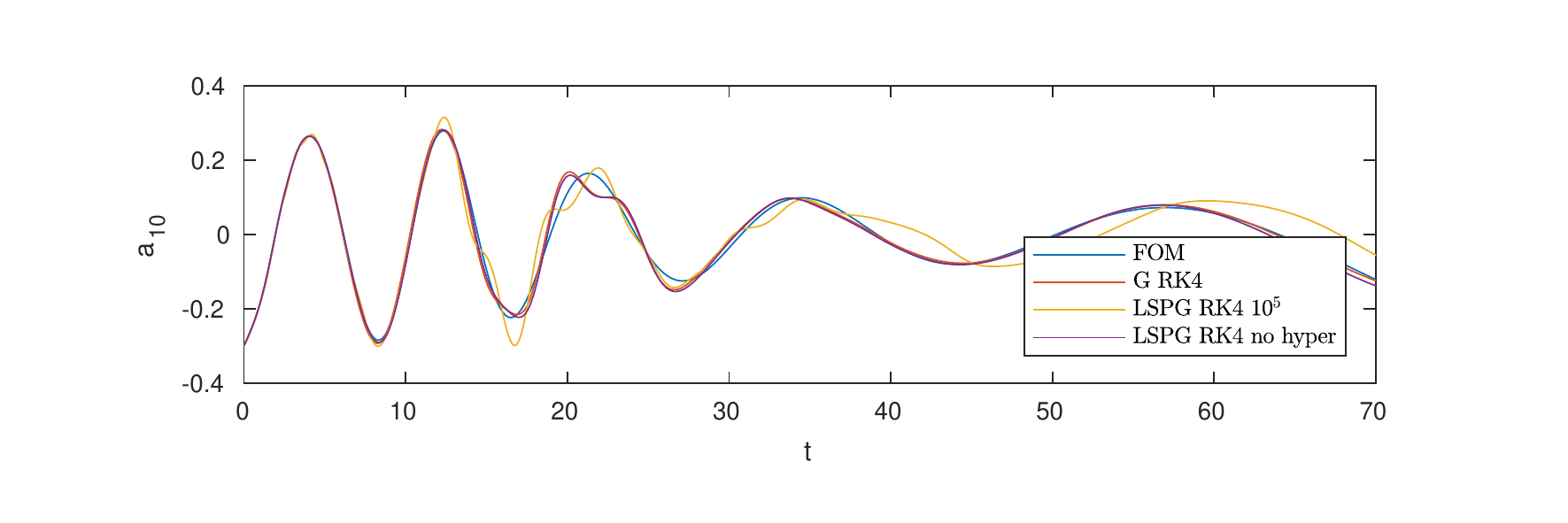}
    \end{subfigure}
    \begin{subfigure}
        \centering
        \includegraphics[width=.8\textwidth,trim={-20mm 5mm 20mm 5mm},clip]{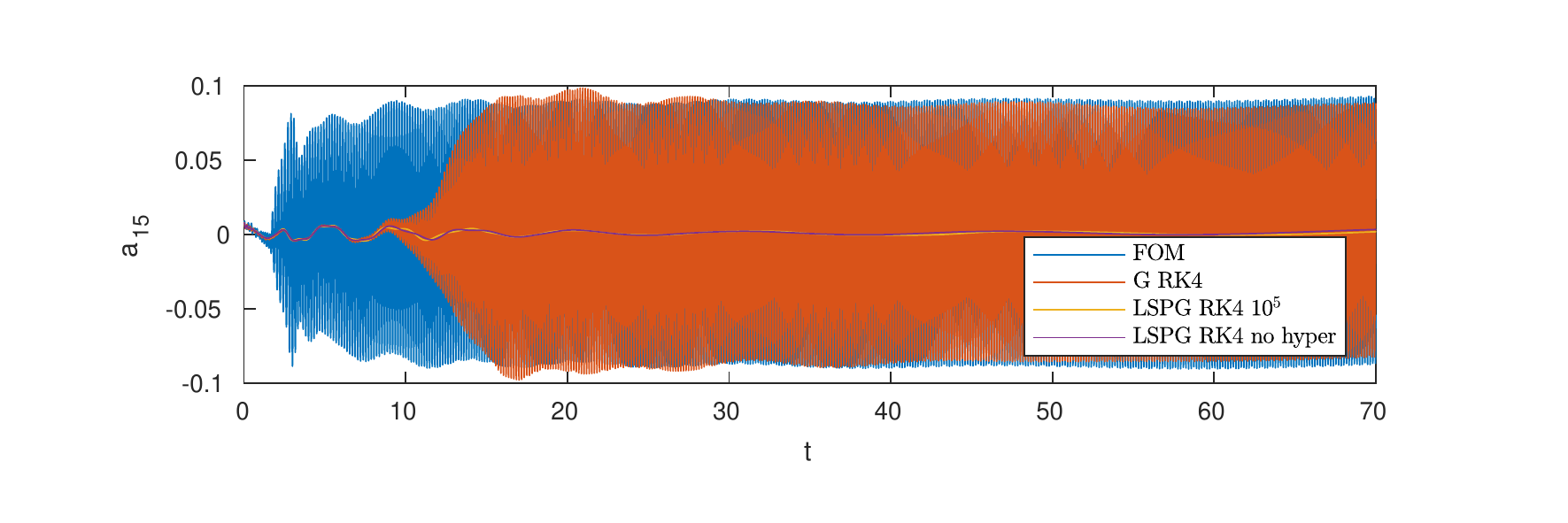}
    \end{subfigure}
    \caption{Temporal dynamics of POD modes $1$, $5$, $10$ and $15$. The LSPG is tested with ($10^5$ sample points) and without hyper-reduction.}
    \label{fig:time_modes_miotto}
\end{figure}

In Fig. \ref{fig:time_modes_miotto}, results are presented for calibrated ROMs computed with the explicit fourth-order Runge-Kutta (RK4) time integrator for the LSPG method with and without hyper-reduction and for the Galerkin method. Without calibration, all ROMs obtained by the RK4 became unstable. The LSPG method is presented for a hyper-reduction including 100,000 sample points. For more aggressive applications of the accelerated greedy-MPE algorithm, the LSPG method could not provide stable ROMs. In the figure, temporal modes 1, 5, 10 and 15 are shown and, as observed before, the first three modes are associated with the shock motion, being well recovered by both Galerkin and LSPG methods without hyper-reduction. When the LSPG method is tested with hyper-reduction, a small phase and amplitude distortion is observed for modes 5 and 10. Mode 1 appears as a moving average since the present flow is transient and a true mean flow is not obtained for the entire period of simulation. Modes 5 and 10 are composed of low frequencies which have some modulation while mode 15 is governed by the vortex shedding dynamics. This latter mode is composed of much higher frequencies (compared to modes 5 and 10) with an initial growth and modulation. For the initial time period $(0<t<10)$, both the Galerkin and LSPG ROMs are not able to capture the initial growth of the vortex shedding oscillations but they capture the low-frequency dynamics of the modulation similarly to a low-pass filter. Beyond $t>10$, the calibrated Galerin model is able recover the periodical dynamics of this mode while the LSPG model has a more dissipative behavior and cannot reproduce the high-frequency oscillations. In the Galerkin model, the delay in the vortex shedding development may occur because the initial flow period coincides with the motion of the shock waves which are associated with more important singular values of the POD decomposition. Once the shocks are reaching their steady-state, the calibrated model is able to capture the less energetic fluctuations associated with vortex shedding.
%



\begin{figure}
    \centering
    \subfigure[Probe location indicated by the purple symbol in Fig. \ref{fig:vort_div}.]{\includegraphics[scale=.75,trim={5mm 5mm 10mm 10mm},clip]{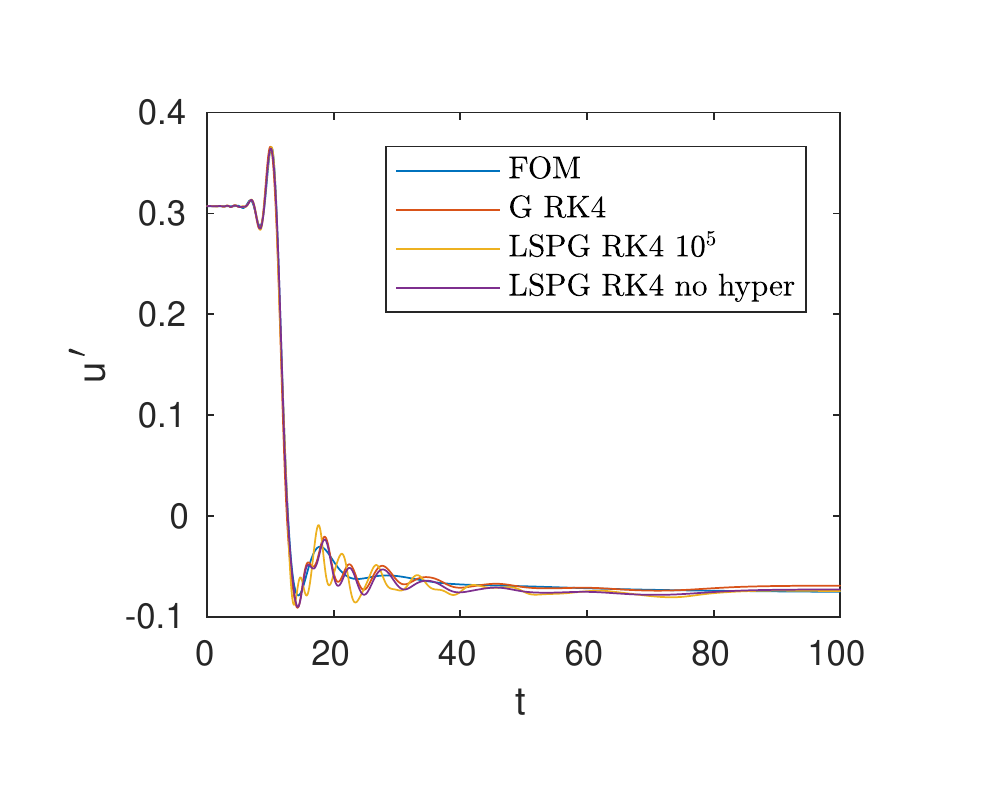}} 
    \subfigure[Probe location indicated by the green symbol in Fig. \ref{fig:vort_div}.]{\includegraphics[scale=.75,trim={5mm 5mm 10mm 10mm},clip]{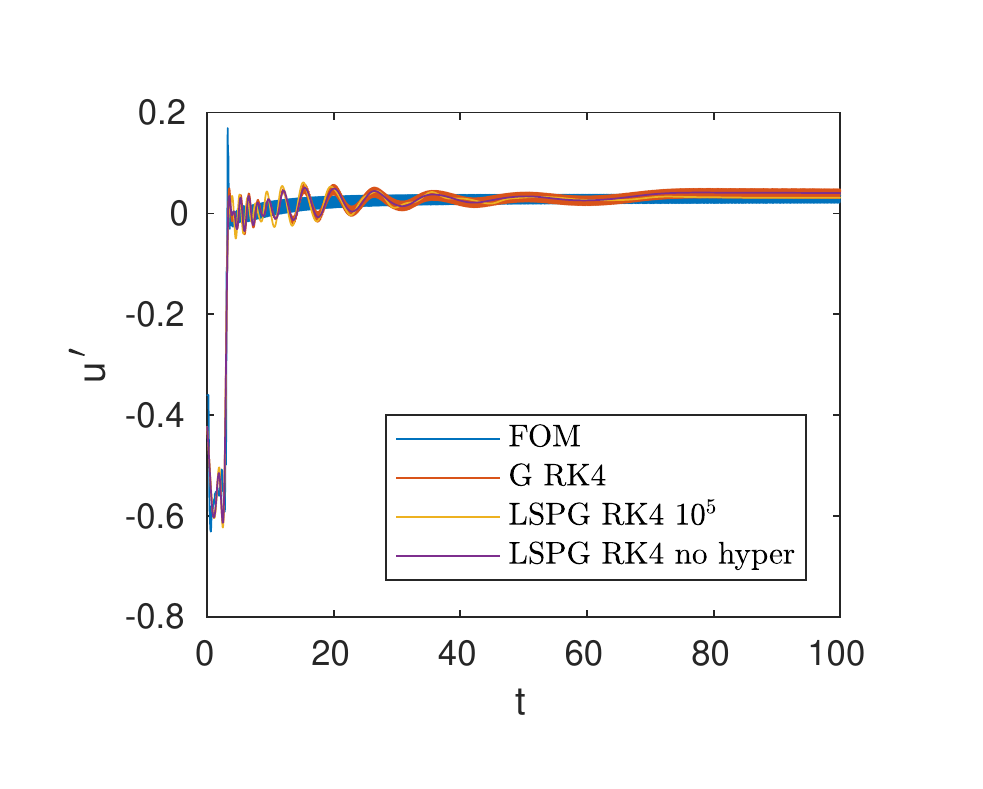}} 
%
    \subfigure[Probe location indicated by the purple symbol in Fig. \ref{fig:vort_div}.]{\includegraphics[scale=.75,trim={5mm 5mm 10mm 8mm},clip]{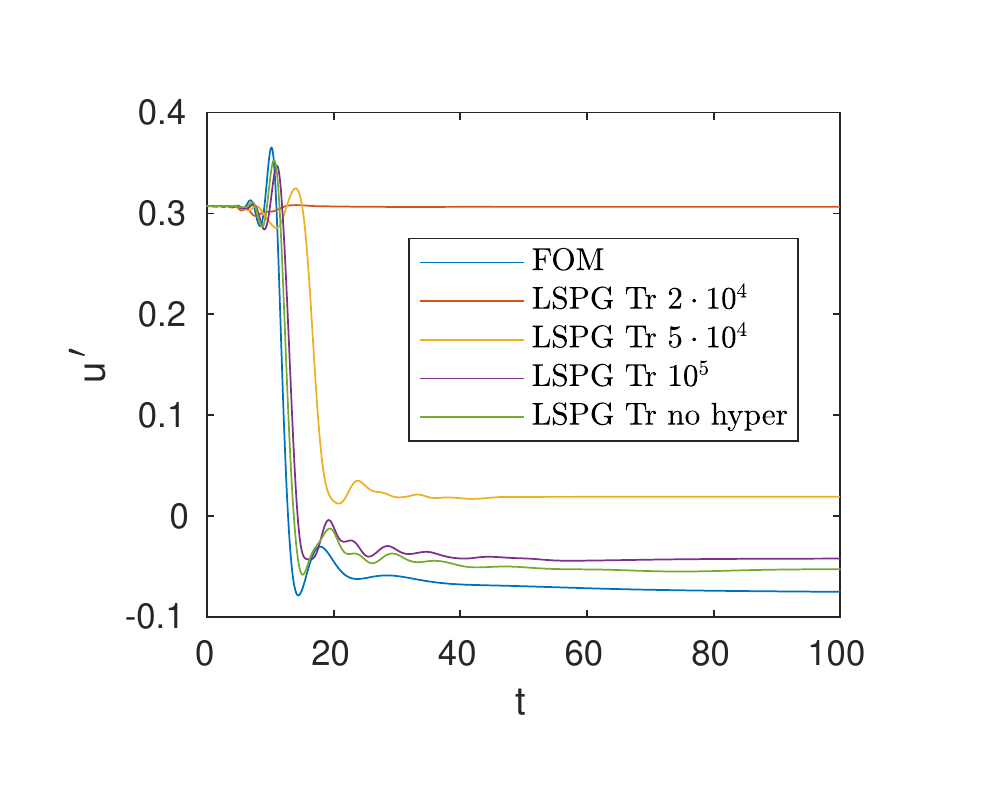}} 
    \subfigure[Probe location indicated by the green symbol in Fig. \ref{fig:vort_div}.]{\includegraphics[scale=.75,trim={5mm 5mm 10mm 10mm},clip]{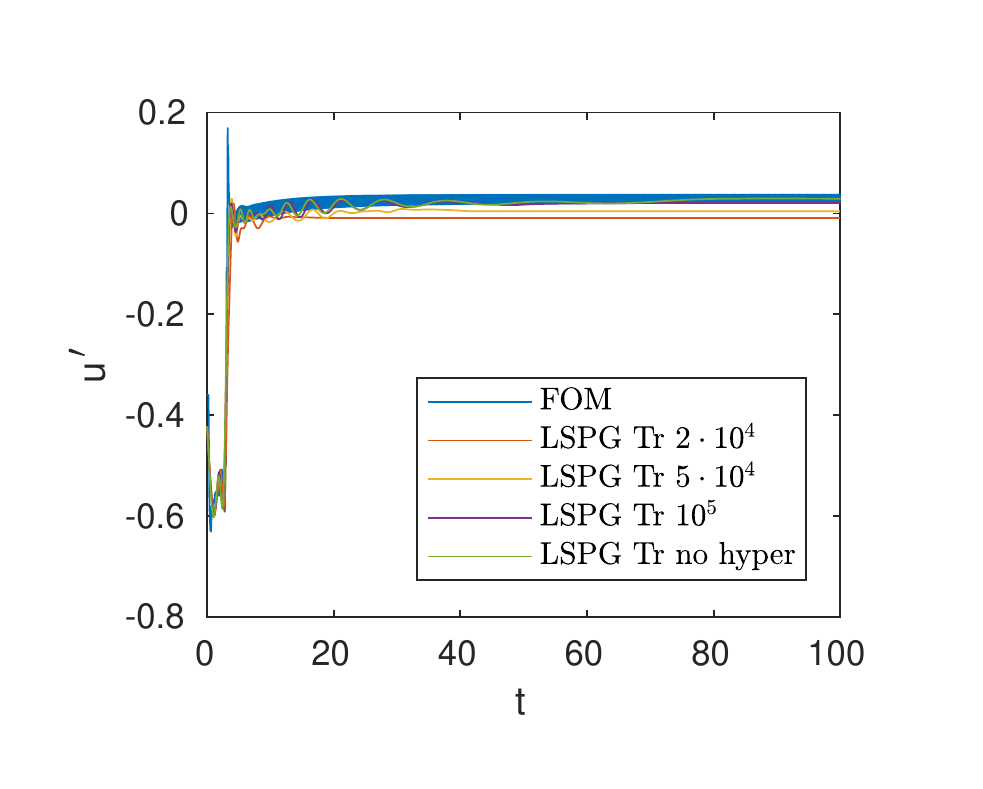}} 
    \caption{Time histories of $u$-velocity fluctuations for probes placed at the shock wave locations.}
    \label{fig:probe_miotto_choque}
\end{figure}

The RK4, implicit Euler and trapezoidal time-marching methods are tested for the present case. We observed that all non-calibrated Galerkin and LSPG ROMs were unstable except by one, the LSPG ROM computed with the implicit Euler method, which was stable but inaccurate. 
Figure \ref{fig:probe_miotto_choque} presents the temporal histories of two probes that capture the $u$-velocity fluctuations of the detached and trailing-edge shock waves. Resuts are provided for the FOM and the Galerkin and LSPG models computed with the RK4 scheme. Solutions obtained by the LSPG with different levels of hyper-reduction are also provided for the trapezoidal scheme. All results are presented for the training region ($0 \leq t \leq 70$) and an extrapolation period for which the FOM solutions are available ($70 < t < 100$). Probe locations are indicated in Fig. \ref{fig:vort_div}(a) by purple and green symbols. From theses figures, one can observe that both RK4 Galerkin and LSPG calibrated ROMs are able to nicely represent the shock motions, despite the strong sharp fluctuations. The LSPG results computed by the trapezoidal scheme show a good comparison with the FOM for the trailing-edge shock. However, the detached shock solution is considerably affected by the hyper-reduction. When 20,000 points are used in the hyper-reduction, the probe misses the detached shock because the points sampled by the accelerated greedy MPE algorithm do not contain the entire pathway of the shock wave. The solution obtained without hyper-reduction is considerably improved but is still less accurate than that computed by the RK4 LSPG.

\begin{figure}
    \centering
    \subfigure[]{\includegraphics[width=.65\textwidth,trim={15mm 5mm 20mm 10mm},clip]{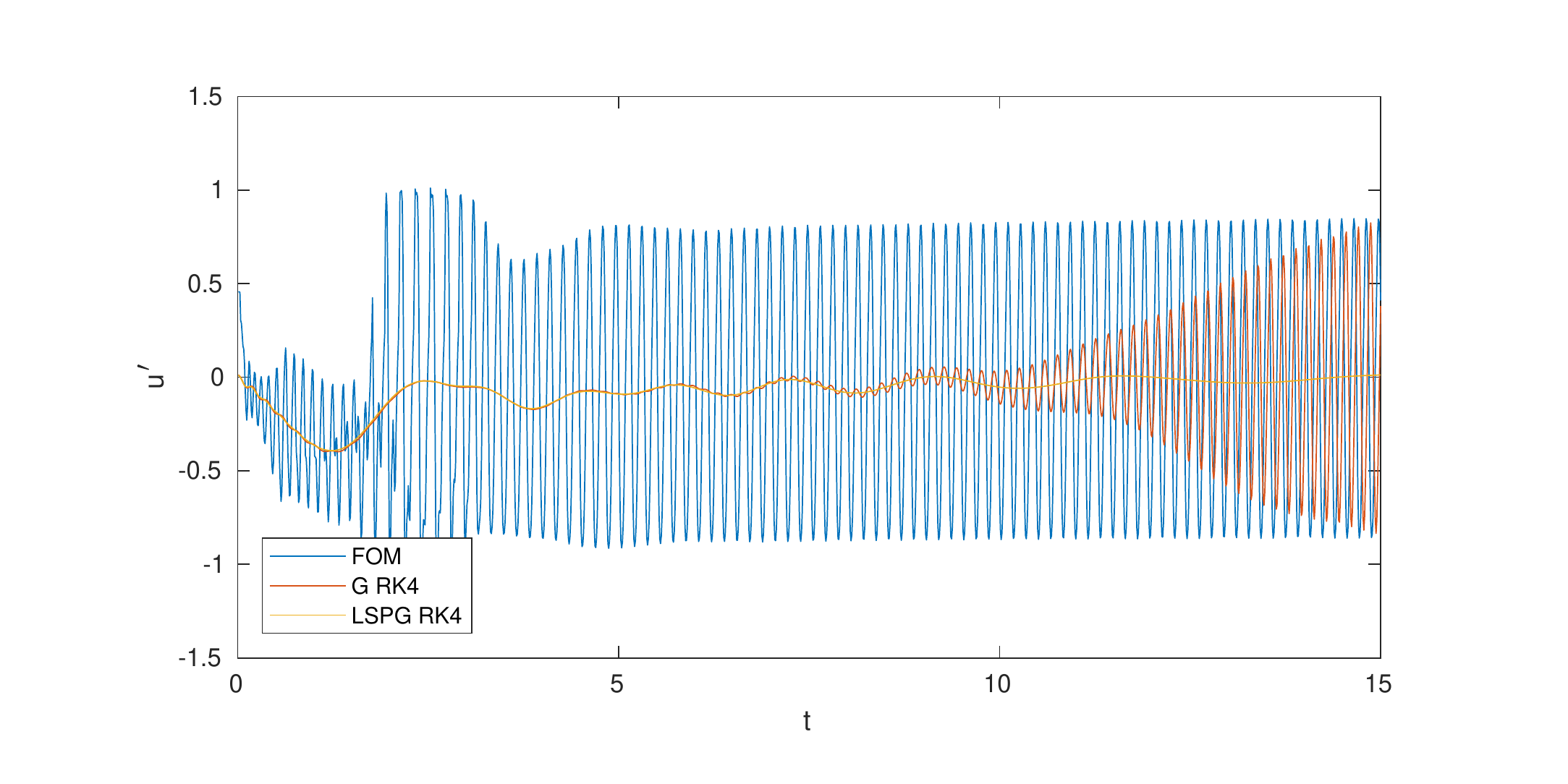}}
    \subfigure[]{\includegraphics[width=.65\textwidth,trim={15mm 5mm 20mm 10mm},clip]{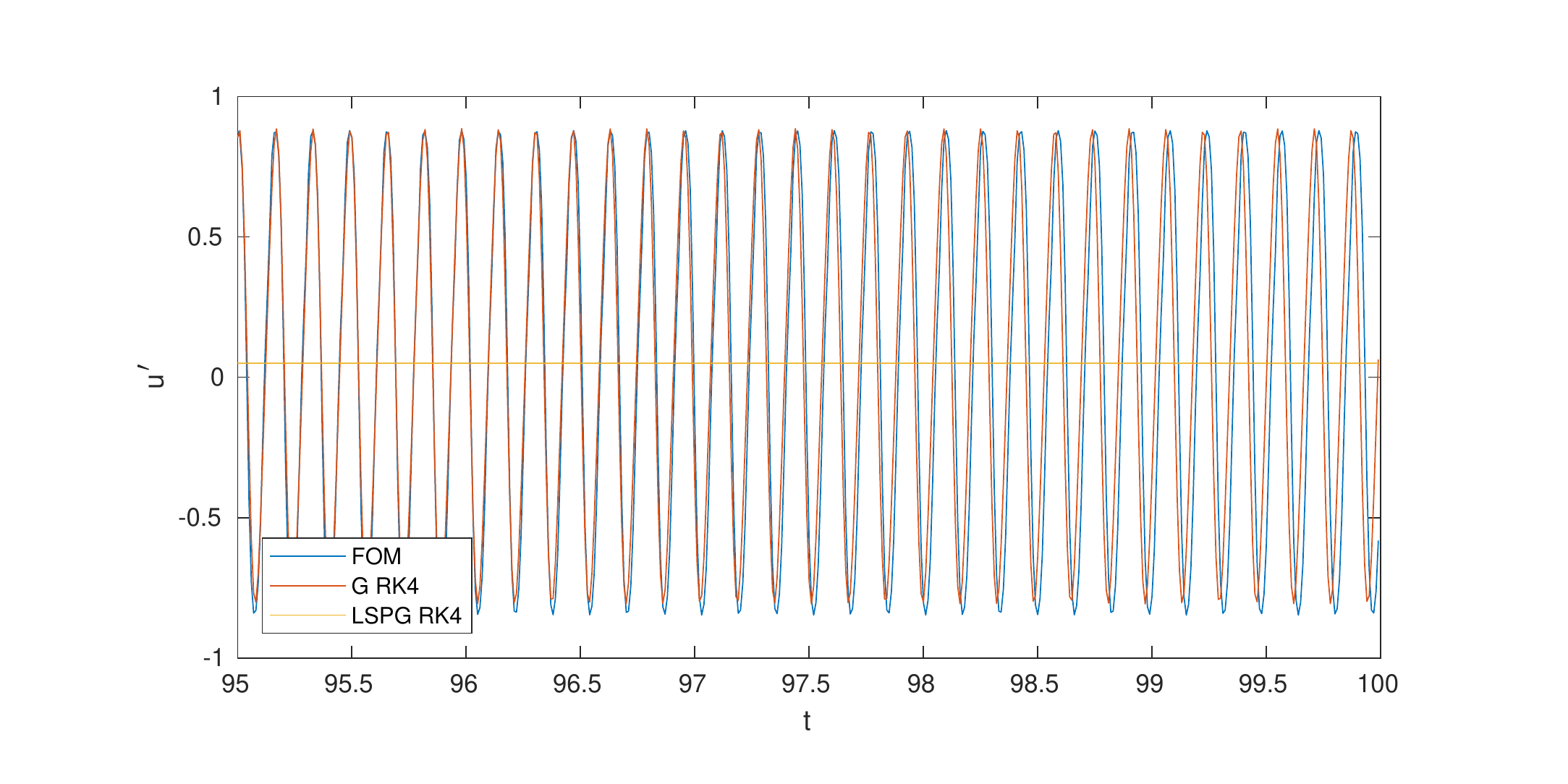}}
    \caption{Time histories of $u$-velocity fluctuations computed for the probe indicated by the yellow symbol in Fig. \ref{fig:vort_div}.} 
    \label{fig:probe_miotto_esteira}
\end{figure}
\begin{figure}
    \centering
    \subfigure[FOM]{\includegraphics[width=.48\textwidth,trim={60mm 30mm 60mm 30mm},clip]{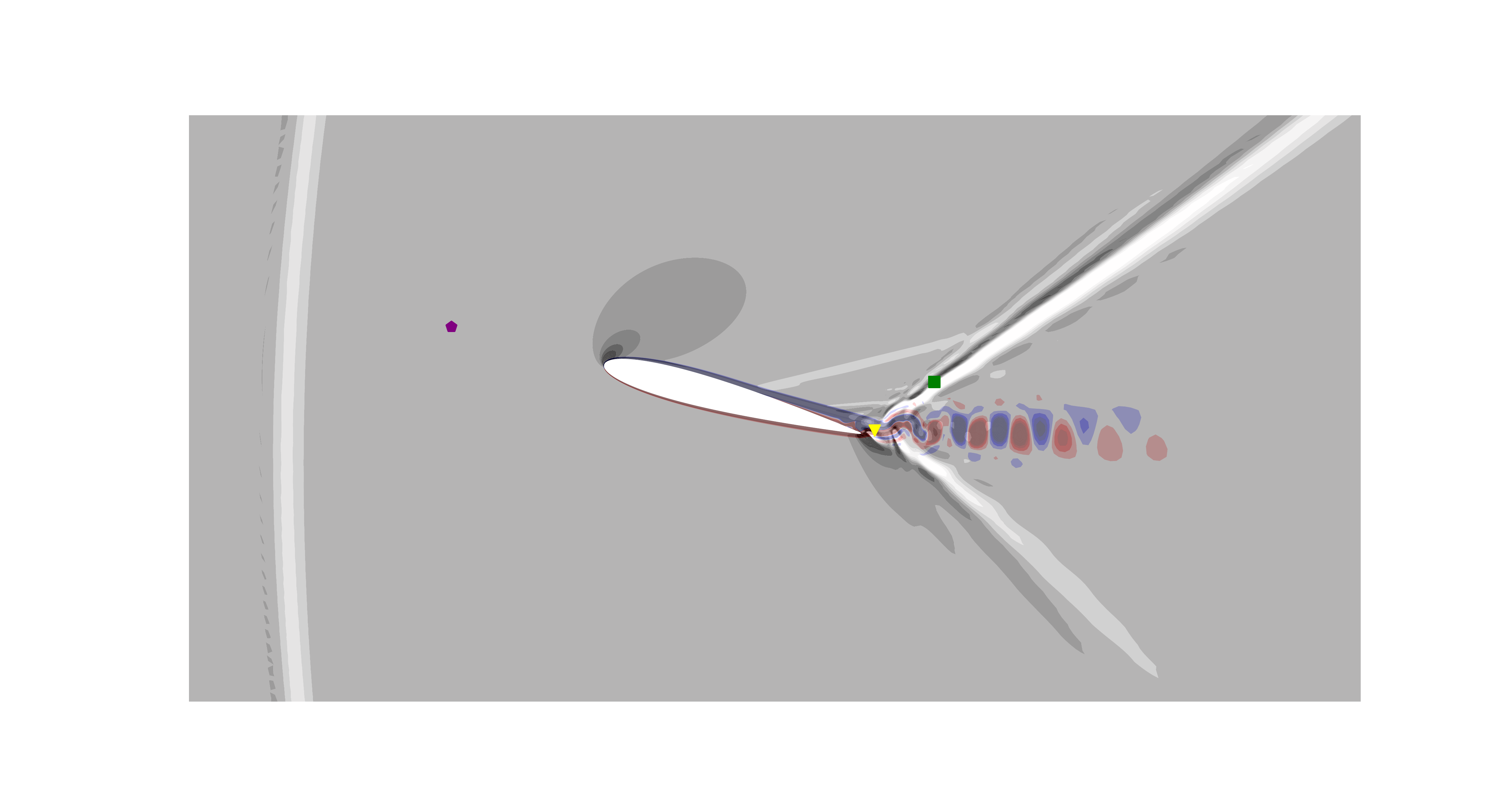}}
    \subfigure[Galerkin RK4  model]{\includegraphics[width=.48\textwidth,trim={60mm 30mm 60mm 30mm},clip]{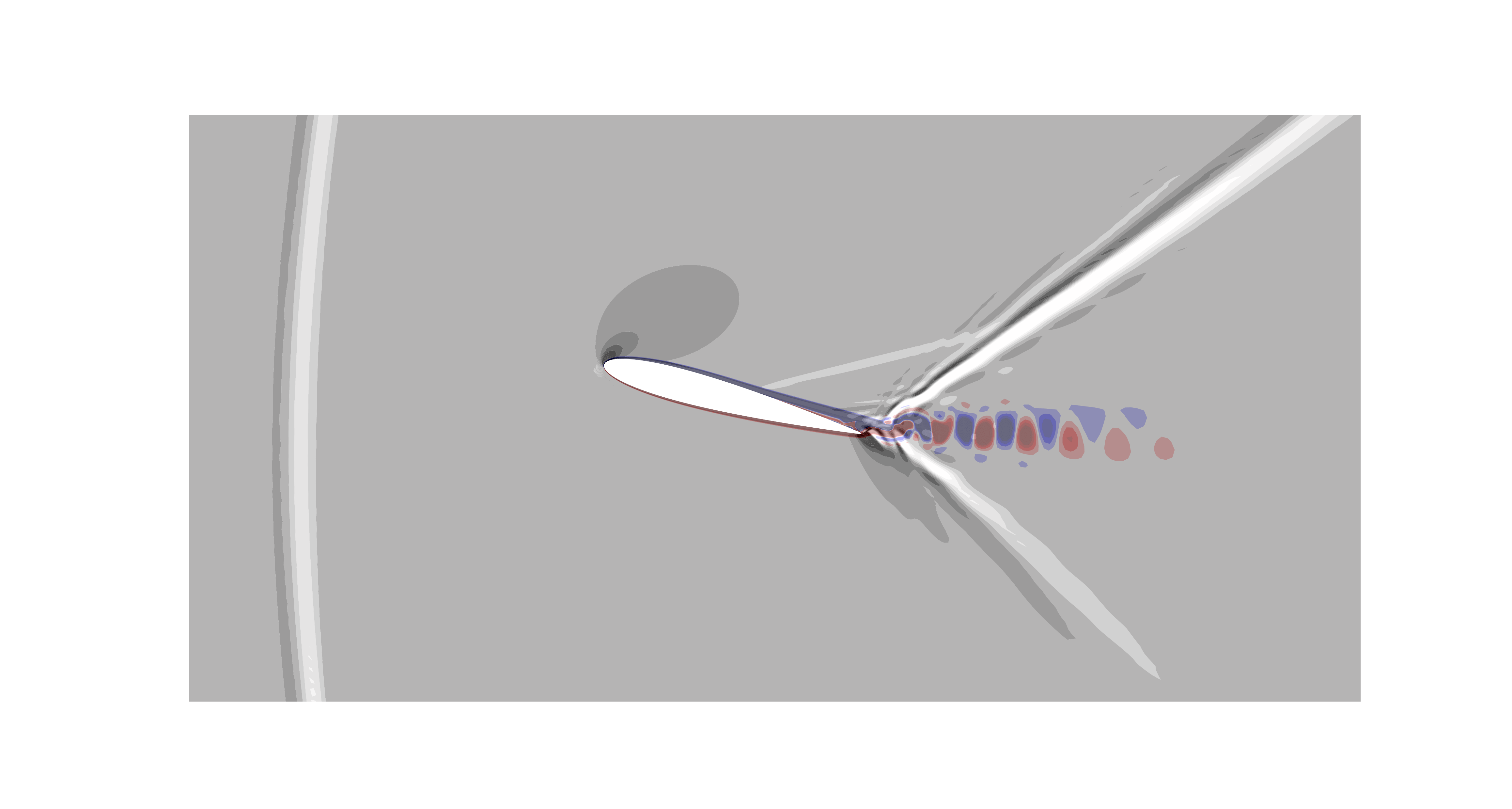}}\\
    \subfigure[LSPG RK4 model without hyper-reduction]{\includegraphics[width=.48\textwidth,trim={60mm 30mm 60mm 30mm},clip]{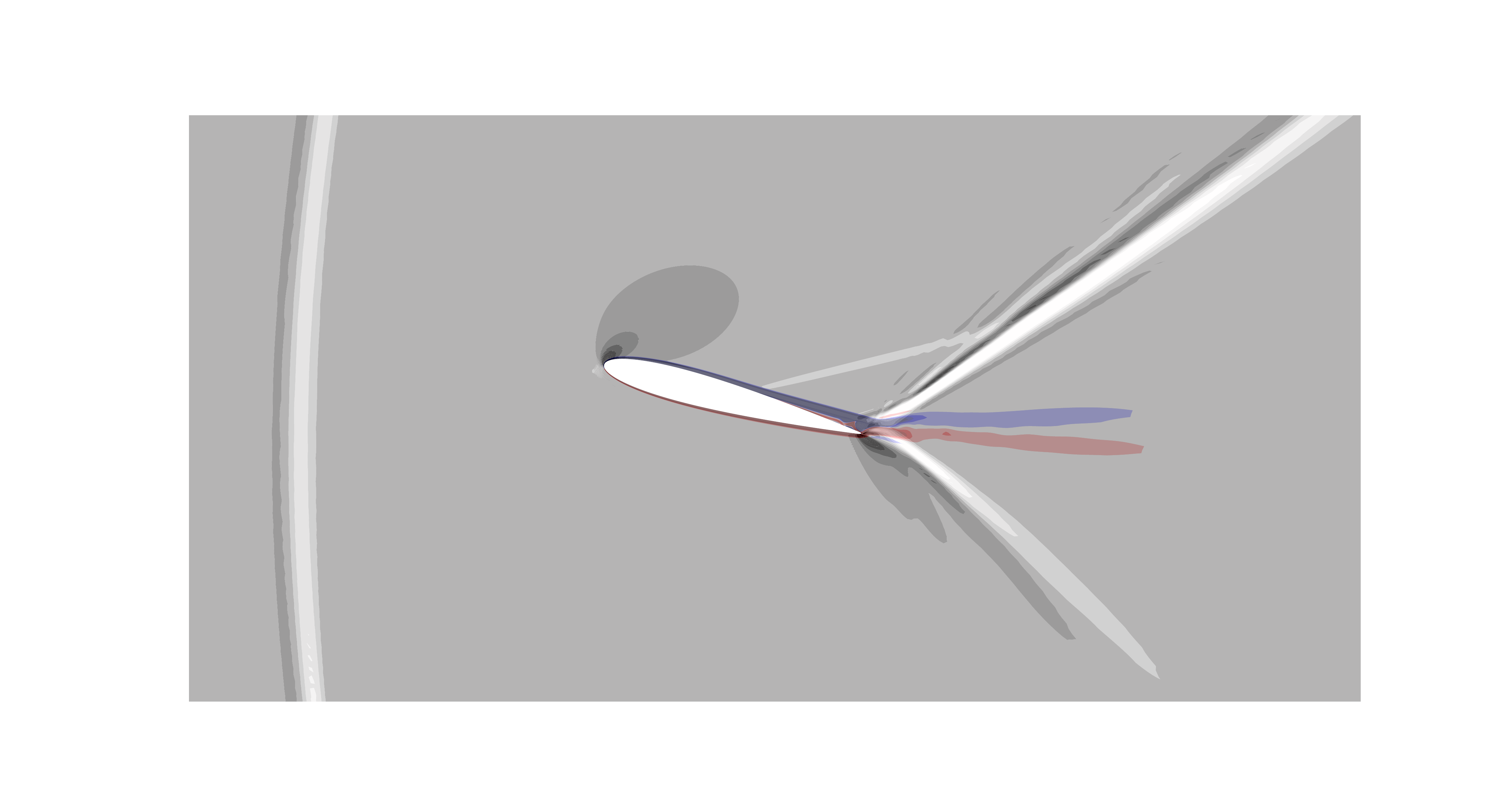}}
    \subfigure[LSPG trapezoidal model with hyper-reduction]{\includegraphics[width=.48\textwidth,trim={60mm 30mm 60mm 30mm},clip]{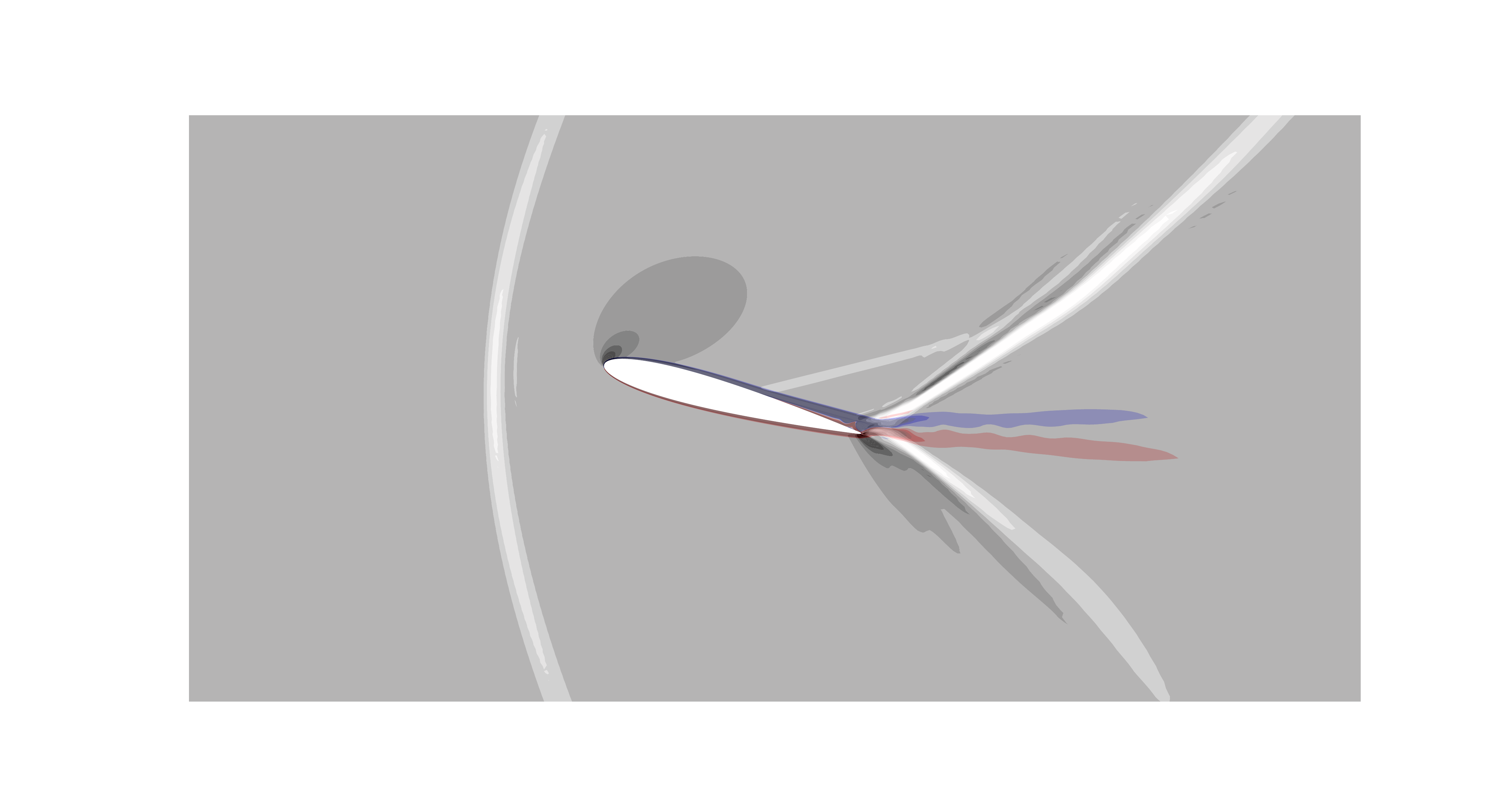}}
    \caption{Contours of dilatation in gray scale and vorticity in color. The purple, green and yellow symbols represent the probe locations from Figs. \ref{fig:probe_miotto_choque} and \ref{fig:probe_miotto_esteira}.}
    \label{fig:vort_div}
\end{figure}

The $u$-velocity time histories of vortex shedding are exhibited in Fig. \ref{fig:probe_miotto_esteira} for the probe location indicated by the yellow symbol in Fig. \ref{fig:vort_div}. This probe captures both the motion of the separation bubble on the suction side of the airfoil and the trailing-edge shock motion induced by vortex shedding. In this case, solutions are presented in detail views for the first 15 and last 5 time units. Here, both the Galerkin and LSPG ROMs are not able to reproduce the initial dynamics of the FOM, as can be seen in Fig. \ref{fig:probe_miotto_esteira}(a). However, such dynamics is captured by the Galerkin model for $t > 10$ leading to vortex shedding with a small phase and amplitude error as shown in Fig. \ref{fig:probe_miotto_esteira}(b). In contrast, the LSPG ROM is not able to reproduce the higher frequency vortex shedding dynamics.

Figure \ref{fig:vort_div} presents contours of divergence of velocity, in gray scale, and vorticity, in color, at $t = 100$. At this time, the flow reached a ``steady-state'' and fluctuations appear only due to vortex shedding and its induced shock motion at the trailing edge. Divergence of velocity allows a better visualization of the shock waves which are well captured by both RK4 calibrated ROMs. Small fluctuations at the trailing-edge shock and vortex shedding are dissipated by the LSPG model but are captured by the Galerkin ROM as can be seen in the figure. The LSPG ROM obtained by the trapezoidal time-marching scheme is also shown and, for this case, hyper-reduction with 20,000 points is employed in the model reconstruction. Although the oblique shocks are well captured by the model, the detached shock wave is positioned downstream compared to the FOM solution. This occurs because the hyper-reduction sampled points do not contain the entire pathway of the shock wave as shown in Fig. \ref{fig:hyper_shock} and, hence, the model is not able to recover upstream information of the shock motion.

\section{Conclusions}

Galerkin and least-squares Petrov-Galerkin reduced-order models are applied to compressible flows involving a disparity in temporal scales. The first case investigated comprises a subsonic flow past a NACA0012 airfoil for which boundary layer hydrodynamic instabilities lead to trailing-edge noise generation. For this case, a separation bubble induces a low-frequency modulation of the higher-frequency boundary layer instabilities that, in turn, leads to the appearance of multiple secondary tones in the acoustic radiation. The second case consists of a supersonic flow over a NACA0012 for which shock-vortex shedding interaction appears at the trailing edge. The dataset employed in the model construction includes a transient period of the flow where a detached shock wave propagates upstream the airfoil leading edge. The initial transient also includes the formation and advection of a starting vortex.

Calibration is employed to construct stable and accurate models and, here, we propose a new form of calibration for the LSPG method. Moreover, differently from other references, we employ the Levemberg-Marquadt method to solve the minimization problem appearing in the LSPG models since it solves the problem of Newton's method when the Hessian is rank-deficient. For both the Galerkin and LSPG methods, calibration is applied only on constant and linear terms appearing in the set of non-linear ODEs resulting from the ROMs. While the present LSPG calibration does not depend on input parameters, the calibration of Galerkin ROMs has a free parameter that balances the model error and the intrusiveness of the calibration terms. For both cases analyzed, an assessment of this free parameter shows that even with maximum intrusiveness the weight of the calibrated coefficients, measured in terms of the Frobenius norm, is still one order of magnitude smaller than that of the original Galerkin coefficients.

Different time-marching schemes are employed with the Galerkin and LSPG ROMs and, for the subsonic airfoil flow, the non-calibrated LSPG method was able to build stable methods using the implicit Euler and trapezoidal schemes. For the supersonic flow, a stable method was also obtained by the non-calibrated LSPG method using the implicit Euler scheme. However, for all these cases, the models were not accurate in comparison with the FOM solutions. Except for the implicit Euler applied to the subsonic flow, the non-calibrated Galerkin models were unstable independently of the time-marching scheme. When calibration was applied to the Galerkin and LSPG methods, solutions became stable and accurate. For the first case investigated, different  implicit and explicit time-marching schemes provided accurate solutions while, for the second case, the fourth-order Runge Kutta presented the most accurate results.

In order to avoid rational functions of the model variables, the non-conservative compressible Navier-Stokes equations are solved in this work. Therefore, the second degree polynomial nonlinearities of the formulation allow pre-computation and storage of constant, linear and non-linear coefficients and, hence, model cost reduction. Since the overall cost of the LSPG models is considerably higher than that of Galerkin models, an accelerated greedy missing point estimation hyper-reduction technique is applied to select the most dynamically relevant points in the flows for model reconstruction.
In the first problem analyzed, the combination of hyper-reduction and calibration allows accurate and stable LSPG models to be reconstructed using only $6\%$ of the total flow information. However, for the second problem, hyper-reduction is not as efficient due to the transient nature of the flow. In this latter case, the sampled points need to capture the entire pathway of the detached shock wave in order to provide an accurate model. Otherwise, its motion is compromised as shown in the results. In order to capture all transient aspects of the flow, at least $30\%$ of the flow information had to be used in the hyper-reduction procedure. For the supersonic flow, calibration was able to account for some of the damage inflicted by hyper-reduction. 

This work shows that calibration is important to construct long-time stable and accurate Galerkin and LSPG methods. In the present studies, the calibrated Galerkin ROMs provide cheaper and more accurate solutions than the LSPG ROMs. Moreover, they do not require hyper-reduction when implemented with the non-conservative form of the Navier-Stokes equations. Both the Galerkin and LSPG models could recover the small hydrodynamic scales and acoustic waves generated in the subsonic airfoil flow. A comparison between ROM and FOM shows that results remain accurate during and beyond the training window. Without hyper-reduction, both models also accurately capture the shock waves present in the airfoil supersonic flow. However, the LSPG models present a more dissipative behavior and could not capture the vortex shedding mechanism along the airfoil wake. On the other hand, the Galerkin models could reproduce the dynamics of the shedding after an initial transient. 









\section*{Acknowledgments}

The authors of this work would like to acknowledge Fun\-da\-\c{c}\~{a}o de Amparo \`{a} Pesquisa do Estado de S\~{a}o Paulo, FAPESP, for supporting the present work under research grants No.\ 2013/08293-7, 2018/11410-9 and 2019/18809-7\@.

\appendix

\section{Galerkin Coefficients}
\label{appendix_galerkin}

Consider the Galerkin coefficients given by the following tensors $\mathbf{e}$, $\mathbf{A}$ and $\mathbf{N}$ from Eq. \ref{galerkin_ode_cav} and computational domain $\mathbf{\Omega}$. These coefficients are functions of the spatial basis $\mathbf{\Phi = [\Phi_{\zeta} \ \Phi_u \ \Phi_v \ \Phi_p]}^{\top}$ obtained by POD, mean flow field $\mathbf{\bar{q}} = \mathbf{[\bar{\zeta} \ \bar{u} \ \bar{v} \ \bar{p}]}^{\top}$, initial conditions $\mathbf{[\zeta_0 \ u_0 \ v_0 \ p_0]}^{\top}$, specific heat ratio $\gamma$, and reference Prandtl ($Pr$), Reynolds ($Re$) and Mach ($Ma$) numbers. 
It is convenient to decompose $\mathbf{e} = e_i = e^{\zeta}_{i} + e^u_{i} + e^v_{i} + e^p_{i}$, which leads to the following terms
\begin{equation}
    e^{\zeta}_{i} = \iint_{\Omega} \mathbf{\Phi_{\zeta_i}} \left(
    \mathbf{\bar{\zeta}} \frac{\partial \mathbf{\bar{u    }}}{\partial x} +
    \mathbf{\bar{\zeta}} \frac{\partial \mathbf{\bar{v    }}}{\partial y} -
    \mathbf{\bar{u    }} \frac{\partial \mathbf{\bar{\zeta}}}{\partial x} -
    \mathbf{\bar{v    }} \frac{\partial \mathbf{\bar{\zeta}}}{\partial y}
    \right) \,dx \,dy
\end{equation}

\begin{equation}
\begin{aligned}
    e^{u}_{i} = \iint_{\Omega} \mathbf{\Phi_{u_i}} \Bigg[ & -
    \mathbf{\bar{u    }} \frac{\partial \mathbf{\bar{u    }}}{\partial x} -
    \mathbf{\bar{v    }} \frac{\partial \mathbf{\bar{u    }}}{\partial y} -
    \mathbf{\bar{\zeta}} \frac{\partial \mathbf{\bar{p}}}{\partial x} + 
    \frac{Ma}{Re} \Bigg( 
    \frac{4}{3} \mathbf{\bar{\zeta}} \frac{\partial^2 \mathbf{\bar{u}}}{\partial x^2} -
    \frac{2}{3} \mathbf{\bar{\zeta}} \frac{\partial^2 \mathbf{\bar{v}}}{\partial x \partial y} +
    \mathbf{\bar{\zeta}} \frac{\partial^2 \mathbf{\bar{v}}}{\partial y \partial x} +
    \mathbf{\bar{\zeta}} \frac{\partial^2 \mathbf{\bar{u}}}{\partial y^2}
    \Bigg) \Bigg] \,dx \,dy
\end{aligned}
\end{equation}

\begin{equation}
\begin{aligned}
    e^{v}_{i} = \iint_{\Omega} \mathbf{\Phi_{v_i}} \Bigg[ & -
    \mathbf{\bar{u    }} \frac{\partial \mathbf{\bar{v}}}{\partial x} -
    \mathbf{\bar{v    }} \frac{\partial \mathbf{\bar{v}}}{\partial y} -
    \mathbf{\bar{\zeta}} \frac{\partial \mathbf{\bar{p}}}{\partial y} + 
    \frac{Ma}{Re} \Bigg( 
    \frac{4}{3}
    \mathbf{\bar{\zeta}} \frac{\partial^2 \mathbf{\bar{v}}}{\partial y^2} -
    \frac{2}{3}
    \mathbf{\bar{\zeta}} \frac{\partial^2 \mathbf{\bar{u}}}{\partial y \partial x} +
    \mathbf{\bar{\zeta}} \frac{\partial^2 \mathbf{\bar{u}}}{\partial x \partial y} +
    \mathbf{\bar{\zeta}} \frac{\partial^2 \mathbf{\bar{v}}}{\partial x^2}
    \Bigg) \Bigg] \,dx \,dy
\end{aligned}
\end{equation}

\begin{equation}
\begin{aligned}
    e^{p}_{i} = \iint_{\Omega} \mathbf{\Phi_{p_i}} \Bigg[ & -
    \mathbf{\bar{u}} \frac{\partial \mathbf{\bar{p}}}{\partial x} -
    \mathbf{\bar{v}} \frac{\partial \mathbf{\bar{p}}}{\partial y} -
    \gamma \Bigg(
    \mathbf{\bar{p}} \frac{\partial \mathbf{\bar{u}}}{\partial x} +
    \mathbf{\bar{p}} \frac{\partial \mathbf{\bar{v}}}{\partial y} \Bigg) + \\ &
    \frac{\gamma Ma}{Re Pr} \Bigg(
    \frac{\partial^2 \mathbf{\bar{p}}}{\partial x^2} \mathbf{\bar{\zeta}} + 2
    \frac{\partial \mathbf{\bar{p}}}{\partial x} \frac{\partial \mathbf{\bar{\zeta}}}{\partial x} +
    \mathbf{\bar{p}} \frac{\partial^2 \mathbf{\bar{\zeta}}}{\partial x^2} + 
    \frac{\partial^2 \mathbf{\bar{p}}}{\partial y^2} \mathbf{\bar{\zeta}} + 2
    \frac{\partial \mathbf{\bar{p}}}{\partial y} \frac{\partial \mathbf{\bar{\zeta}}}{\partial y} +
    \mathbf{\bar{p}} \frac{\partial^2 \mathbf{\bar{\zeta}}}{\partial y^2} \Bigg) + \\ &
    \frac{(\gamma - 1) Ma}{Re} \Bigg(
    \frac{4}{3}
    \frac{\partial \mathbf{\bar{u}}}{\partial x} \frac{\partial \mathbf{\bar{u}}}{\partial x} -
    \frac{2}{3}
    \frac{\partial \mathbf{\bar{u}}}{\partial x} \frac{\partial \mathbf{\bar{v}}}{\partial y} +
    \frac{4}{3}
    \frac{\partial \mathbf{\bar{v}}}{\partial y} \frac{\partial \mathbf{\bar{v}}}{\partial y} -
    \frac{2}{3}
    \frac{\partial \mathbf{\bar{v}}}{\partial y} \frac{\partial \mathbf{\bar{u}}}{\partial x} +
    \frac{\partial \mathbf{\bar{v}}}{\partial x} \frac{\partial \mathbf{\bar{v}}}{\partial x} + 2
    \frac{\partial \mathbf{\bar{v}}}{\partial x} \frac{\partial \mathbf{\bar{u}}}{\partial y} +
    \frac{\partial \mathbf{\bar{u}}}{\partial y} \frac{\partial \mathbf{\bar{u}}}{\partial y}
    \Bigg)
    \Bigg] \,dx \,dy
\end{aligned}
\end{equation}

The term $\mathbf{A}$ is also conveniently decomposed as $\mathbf{A} = A_{ij} = A^{\zeta}_{ij} + A^u_{ij} + A^v_{ij} + A^p_{ij}$
\begin{equation}
\begin{aligned}
    A^{\zeta}_{ij} = \iint_{\Omega} \mathbf{\Phi_{\zeta_i}} \Bigg(
    \mathbf{\bar{\zeta}} \frac{\partial \mathbf{\Phi_{u_j    }}}{\partial x} + &
    \mathbf{\bar{\zeta}} \frac{\partial \mathbf{\Phi_{v_j   }}}{\partial y} -
    \mathbf{\bar{u    }} \frac{\partial \mathbf{\Phi_{\zeta_j}}}{\partial x} -
    \mathbf{\bar{v    }} \frac{\partial \mathbf{\Phi_{\zeta_j}}}{\partial y} + 
    \frac{\partial \bar{\mathbf{u    }}}{\partial x} \mathbf{\Phi_{\zeta_j}} + 
    \frac{\partial \bar{\mathbf{v    }}}{\partial y} \mathbf{\Phi_{\zeta_j}} -
    \frac{\partial \bar{\mathbf{\zeta}}}{\partial x} \mathbf{\Phi_{u_j    }} -
    \frac{\partial \bar{\mathbf{\zeta}}}{\partial y} \mathbf{\Phi_{v_j    }}
    \Bigg) \,dx \,dy
\end{aligned}
\end{equation}

\begin{equation}
\begin{aligned}
    A^{u}_{ij} = \iint_{\Omega} \mathbf{\Phi_{u_i}} \Bigg\{ & -
    \mathbf{\bar{u    }} \frac{\partial \mathbf{\Phi_{u_j}}}{\partial x} -
    \mathbf{\bar{v    }} \frac{\partial \mathbf{\Phi_{u_j}}}{\partial y} -
    \mathbf{\bar{\zeta}} \frac{\partial \mathbf{\Phi_{p_j}}}{\partial x} -
    \frac{\partial \bar{\mathbf{u}}}{\partial x} \mathbf{\Phi_{u_j    }} -
    \frac{\partial \bar{\mathbf{u}}}{\partial y} \mathbf{\Phi_{v_j    }} -
    \frac{\partial \bar{\mathbf{p}}}{\partial x} \mathbf{\Phi_{\zeta_j}} + \\ &
    \frac{Ma}{Re} \Bigg[
    \frac{4}{3}  \Bigg(
    \mathbf{\bar{\zeta}} \frac{\partial^2 \mathbf{\Phi_{u_j}}}{\partial x^2} +
    \frac{\partial^2 \bar{\mathbf{u}}}{\partial x^2} \mathbf{\Phi_{\zeta_j}} \Bigg) - 
    \frac{2}{3}  \Bigg(
    \mathbf{\bar{\zeta}} \frac{\partial^2 \mathbf{\Phi_{v_j}}}{\partial x \partial y} +
    \frac{\partial^2 \bar{\mathbf{v}}}{\partial x \partial y} \mathbf{\Phi_{\zeta_j}} \Bigg) + \\ &
    \mathbf{\bar{\zeta}} \frac{\partial^2 \mathbf{\Phi_{v_j}}}{\partial y \partial x} +
    \frac{\partial^2 \bar{\mathbf{v}}}{\partial y \partial x} \mathbf{\Phi_{\zeta_j}} + 
    \mathbf{\bar{\zeta}} \frac{\partial^2 \mathbf{\Phi_{u_j}}}{\partial y^2} +
    \frac{\partial^2 \bar{\mathbf{u}}}{\partial y^2} \mathbf{\Phi_{\zeta_j}}
    \Bigg] \Bigg\} \,dx \,dy
\end{aligned}
\end{equation}

\begin{equation}
\begin{aligned}
    A^{v}_{ij} = \iint_{\Omega} \mathbf{\Phi_{v_i}} \Bigg\{ & -
    \mathbf{\bar{u    }} \frac{\partial \mathbf{\Phi_{v_j}}}{\partial x} -
    \mathbf{\bar{v    }} \frac{\partial \mathbf{\Phi_{v_j}}}{\partial y} -
    \mathbf{\bar{\zeta}} \frac{\partial \mathbf{\Phi_{p_j}}}{\partial y} -
    \frac{\partial \bar{\mathbf{v}}}{\partial x} \mathbf{\Phi_{u_j    }} -
    \frac{\partial \bar{\mathbf{v}}}{\partial y} \mathbf{\Phi_{v_j    }} -
    \frac{\partial \bar{\mathbf{p}}}{\partial y} \mathbf{\Phi_{\zeta_j}} + \\ &
    \frac{Ma}{Re} \Bigg[
    \frac{4}{3}  \Bigg(
    \mathbf{\bar{\zeta}} \frac{\partial^2 \mathbf{\Phi_{v_j}}}{\partial y^2} +
    \frac{\partial^2 \bar{\mathbf{v}}}{\partial y^2} \mathbf{\Phi_{\zeta_j}} \Bigg) - 
    \frac{2}{3}  \Bigg(
    \mathbf{\bar{\zeta}} \frac{\partial^2 \mathbf{\Phi_{u_j}}}{\partial y \partial x} +
    \frac{\partial^2 \bar{\mathbf{u}}}{\partial y \partial x} \mathbf{\Phi_{\zeta_j}} \Bigg) + \\ &
    \mathbf{\bar{\zeta}} \frac{\partial^2 \mathbf{\Phi_{u_j}}}{\partial x \partial y} +
    \frac{\partial^2 \bar{\mathbf{u}}}{\partial x \partial y} \mathbf{\Phi_{\zeta_j}} + 
    \mathbf{\bar{\zeta}} \frac{\partial^2 \mathbf{\Phi_{v_j}}}{\partial x^2} +
    \frac{\partial^2 \bar{\mathbf{v}}}{\partial x^2} \mathbf{\Phi_{\zeta_j}}
    \Bigg] \Bigg\} \,dx \,dy
\end{aligned}
\end{equation}

\begin{equation}
\begin{aligned}
    A^{p}_{ij} = \iint_{\Omega} \mathbf{\Phi_{p_i}} \Bigg\{ & -
    \mathbf{\bar{u}} \frac{\partial \mathbf{\Phi_{p_j}}}{\partial x} -
    \mathbf{\bar{v}} \frac{\partial \mathbf{\Phi_{p_j}}}{\partial y} -
    \frac{\partial \bar{\mathbf{p}}}{\partial x} \mathbf{\Phi_{u_j}} - 
    \frac{\partial \bar{\mathbf{p}}}{\partial y} \mathbf{\Phi_{v_j}} - 
    \gamma \Bigg(
    \mathbf{\bar{p}} \frac{\partial \mathbf{\Phi_{u_j}}}{\partial x} +
    \mathbf{\bar{p}} \frac{\partial \mathbf{\Phi_{v_j}}}{\partial y} +
    \frac{\partial \bar{\mathbf{u}}}{\partial x} \mathbf{\Phi_{p_j}} +
    \frac{\partial \bar{\mathbf{v}}}{\partial y} \mathbf{\Phi_{p_j}} \Bigg) + \\ &
    \frac{\gamma Ma}{Re Pr} \Bigg[
    \frac{\partial^2 \mathbf{\bar{p}}}{\partial x^2} \mathbf{\Phi_{\zeta_j}} +
    \mathbf{\bar{\zeta}} \frac{\partial^2 \mathbf{\Phi_{p_j}}}{\partial x^2} + 2
    \Bigg(
    \frac{\partial \bar{\mathbf{p}}}{\partial x} \frac{\partial \mathbf{\Phi_{\zeta_j}}}{\partial x} +
    \frac{\partial \bar{\mathbf{\zeta}}}{\partial x} \frac{\partial \mathbf{\Phi_{p_j}}}{\partial x} \Bigg) + 
    \mathbf{\bar{p}} \frac{\partial^2 \mathbf{\Phi_{\zeta_j}}}{\partial x^2} +
    \frac{\partial^2 \bar{\mathbf{\zeta}}}{\partial x^2} \mathbf{\Phi_{p_j}} + \\ &
    \frac{\partial^2 \mathbf{\bar{p}}}{\partial y^2} \mathbf{\Phi_{\zeta_j}} +
    \mathbf{\bar{\zeta}} \frac{\partial^2 \mathbf{\Phi_{p_j}}}{\partial y^2} + 2
    \Bigg(
    \frac{\partial \bar{\mathbf{p}}}{\partial y} \frac{\partial \mathbf{\Phi_{\zeta_j}}}{\partial y} +
    \frac{\partial \bar{\mathbf{\zeta}}}{\partial y} \frac{\partial \mathbf{\Phi_{p_j}}}{\partial y} \Bigg) + 
    \mathbf{\bar{p}} \frac{\partial^2 \mathbf{\Phi_{\zeta_j}}}{\partial y^2} +
    \frac{\partial^2 \bar{\mathbf{\zeta}}}{\partial y^2} \mathbf{\Phi_{p_j}}\Bigg] + \\ &
    \frac{(\gamma - 1)Ma}{Re} \Bigg[
    \frac{4}{3} \Bigg( 
    \frac{\partial \bar{\mathbf{u}}}{\partial x} \frac{\partial \mathbf{\Phi_{u_j}}}{\partial x} + 
    \frac{\partial \bar{\mathbf{u}}}{\partial x} \frac{\partial \mathbf{\Phi_{u_j}}}{\partial x} \Bigg) -
    \frac{2}{3} \Bigg(
    \frac{\partial \bar{\mathbf{u}}}{\partial x} \frac{\partial \mathbf{\Phi_{v_j}}}{\partial y} + 
    \frac{\partial \bar{\mathbf{v}}}{\partial y} \frac{\partial \mathbf{\Phi_{u_j}}}{\partial x} \Bigg) + \\ &
    \frac{4}{3} \Bigg(
    \frac{\partial \bar{\mathbf{v}}}{\partial y} \frac{\partial \mathbf{\Phi_{v_j}}}{\partial y} +
    \frac{\partial \bar{\mathbf{v}}}{\partial y} \frac{\partial \mathbf{\Phi_{v_j}}}{\partial y} \Bigg) - 
    \frac{2}{3} \Bigg(
    \frac{\partial \bar{\mathbf{v}}}{\partial y} \frac{\partial \mathbf{\Phi_{u_j}}}{\partial x} + 
    \frac{\partial \bar{\mathbf{u}}}{\partial x} \frac{\partial \mathbf{\Phi_{v_j}}}{\partial y} \Bigg) + \\ &
2\, \Bigg(\frac{\partial \bar{\mathbf{v}}}{\partial x} \frac{\partial \mathbf{\Phi_{v_j}}}{\partial x} + 
    \frac{\partial \bar{\mathbf{v}}}{\partial x} \frac{\partial \mathbf{\Phi_{u_j}}}{\partial y} +
    \frac{\partial \bar{\mathbf{u}}}{\partial y} \frac{\partial \mathbf{\Phi_{v_j}}}{\partial x} + 
    \frac{\partial \bar{\mathbf{u}}}{\partial y} \frac{\partial \mathbf{\Phi_{u_j}}}{\partial y}\Bigg)
    \Bigg] \Bigg\} \,dx \,dy
\end{aligned}
\end{equation}

\noindent
and, finally, the third order tensor is written as $\mathbf{N} = N_{ijk} = N^{\zeta}_{ijk} + N^u_{ijk} + N^v_{ijk} + N^p_{ijk}$

\begin{equation}
    N^{\zeta}_{ijk} = \iint_{\Omega} \mathbf{\Phi_{{\zeta}_i}} \left(
    \mathbf{\Phi_{\zeta_j}} \frac{\partial \mathbf{\Phi_{u_k    }}}{\partial x} +
    \mathbf{\Phi_{\zeta_j}} \frac{\partial \mathbf{\Phi_{v_k    }}}{\partial y} -
    \mathbf{\Phi_{u_j    }} \frac{\partial \mathbf{\Phi_{\zeta_k}}}{\partial x} -
    \mathbf{\Phi_{v_j    }} \frac{\partial \mathbf{\Phi_{\zeta_k}}}{\partial y}
    \right) \,dx \,dy
\end{equation}

\begin{equation}
\begin{aligned}
    N^{u}_{ijk} = \iint_{\Omega} \mathbf{\Phi_{{u}_i}} \Bigg[ - &
    \mathbf{\Phi_{u_j    }} \frac{\partial \mathbf{\Phi_{u_k}}}{\partial x} -
    \mathbf{\Phi_{v_j    }} \frac{\partial \mathbf{\Phi_{u_k}}}{\partial y} -
    \mathbf{\Phi_{\zeta_j}} \frac{\partial \mathbf{\Phi_{p_k}}}{\partial x} + \\ &
    \frac{Ma}{Re} \Bigg(
    \frac{4}{3}
    \mathbf{\Phi_{\zeta_j}} \frac{\partial^2 \mathbf{\Phi_{u_k}}}{\partial x^2} - 
    \frac{2}{3} 
    \mathbf{\Phi_{\zeta_j}} \frac{\partial^2 \mathbf{\Phi_{v_k}}}{\partial x \partial y} + 
    \mathbf{\Phi_{\zeta_j}} \frac{\partial^2 \mathbf{\Phi_{v_k}}}{\partial y \partial x} +
    \mathbf{\Phi_{\zeta_j}} \frac{\partial^2 \mathbf{\Phi_{u_k}}}{\partial y^2}
    \Bigg) \Bigg] \,dx \,dy
\end{aligned}
\end{equation}

\begin{equation}
\begin{aligned}
    N^{v}_{ijk} = \iint_{\Omega} \mathbf{\Phi_{{v}_i}} \Bigg[ - &
    \mathbf{\Phi_{u_j    }} \frac{\partial \mathbf{\Phi_{v_k}}}{\partial x} -
    \mathbf{\Phi_{v_j    }} \frac{\partial \mathbf{\Phi_{v_k}}}{\partial y} -
    \mathbf{\Phi_{\zeta_j}} \frac{\partial \mathbf{\Phi_{p_k}}}{\partial y} + \\ &
    \frac{Ma}{Re} \Bigg(
    \frac{4}{3}
    \mathbf{\Phi_{\zeta_j}} \frac{\partial^2 \mathbf{\Phi_{v_k}}}{\partial y^2} - 
    \frac{2}{3} 
    \mathbf{\Phi_{\zeta_j}} \frac{\partial^2 \mathbf{\Phi_{u_k}}}{\partial y \partial x} + 
    \mathbf{\Phi_{\zeta_j}} \frac{\partial^2 \mathbf{\Phi_{u_k}}}{\partial x \partial y} +
    \mathbf{\Phi_{\zeta_j}} \frac{\partial^2 \mathbf{\Phi_{v_k}}}{\partial x^2}
    \Bigg) \Bigg] \,dx \,dy
\end{aligned}
\end{equation}

\begin{equation}
\begin{aligned}
    N^{p}_{ijk} = \iint_{\Omega} \mathbf{\Phi_{{p}_i}} \Bigg[ - &
    \mathbf{\Phi_{u_j}} \frac{\partial \mathbf{\Phi_{p_k}}}{\partial x} -
    \mathbf{\Phi_{v_j}} \frac{\partial \mathbf{\Phi_{p_k}}}{\partial y} -
    \gamma \Bigg(
    \mathbf{\Phi_{p_j}} \frac{\partial \mathbf{\Phi_{u_k}}}{\partial x} + 
    \mathbf{\Phi_{p_j}} \frac{\partial \mathbf{\Phi_{v_k}}}{\partial y} \Bigg) + \\ &
    \frac{\gamma Ma}{Re Pr} \Bigg(
    \mathbf{\Phi_{\zeta_j}} \frac{\partial^2 \mathbf{\Phi_{p_k}}}{\partial x^2} + 2
    \frac{\partial \mathbf{\Phi_{p_j}}}{\partial x} \frac{\partial \mathbf{\Phi_{\zeta_k}}}{\partial x} +
    \mathbf{\Phi_{p_j}} \frac{\partial^2 \mathbf{\Phi_{\zeta_k}}}{\partial x^2} + \\ &
    \frac{\partial^2 \mathbf{\Phi_{p_j}}}{\partial y^2} \mathbf{\Phi_{\zeta_k}} + 2
    \frac{\partial \mathbf{\Phi_{p_j}}}{\partial y} \frac{\partial \mathbf{\Phi_{\zeta_k}}}{\partial y} +
    \mathbf{\Phi_{p_j}} \frac{\partial^2 \mathbf{\Phi_{\zeta_k}}}{\partial y^2}
    \Bigg) + \\ &
    \frac{(\gamma - 1) Ma}{Re} \Bigg(
    \frac{4}{3}
    \frac{\partial \mathbf{\Phi_{u_j}}}{\partial x} \frac{\partial \mathbf{\Phi_{u_k}}}{\partial x} - 
    \frac{2}{3}
    \frac{\partial \mathbf{\Phi_{u_j}}}{\partial x} \frac{\partial \mathbf{\Phi_{v_k}}}{\partial y} +
    \frac{4}{3}
    \frac{\partial \mathbf{\Phi_{v_j}}}{\partial y} \frac{\partial \mathbf{\Phi_{v_k}}}{\partial y} - 
    \frac{2}{3}
    \frac{\partial \mathbf{\Phi_{v_j}}}{\partial y} \frac{\partial \mathbf{\Phi_{u_k}}}{\partial x} + \\ &
    \frac{\partial \mathbf{\Phi_{v_j}}}{\partial x} \frac{\partial \mathbf{\Phi_{v_k}}}{\partial x} +  
    \frac{\partial \mathbf{\Phi_{v_j}}}{\partial x} \frac{\partial \mathbf{\Phi_{u_k}}}{\partial y} + 
    \frac{\partial \mathbf{\Phi_{u_j}}}{\partial y} \frac{\partial \mathbf{\Phi_{v_k}}}{\partial x} + 
    \frac{\partial \mathbf{\Phi_{u_j}}}{\partial y} \frac{\partial \mathbf{\Phi_{u_k}}}{\partial y}
    \Bigg) \Bigg] \,dx \,dy
\end{aligned}
\end{equation}

The initial condition $\mathbf{a^0}$ of Eq. \ref{galerkin_ode_cav} is obtained by projection of the first snapshot on the basis vector as
\begin{equation}
a_{i}^0 = \iint_{\Omega} \left(
\mathbf{(\zeta_0 - \bar{\zeta}) \Phi_{\zeta_i}} +
\mathbf{(u_0 - \bar{u}) \Phi_{u_i}} +
\mathbf{(v_0 - \bar{v}) \Phi_{v_i}} +
\mathbf{(p_0 - \bar{p}) \Phi_{p_i}}
\right) \,dx \,dy  \mbox{ .}
\end{equation}




\bibliographystyle{model1-num-names}
\bibliography{mybibfile}

\end{document}